%% file: ANA-EXOT-2021-36-PAPER.tex
\pdfoutput=1
\documentclass[atlasdraft=false, cernpreprint, UKenglish, texmf, orcidlogo]{atlasdoc}
 
\usepackage{atlaspackage}
\usepackage{atlasbiblatex}
 
\usepackage{atlasphysics}

\graphicspath{{logos/}{figures/}}
 
\usepackage{ANA-EXOT-2021-36-PAPER-defs}

\AtlasTitle{Search for vector-boson resonances decaying into a top quark and
a bottom quark using $pp$ collisions at $\sqrt{s} = 13$~TeV with the ATLAS detector}

\AtlasAbstract{
A search for a new massive charged gauge boson, $W'$, is performed
with the ATLAS detector at the LHC. The dataset used in this analysis was collected from
proton--proton collisions at a centre-of-mass energy of
$\sqrt{s} =13$~TeV, and corresponds to an integrated luminosity of
139~fb$^{-1}$. The reconstructed $tb$ invariant mass is used to search for a $W'$ boson decaying into a top quark and a bottom quark.
The result is interpreted in terms of a $W'$ boson with purely
right-handed or left-handed chirality in a mass range of 0.5--6~TeV.
Different values for the coupling of the $W'$ boson to the top and bottom
quarks are considered, taking into account interference with
single-top-quark production in the $s$-channel. No significant deviation from the background
prediction is observed. The results are expressed as upper limits on the $W' \rightarrow
tb$ production cross-section times branching ratio as a
function of the $W'$-boson mass and in the plane of the coupling vs the $W'$-boson mass.
}

\AtlasRefCode{EXOT-2021-36}
 
\PreprintIdNumber{CERN-EP-2023-149}

\AtlasJournal{\JHEP}
\AtlasJournalRef{\JHEP 12 (2023) 073}
\AtlasDOI{10.1007/JHEP12(2023)073}

\hypersetup{pdftitle={ATLAS document},pdfauthor={The ATLAS Collaboration}}
 
\newcommand{\AtlasCoordFootnote}{
ATLAS uses a right-handed coordinate system with its origin at the nominal interaction point (IP)
in the centre of the detector and the \(z\)-axis along the beam pipe.
The \(x\)-axis points from the IP to the centre of the LHC ring,
and the \(y\)-axis points upwards.
Cylindrical coordinates \((r,\phi)\) are used in the transverse plane,
\(\phi\) being the azimuthal angle around the \(z\)-axis.
The pseudorapidity is defined in terms of the polar angle \(\theta\) as \(\eta = -\ln \tan(\theta/2)\).
Angular distance is measured in units of \(\Delta R \equiv \sqrt{(\Delta\eta)^{2} + (\Delta\phi)^{2}}\).}
 
\begin{document}
 
\maketitle
 
\tableofcontents

% The next lines are included from the .//TeX/introduction.tex input file
\section{Introduction}
\label{sec:intro}
 
Multiple theories beyond the Standard Model (SM) involve enhanced symmetries that predict new gauge bosons, usually referred to as $W'$ or $Z'$~bosons. The $W'$~boson is the mediator of a new charged vector current and can be massive enough to decay into a top quark and a bottom quark. Many models, such as those with extra dimensions~\cite{Dienes_1999}, strong dynamics~\cite{PhysRevD.19.1277,PhysRevD.20.2619,DIMOPOULOS1979237,EICHTEN1980125}, or a composite Higgs boson~\cite{Muller_1996}, predict new vector charged-current interactions. Some models predict $W'$~bosons that preferentially couple to third-generation particles~\cite{Burdman_2006,Malkawi_1996,PhysRevD.10.275,Hill_1995} and are only observable in third-generation decay modes. Some of those models predict $W'$~bosons that can only couple to quarks and are therefore not observable in leptonic decay modes~\cite{Malkawi_1996,PhysRevD.10.275}.
 
In the Sequential Standard Model (SSM)~\cite{Altarelli:1989ff}, an effective Lagrangian is used to capture the phenomenology of a $W'$~boson decaying into a top quark and bottom quark ($W'\rightarrow tb$), which includes $W'^+ \to t\bar{b}$ and  $W'^- \to \bar{t}b$~\cite{Sullivan_2002,Duffty_2012}. This effective Lagrangian has a $W'$~boson with the same coupling structure as the SM $W$~boson. It has three free parameters: the mass of the $W'$~boson, the chirality of the interaction, and an overall strength parameter that multiplies the fermion couplings of the new boson, which makes it possible to study $W'$~bosons with different widths. This choice of Lagrangian has a wide applicability: in many beyond-the-SM theories predicting a $W'$~boson, the top-quark phenomenology is independent of the light quarks because of its high mass.
 
\begin{figure}[htbp]
\centering
\subfloat[]{\includegraphics[width=0.45\textwidth]{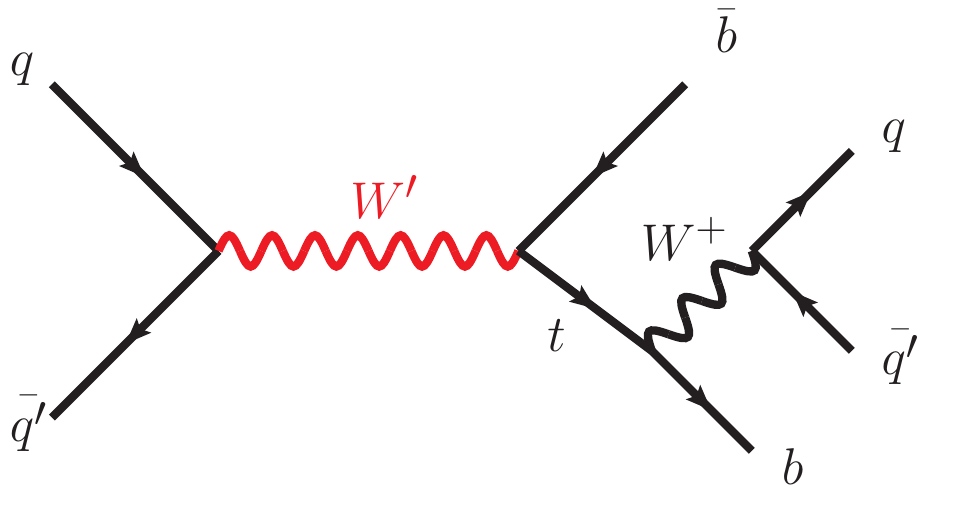}}
\subfloat[]{\includegraphics[width=0.45\textwidth]{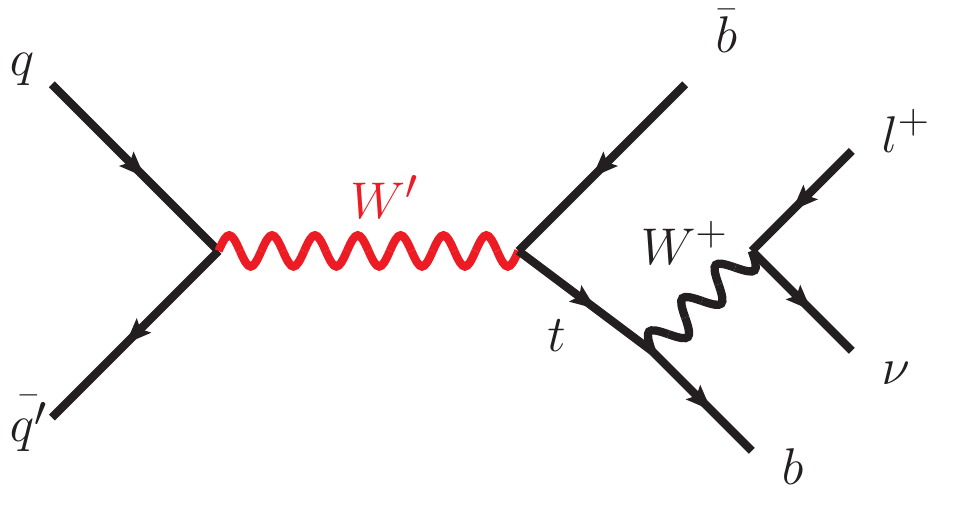}}
\caption{Representative leading-order Feynman diagrams for $s$-channel $W'$-boson production with decay into $tb$, for (a) a top quark decaying into a $W$~boson that decays hadronically (all-hadronic decay mode) and (b) a top quark decaying into a $W$~boson that decays into a lepton and neutrino (lepton+jets decay mode).}
\label{fig:fey}
\end{figure}

Figure~\ref{fig:fey} shows the leading-order (LO) Feynman diagram for $W'$-boson production and its decay into $tb$. The top quark decays into a $W$~boson and a bottom quark, with the $W$~boson subsequently decaying either into quarks (Figure~\ref{fig:fey}(a), all-hadronic decay mode) or into a lepton and a neutrino (Figure~\ref{fig:fey}(b), lepton+jets decay mode).
Two chirality scenarios are considered for the $W'$~boson: right-handed chirality and left-handed chirality. A $W'$~boson with  right-handed chirality couples only to right-handed fermions. Its production and decay does not interfere with any SM processes. A $W'$~boson with left-handed chirality couples only to left-handed fermions. Its production and decay interferes with the SM $s$-channel single-top-quark process, where the $W$~boson replaces the $W'$~boson in Figure~\ref{fig:fey}. In this paper, the reconstructed mass of the $tb$ system is used to search for the $W'$-boson signal in both of these scenarios. In the absence of a signal, limits are set on the $W'$-boson production cross-section times branching ratio $W' \rightarrow tb$ as a function of the mass of the $W'$~boson. For the scenario with right-handed chirality, the mass of the right-handed neutrino is assumed to be much higher than that of the right-handed $W'$~boson, so the $W'$~boson cannot decay leptonically~\cite{Duffty_2012}. As a result, the branching ratio for decay of a right-handed $W'$~boson into $tb$ is about 10\% higher than that for a left-handed $W'$~boson with the same mass, which can also decay into a lepton and a neutrino.

Searches for a $W'$~boson decaying into $tb$ have been performed at the Tevatron~\cite{Abazov:2011xs,CDF:2015aej} and the Large Hadron Collider (LHC)~\cite{CMS-EXO-12-001,CMS-B2G-12-010,CMS-B2G-17-010,TOPQ-2012-19,EXOT-2013-14,EXOT-2017-02,EXOT-2016-18,CMS-B2G-20-005}. The most recent search by the CMS Collaboration, using $\sqrt{s}=13$~\TeV\ proton--proton ($pp$) collision data with an integrated luminosity of 137~fb$^{-1}$ and targeting the all-hadronic decay mode, excluded a right-handed or left-handed $W'$~boson with a mass below 3.4~\TeV~\cite{CMS-B2G-20-005}. Previous searches by the ATLAS Collaboration for $W' \rightarrow tb$, using 36.1~fb$^{-1}$ of $pp$ collision data at $\sqrt{s}=13$~\TeV, excluded a right-handed $W'$~boson with a mass below 3.25~\TeV\ by combining the all-hadronic and lepton+jets decay modes~\cite{EXOT-2017-02,EXOT-2016-18} and assuming a $W'$-boson coupling equal to the SM $W$ coupling.
 
This paper presents a search for $W'$~bosons using the full Run~2 dataset collected by the ATLAS detector. Compared to the previous ATLAS analyses~\cite{EXOT-2017-02,EXOT-2016-18}, this search has improved top-quark and $b$-quark identification, better multi-jet background estimation and a refined selection strategy. The search is performed in both the all-hadronic (\Zerol) channel and the lepton+jets (\Onel, either electron or muon) channel. Tau-leptons are not considered explicitly in either channel, and the electrons and muons are simply referred to as \enquote{leptons} in this paper. The \Onel analysis allows studies of the lower transverse momentum (\pt) region, which is out of reach for the \Zerol channel's trigger selection. In exchange, the \Zerol channel provides optimal sensitivity to high $W'$ masses.
 
The paper is organised as follows. The ATLAS detector at the LHC is described in Section~\ref{sec:detector}. Section~\ref{sec:samples} provides details of the data and simulated event samples. Object reconstruction is described in Section~\ref{sec:objects}. The analysis strategy, including event selection and categorisation, is described in Sections~\ref{sec:strategy_0l} and~\ref{sec:strategy_1l} for the \Zerol and \Onel channels, respectively. The background estimation for the two channels is described in Sections~\ref{sec:background_0l} and~\ref{sec:background_1l}. Systematic uncertainties considered in the statistical analysis are discussed in Section~\ref{sec:syst}. The results of the fit to data are presented in Section~\ref{sec:results}, and Section~\ref{sec:summary} provides the conclusions.

% End of text imported from the .//TeX/introduction.tex input file

% The next lines are included from the .//TeX/detector.tex input file
\section{ATLAS detector}
\label{sec:detector}
 
The ATLAS detector~\cite{PERF-2007-01} at the LHC covers nearly the entire solid angle around the collision point.\footnote{\AtlasCoordFootnote}
It consists of an inner tracking detector surrounded by a thin superconducting solenoid, electromagnetic and hadron calorimeters,
and a muon spectrometer incorporating three large superconducting air-core toroidal magnets.
 
The inner-detector system (ID) is immersed in a \SI{2}{\tesla} axial magnetic field
and provides charged-particle tracking in the range \(|\eta| < 2.5\).
The high-granularity silicon pixel detector covers the vertex region and typically provides four measurements per track,
the first hit normally being in the insertable B-layer (IBL) installed before Run~2~\cite{ATLAS-TDR-19,PIX-2018-001}.
It is followed by the silicon microstrip tracker, which usually provides eight measurements per track.
These silicon detectors are complemented by the transition radiation tracker (TRT),
which enables radially extended track reconstruction up to \(|\eta| = 2.0\).
The TRT also provides electron identification information
based on the fraction of hits (typically 30 in total) above a higher energy-deposit threshold corresponding to transition radiation.
 
The calorimeter system covers the pseudorapidity range \(|\eta| < 4.9\).
Within the region \(|\eta|< 3.2\), electromagnetic calorimetry is provided by barrel and
endcap high-granularity lead/liquid-argon (LAr) calorimeters,
with an additional thin LAr presampler covering \(|\eta| < 1.8\)
to correct for energy loss in material upstream of the calorimeters.
Hadron calorimetry is provided by the steel/scintillator-tile calorimeter,
segmented into three barrel structures within \(|\eta| < 1.7\), and two copper/LAr hadron endcap calorimeters.
The solid angle coverage is completed with forward copper/LAr and tungsten/LAr calorimeter modules
optimised for electromagnetic and hadronic energy measurements respectively.
 
The muon spectrometer comprises separate trigger and
high-precision tracking chambers measuring the deflection of muons in a magnetic field generated by the superconducting air-core toroidal magnets.
The field integral of the toroids ranges between \num{2.0} and \SI{6.0}{\tesla\metre}
across most of the detector.
Three layers of precision chambers, each consisting of layers of monitored drift tubes, cover the region \(|\eta| < 2.7\),
complemented by cathode-strip chambers in the forward region, where the background is highest.
The muon trigger system covers the range \(|\eta| < 2.4\) with resistive-plate chambers in the barrel, and thin-gap chambers in the endcap regions.
 
Interesting events are selected by the first-level trigger system implemented in custom hardware,
followed by selections made by algorithms implemented in software in the high-level trigger~\cite{TRIG-2016-01}.
The first-level trigger accepts events from the \SI{40}{\MHz} bunch crossings at a rate below \SI{100}{\kHz},
which the high-level trigger reduces in order to record events to disk at about \SI{1}{\kHz}.
 
An extensive software suite~\cite{ATL-SOFT-PUB-2021-001} is used in data simulation, in the reconstruction and analysis of real
and simulated data, in detector operations, and in the trigger and data acquisition systems of the experiment.

% End of text imported from the .//TeX/detector.tex input file

% The next lines are included from the .//TeX/samples.tex input file
\section{Data and simulated event samples}
\label{sec:samples}
 
This analysis is performed using data from $pp$ collisions at $\sqrt{s}=13~\tev$ collected with the ATLAS detector during Run~2, from 2015 to 2018. After applying a number of criteria to ensure that the detector was in good operating condition~\cite{DAPR-2018-01}, the data used in this analysis have an integrated luminosity of 139~$\ifb$.
 
Monte Carlo (MC) event generators were used to simulate signal and background events.
The generation of all simulated event samples includes the effect of multiple $pp$ interactions per bunch crossing, as well as changes in detector response due to interactions in bunch crossings before or after the one containing the hard interaction, modelled by overlaying simulated inelastic events on the physics event. These two effects are referred to as pile-up. The simulated event samples were processed with the \GEANT-based ATLAS detector simulation~\cite{Agostinelli:2002hh,SOFT-2010-01}.
 
All samples are weighted to match the pile-up distribution observed in data and are processed with the same reconstruction algorithms as data~\cite{ATLAS-CONF-2019-021}.

\subsection{Signal and interference samples}
\label{sec:sigsamples}
 
Signal events were generated at LO in QCD with \MGNLO[2.6.7]~\cite{Alwall:2014hca}, using a chiral $W'$-boson model that implements the effective Lagrangian described in Section~\ref{sec:intro}. In this model the coupling strength of the $W'$~boson to right- or left-handed fermions ($g'$) can be freely scaled relative to the SM coupling ($g$) by an arbitrary factor. Only purely right-handed or purely left-handed $W'$~bosons are considered. The right-handed $W'$~boson cannot decay to leptons because the right-handed neutrino is assumed to be more massive than the $W'$~boson. \MGNLO was also used to decay the top quark and $W$~boson, with spin correlations taken into account. \PYTHIA[8.244]~\cite{Sjostrand:2014zea} was used for the modelling of the parton shower, fragmentation and underlying event. The \textsc{PDF4LHC15} set of parton distribution functions (PDF)~\cite{Butterworth:2015oua} and a set of tuned parameters called the A14 tune~\cite{ATL-PHYS-PUB-2014-021} were used for the event generation. Signal samples were normalised to the next-to-leading-order (NLO) cross-section computed by \textsc{ZTOP}~\cite{Duffty_2012}. Several contributions to the uncertainty in the NLO cross-section are considered for each coupling value.  An uncertainty accounting for missing higher-order terms is estimated by doubling and halving both the renormalisation and factorisation scales independently. Uncertainties associated with the choice of PDF set and strong coupling constant value are obtained using the \textsc{PDF4LHC15} PDF set. Finally, an uncertainty due to the choice of top-quark mass value (172.5~\GeV) is obtained by raising and lowering the chosen value by 1~\GeV. The NLO/LO cross-section normalisation ratios ($K$-factors) range from 1.3 to 1.4, depending on the mass of the $W'$~boson. The width of the $W'$~boson is about 3\% of its mass for a coupling equal to the SM coupling ($g'/g=1$) and scales with $(g'/g)^2$; it was calculated at NLO with \textsc{ZTOP}.
 
Signal samples were generated in 0.5~\TeV\ steps for $W'$-boson masses between 0.5~and 6.0~\TeV\ for lepton+jets top-quark decays and between 1.5~and 6.0~\TeV\ for all-hadronic top-quark decays. Samples corresponding to right-handed and left-handed chiralities were produced separately. The coupling in the event generation was set to $g'/g=2.0$. Weights were computed by \MGNLO during the parton-level event generation to reweight each sample to coupling values between $g'/g=0.1$ and $g'/g=0.5$ in steps of 0.1 and between $g'/g=0.5$ and $g'/g=5.0$ in steps of 0.5. Masses below 1.5~\TeV\ were not generated in the all-hadronic case because the trigger selection utilised in the \Zerol channel is completely inefficient in that region of phase space.
 
Interference between left-handed $W'$-boson production and SM single-top-quark production in the $s$-channel was modelled by reweighting the nominal signal samples using a parameterisation of the ratio of $W'$~boson production to the interference contributions as a function of the parton-level invariant mass of the $tb$ system~\cite{Boos:2006xe}. The interference effects are destructive on the low-mass side of the $W'$ mass peak and constructive on the high-mass side. Their size and relative importance depends strongly on the mass and coupling values considered, but their effect on the results shown in this paper is small.

\subsection{Background samples for the \Zerol channel}
\label{sec:samples-0lep}
 
The dominant SM background process for all-hadronic events is QCD multi-jet production. This background is estimated with data-driven methods as described in Section~\ref{sec:background_0l}.
The second most important background is top-quark-pair production ($\ttbar$), with an inclusive cross-section of $\mathrm{832\pm51~pb}$ for a top-quark mass of $172.5$~\GeV, as obtained from calculations at next-to-next-to-leading order (NNLO) in QCD including the resummation of next-to-next-to-leading logarithmic (NNLL) soft-gluon terms with \TOPpp[2.0]~\cite{Beneke:2011mq,Cacciari:2011hy,Baernreuther:2012ws,Czakon:2012zr,Czakon:2012pz,Czakon:2013goa,Czakon:2011xx}. Other small backgrounds, such as $V$+jets ($V=W$~or~$Z$~boson) or single-top production, are accounted for in the data-driven multi-jet estimate.
 
The production of \ttbar events was modelled using the \POWHEGBOX[v2]~\cite{Frixione:2007nw,Nason:2004rx,Frixione:2007vw,Alioli:2010xd} generator at NLO with the \NNPDF[3.0nlo]~\cite{Ball:2014uwa} PDF set and the \hdamp parameter\footnote{The \hdamp parameter is a resummation damping factor and one of the parameters that controls the matching of \POWHEG matrix elements to the parton shower and thus effectively regulates the high-\pt radiation against which the \ttbar system recoils.} set to 1.5\ times the mass of the top quark~\cite{ATL-PHYS-PUB-2016-020}. The events were passed to \PYTHIA[8.230] to model the parton shower, hadronisation, and underlying event, with parameter values set according to the A14 tune and using the \NNPDF[2.3lo] PDF set~\cite{Ball:2012cx}. The decays of bottom and charm hadrons were performed by \EVTGEN[1.6.0]~\cite{Lange:2001uf}.
 
Two $\ttbar$ background contributions are considered: all-hadronic $\ttbar$ events where both $W$~bosons decay into quarks, resulting in a signature that is similar to the signal, and non-all-hadronic $\ttbar$ events where at least one of the $W$~bosons decays leptonically. These events can contribute to the background in two ways: when none of the charged leptons from $W$-boson decays are identified or through hadronic tau decays in $W \rightarrow \tau\nu$ events.
 
The modelling of the $\ttbar$ background is improved by correcting the $\ttbar$ samples so that the top-quark $\pT$ distribution matches that predicted at NNLO in QCD and NLO EW accuracy~\cite{Czakon:2017wor}. The corrections entail an implicit change in the PDF set and top-quark mass value considered with respect to those used in sample generation. The NNLO differential calculations are performed using the \NNPDF[3.0qed] PDF set and a top-quark mass of $173.3$~\GeV.

\subsection{Background samples for the \Onel channel}
 
The largest background in the lepton+jets channel is $\ttbar$ production, already described for the \Zerol channel in Section~\ref{sec:samples-0lep}. Other important backgrounds arise from $V$+jets production, especially $W$+jets in which the $W$~boson decays leptonically. Other subdominant backgrounds such as single-top-quark and multi-boson production were also considered. Finally, a small multi-jet contribution was also taken into account and was estimated with data-driven methods as described in Section~\ref{sec:background_1l}.
 
The production of \(V+\)jets was simulated with the \SHERPA[2.2.11]~\cite{Bothmann:2019yzt} generator using NLO matrix elements for up to two partons, and LO matrix elements for up to four partons, calculated with the Comix~\cite{Gleisberg:2008fv} and \OPENLOOPS~\cite{Buccioni:2019sur,Cascioli:2011va,Denner:2016kdg} libraries. They were matched with the \SHERPA parton shower~\cite{Schumann:2007mg} using the \MEPSatNLO prescription~\cite{Hoeche:2011fd,Hoeche:2012yf,Catani:2001cc,Hoeche:2009rj} with the set of tuned parameters developed by the \SHERPA authors. The \NNPDF[3.0nnlo] set of PDFs was used and the samples were normalised to the NNLO prediction~\cite{Anastasiou:2003ds}.
 
The three single-top production modes ($s$-channel, $t$-channel, and $tW$-channel) were considered. They were modelled with the \POWHEGBOX[v2]~\cite{Frederix:2012dh,Nason:2004rx,Frixione:2007vw,Alioli:2010xd} generator at NLO in QCD, using the five-flavour scheme (four-flavour scheme for $t$-channel production) and the corresponding \NNPDF[3.0nlo] set of PDFs. The events were interfaced with \PYTHIA[8.230], which used the A14 tune and the \NNPDF[2.3lo] set of PDFs. The diagram removal scheme~\cite{Frixione:2008yi} was used to remove interference and overlap between the $tW$-channel and \ttbar production.
 
Samples of diboson final states (\(VV\)) were simulated with the \SHERPA[2.2.1]
generator, including off-shell effects and Higgs boson contributions where appropriate. Fully leptonic final states  (where both bosons decay leptonically) and lepton+jets final states (where one decays leptonically and the other hadronically) were generated using matrix elements at NLO accuracy in QCD for up to one additional parton emission and at LO accuracy for up to three additional parton emissions. Samples for loop-induced \(gg \to VV\)  processes were generated using matrix elements calculated at LO accuracy for up to one additional parton emission.
The matrix element calculations were matched and merged with the \SHERPA parton shower based on Catani--Seymour dipole factorisation~\cite{Gleisberg:2008fv,Schumann:2007mg} using the MEPS@NLO prescription. The virtual QCD corrections were provided by the \OPENLOOPS library. The \NNPDF[3.0nnlo] set of PDFs was used, along with the dedicated set of tuned parton-shower parameters developed by the \SHERPA authors.

% End of text imported from the .//TeX/samples.tex input file

% The next lines are included from the .//TeX/objects.tex input file
\section{Object reconstruction}
\label{sec:objects}
 
The signal process $W'\rightarrow tb$ targeted in this search results in a final state with a high-$\pt$ top quark and a high-$\pt$ $b$-quark. The $b$-quark is reconstructed as a small-radius (small-$R$) jet, while the reconstruction of the top quark depends on the decay mode of the $W$~boson from the top-quark decay. In the \Zerol channel, the $W$~boson decays hadronically, and the top quark is reconstructed as a high-$\pt$ large-radius (large-$R$) jet. In the \Onel channel, the $W$~boson decays leptonically, and the top quark is reconstructed from the lepton  (electron or muon), the missing transverse momentum, and a small-$R$ jet.
 
For each event, collision vertices are reconstructed from inner-detector tracks with $\pT > 0.5$~\GeV. The primary vertex in each event is chosen to be the one with the largest sum of the squared transverse momenta of all associated tracks.
 
Large-$R$ jets are built from three-dimensional topological clusters of energy deposits in the calorimeter, which are calibrated to the hadronic energy scale with the local cluster weighting (LCW) procedure~\cite{PERF-2014-07}. The \antikt~\cite{Cacciari:2008gp,Fastjet} algorithm with radius parameter $R=1.0$ is used to reconstruct large-$R$ jets. These jets are trimmed~\cite{Krohn_2010} to reduce contributions from pile-up and soft interactions by reclustering the jet constituents into subjets using the \kt~algorithm~\cite{CATANI1993187,Ellis_1993} with a radius parameter $R=0.2$ and discarding constituents belonging to subjets with $\pT$ less than 5\% of the $\pT$ of the parent jet. The large-$R$ jet four-momentum is then recomputed from the four-momenta of the remaining constituents and corrected using simulation and data~\cite{JETM-2018-02}. Only large-$R$ jets with $|\eta| < 2.0$ and $\pt > 500$~\GeV\ are considered in this analysis.
 
Large-$R$ jets are identified as containing a hadronically decaying top quark (henceforth called a top-tagged jet) using a multivariate classification algorithm implemented as a deep neural network (DNN)~\cite{JETM-2018-03}. In most of the kinematic region of interest in the \Zerol channel, a single large-$R$ jet captures the top-quark decay products, resulting in a characteristic three-prong substructure within the jet, in contrast to a typical one-prong substructure associated with jets in multi-jet background processes. The DNN uses multiple features of the jet as inputs, e.g.\ calibrated jet \pt and mass, information about the dispersion of the jet constituents such as N-subjettiness~\cite{Thaler:2010tr}, splitting scales~\cite{Thaler:2008ju}, and energy correlation functions~\cite{Larkoski:2013eya}. A DNN score between zero and one is obtained, with top-quark-initiated jets having values close to one and light-parton-initiated jets (including gluon jets) having values close to zero.
 
The top-tagging algorithm used in this analysis is optimised for top-quark-initiated jets that satisfy the \enquote{contained} criteria, where most of the top-quark decay products are contained inside the large-$R$ jet~\cite{ATL-PHYS-PUB-2020-017}. The criteria are defined in the simulation as follows. First, trimmed large-$R$ jets are built at particle level from all stable particles (with $c\tau > 10$~mm), excluding muons and neutrinos, and using a radius parameter $R=1.0$. This trimmed particle-level jet must be matched to a generator-level top quark within $\Delta R < 0.75$ and have a mass larger than 140~\GeV. At least one $b$-hadron must be associated with the jet~\cite{Cacciari_2008}. Finally, a detector-level large-$R$ jet is considered contained if it is within $\Delta R =0.75$ of such a particle-level jet.
 
Two different efficiency working points, based on the DNN score, are used to define the signal regions in the \Zerol channel: one in which the requirements correspond to a top-tagging efficiency of 80\% (DNN score cut of $\sim$0.6--0.7, depending on \pT, to keep the efficiency constant), and a tighter one in which they correspond to an efficiency of 50\% (DNN score cut of $\sim$0.9, depending on \pT, to keep the efficiency constant). Both efficiencies are calculated using simulated \ttbar~events. The corresponding light-jet rejection factors are between 10 and 40 (80\% working point) and between 30 and 150 (50\% working point) depending on \pT. Scale factors are used to correct for possible efficiency differences between simulated event samples and data~\cite{JETM-2018-03}. Two additional efficiency working points are used to define control regions and to estimate the multi-jet background in the \Zerol channel. A DNN score boundary of $\text{e}^{-4}$ is used to divide the events in the control regions used for background estimation into two roughly equal-size samples, and a very loose DNN score cut of $\text{e}^{-7}$ is used in the definition of the top-proxy jets that are used in the multi-jet background estimation (Section~\ref{sec:regions_0l}). These working points have efficiencies higher than 95\% and light-jet rejection factors ranging between 1.5 and 2.5 approximately.
 
Small-$R$ jets are reconstructed by applying the \antikt algorithm with a radius parameter $R=0.4$ to inner-detector tracks associated with the primary vertex and calorimeter clusters selected by a particle-flow reconstruction algorithm~\cite{PERF-2015-09}. An energy calibration is applied to both the input calorimeter clusters~\cite{PERF-2014-07} and the final reconstructed jets~\cite{JETM-2018-05}. The latter takes into account both pile-up effects and flavour dependencies. Only small-$R$ jets with $|\eta| < 2.5$ and $\pt > 25$~\GeV\ are considered in this analysis. To reject jets arising from pile-up, a jet-vertex-tagging technique using a multivariate likelihood~\cite{PERF-2014-03} is applied to jets with $\pt < 60$~\GeV, ensuring that selected jets are matched to the primary vertex.
 
Small-$R$ jets are identified as containing a $b$-hadron (henceforth called $b$-tagged) using the \enquote{DL1r} algorithm~\cite{FTAG-2018-01,FTAG-2019-07}. This algorithm is based on a multivariate classification technique with a DNN combining information about the impact parameters of tracks and topological properties of secondary and tertiary decay vertices reconstructed from the tracks associated with the jet. In this analysis, the $b$-tagged jets are selected by using a working point corresponding to an efficiency of 85\% for identifying true $b$-jets in simulated \ttbar~events. Light-jet rejection factors range between 20 and 50, depending on \pT~\cite{FTAG-2019-07}. Scale factors are used to correct for possible differences between the $b$-tagging efficiencies in simulated events and data events~\cite{FTAG-2018-01,FTAG-2020-08,FTAG-2019-02}.
 
A third kind of jet is used in the \Onel channel to reject \ttbar~events containing hadronically decaying top quarks without using the previously defined traditional large-$R$ jets. They are obtained by reclustering~\cite{Nachman:2014kla} small-$R$ jets (passing the aforementioned selection) with a variable-$R$ \antikt algorithm with a density parameter $\rho$ of 350~\GeV\ and a maximum radius of 1.0~\cite{Krohn:2009zg}. Since the inputted constituent small-$R$ jets are fully calibrated, their calibration and uncertainties can be propagated directly to the reclustered jet, and no further calibration step is necessary. In order to suppress contributions from pile-up and soft radiation, the reclustered variable-$R$ jets are trimmed by removing all associated small-$R$ jets that have $\pt$ below 5\% of the $\pt$ of the reclustered jet. These jets, henceforth referred to as vRC-jets, are required to have $\pt > 100$~\GeV\ and $|\eta|<2.0$.
 
Electron candidates are reconstructed from energy deposits in the electromagnetic (EM) calorimeter that are matched to charged-particle tracks in the ID~\cite{EGAM-2018-01}. They are required to satisfy $\pt > 25$~\GeV\ and $|\eta| < 2.47$, excluding the transition region between the barrel and endcap EM calorimeters $(1.37 < |\eta| < 1.52)$. They are identified using the \enquote{tight} likelihood identification operating point~\cite{EGAM-2018-01}. The number of hits in the innermost pixel layer, the IBL, is used to discriminate between electrons and converted photons, and the longitudinal impact parameter $z_{0}$ relative to the primary vertex is required to satisfy $|\zzsth| < \SI[parse-numbers=false]{0.5}{\mm}$. The significance of the transverse impact parameter $d_{0}$ must satisfy $|d_{0}/ \sigma_{d_{0}}| < 5 $. Electrons are also required to be isolated from other activity in the tracking and calorimeter systems, using the \enquote{FCTight} isolation working point~\cite{EGAM-2018-01}. The isolation criteria must be satisfied in a cone of size $\Delta R = 0.2$ around the electron in the calorimeter and a cone of $\pt$-dependent size in the ID. The latter choice improves the performance for electrons produced in the decay of high-\pt particles.
 
Muon candidates are reconstructed from matching tracks in the ID and the muon spectrometer, refined by a global fit which makes use of the hits in both subdetectors~\cite{MUON-2018-03}. Muons must have $\pt > 25$~\GeV\ and $|\eta| < 2.5$, and satisfy the \enquote{medium} identification criteria~\cite{MUON-2018-03}. Like the electrons, their longitudinal impact parameter is required to satisfy $|\zzsth| < \SI[parse-numbers=false]{0.5}{\mm}$. The significance of the transverse impact parameter $d_{0}$ must satisfy $|d_{0}/\sigma(d_{0})| < 3$. Muons are required to be isolated from other activity in the tracking system, using the \enquote{TightTrackOnly} isolation working point~\cite{MUON-2018-03}. Similarly to the electrons, the isolation criterion must be satisfied in a cone of $\pt$-dependent size around the muon in the ID.
 
For both the electrons and muons, correction factors are applied to compensate for differences between data and simulation in trigger, reconstruction efficiency, particle identification, and isolation, usually as a function of relevant kinematic variables.
 
To resolve any reconstruction ambiguities between electrons, muons and jets, an overlap removal procedure is applied in a prioritised sequence as follows. First, if an electron shares the same ID track with another electron, the electron with lower \pt is discarded. Any electron sharing the same ID track with a muon is rejected. Next, jets are rejected if they lie within $\Delta R = 0.2$ of an electron. Similarly, jets within $\Delta R = 0.2$ of a muon are rejected if the jet has fewer than three associated tracks or if the muon is matched to the jet through ghost association~\cite{Cacciari_2008}. Finally, electrons that are close to a remaining jet  are discarded if their distance from the jet is $\Delta R < 0.4$, while for muons the distance is $\Delta R < \min(0.4, 0.04 + \SI{10}{\GeV}/\pT)$.
 
The missing transverse momentum $\vec{p}_{\text{T}}^{\text{miss}}$, with magnitude \met, is calculated as the negative vectorial sum of the transverse momenta of all reconstructed physics objects (electrons, muons and jets)~\cite{ATLAS-CONF-2018-023} and the soft term. The soft term includes all tracks associated with the primary vertex but not matched to any reconstructed physics object. Only tracks associated with the primary vertex are considered, improving the \met resolution by suppressing the effect of pile-up.

% End of text imported from the .//TeX/objects.tex input file

% The next lines are included from the .//TeX/selection.tex input file
\section{Analysis strategy in the \Zerol channel}
\label{sec:strategy_0l}
 
Events containing at least one high-$\pt$ large-$R$ jet and one high-$\pt$ small-$R$ jet that do not overlap are selected, according to the decay products in the all-hadronic decay mode of the $tb$ final state. Events are separated into signal, control, validation and template regions based on the properties of the event's large-$R$ and small-$R$ jets. Signal regions are signal-enriched regions, with a top-tagged large-$R$ jet and a $b$-tagged small-$R$ jet. Template and control regions are used to estimate the multi-jet background. Template regions are used to obtain the initial shape of the reconstructed $tb$ mass distribution of the multi-jet background. Control regions are used to normalise those templates and obtain the final background distributions as well as to estimate their uncertainty. The validation region is used to validate the background estimation method.
 
\subsection{Event selection}
\label{sec:selection_0l}
Events are first selected at the trigger level by requiring at least one large-$R$ jet with $\pt$ exceeding a threshold which depends on the data-taking year: 360~\GeV\ for 2015, 420~\GeV\ for 2016, and 460~\GeV\ for 2017 and 2018. In order to perform the analysis in the regime where the trigger selection is fully efficient, events are required to have at least one reconstructed large-$R$ jet with $\pT> 500$~\GeV.
 
Events with noise bursts or coherent noise in the calorimeters are removed, as are events containing large energy deposits from non-collision or cosmic sources of background. Events without a reconstructed primary vertex are rejected.
 
Events containing charged leptons (electron or muon) are removed to ensure orthogonality to the \Onel channel. The definition of lepton candidates described in Section~\ref{sec:objects} is used for this veto, with the exception of the isolation requirement, which is dropped.

\subsection{Event categorisation}
\label{sec:regions_0l}
 
Selected events are categorised into regions according to the procedure outlined in Figure~\ref{fig:flowChart}, separately for events where the large-$R$ jet is $b$-tagged and for those where it is not, as described in Section~\ref{sec:categories_0l}.
 
\begin{figure}[ht]
\centering
\includegraphics[width=0.9\textwidth]{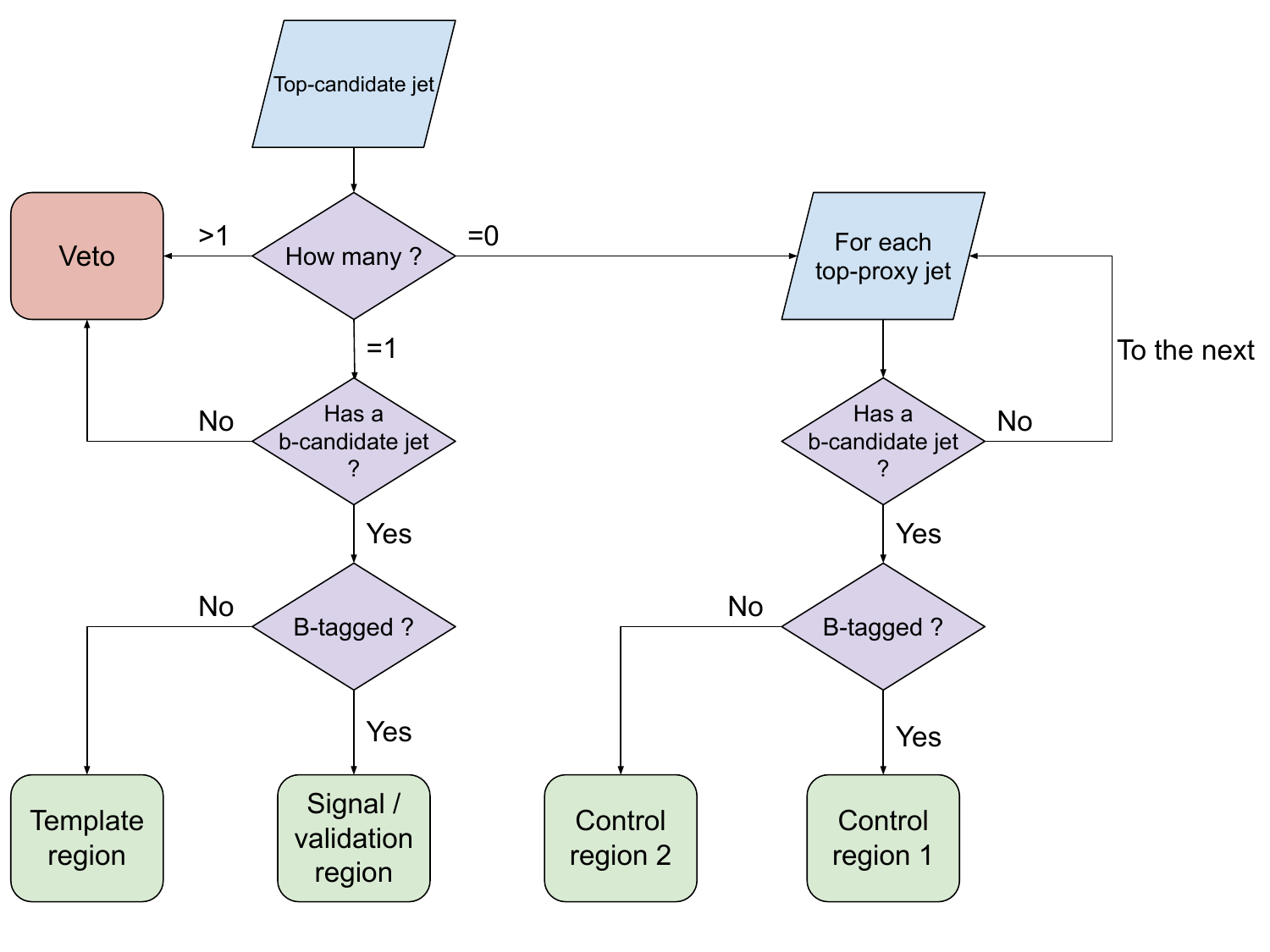}
\caption{Flow chart of the event categorisation, starting from events containing one or more large-$R$ jets that meet the preselection criteria. If that jet is top-tagged, using the 80\% efficiency working point, it is a top-candidate jet. If no top-candidate jet is found in the event, each large-$R$ jet becomes a top-proxy jet as long as it has a DNN score of at least $\text{e}^{-7}$. The $b$-candidate jet is the small-$R$ jet that is back-to-back with a top-candidate jet or top-proxy jet. Top-proxy jets not back-to-back with a $b$-candidate jet are skipped. }
\label{fig:flowChart}
\end{figure}

First, the number of top-candidate jets in an event is checked. A top-candidate jet is defined as a large-$R$ jet with $\pT>500$~\GeV\ that is top-tagged using the 80\% efficiency working point. Events with more than one such jet are vetoed, which reduces $\ttbar$ contamination. Events with exactly one top-candidate jet are kept and are considered for the signal regions, the validation region, and the template regions.
 
Events with no top-candidate jet are considered for the control regions. For this type of event, all large-$R$ jets with $\pT > 500$~\GeV\ and a top-tagging DNN score higher than $\text{e}^{-7}$ are considered. This minimal top-tagging DNN score requirement removes less than 5\% of events in data. The removed events show a flavour composition different from the rest when studied in simulation. Removing these events makes the flavour composition of jets more uniform in the control regions. The large-$R$ jets thus defined are referred to as top-proxy jets in the following. They are a good representation of the light-flavour and gluon-initiated jets that form the multi-jet background in the signal region.
 
Next, an attempt to find a $b$-candidate jet is made. Pairs consisting of a large-$R$ jet and a small-$R$ jet are formed using all of the top-candidate and top-proxy jets selected in the previous step. For each top-candidate or top-proxy jet ($J$), the leading small-$R$ jet ($j$) with $\Delta\phi(j,J) > 2.0$ and $\pT(j)>500$~\GeV\ is found. If an event has no pair formed in this way, it is rejected. The small-$R$ jet thus selected is the $b$-candidate jet for that specific top-candidate or top-proxy jet.
 
Events are also rejected if the pseudorapidity difference, $|\Delta\eta|$, between the top-candidate (or top-proxy) jet and its associated $b$-candidate jet is greater than 2.0. This selection requirement reduces the background from $t$-channel multi-jet processes, which are important in the high-$\pT$ regime.
 
Events with a top-candidate jet and a $b$-candidate jet that is $b$-tagged are assigned to the signal region  SR1, SR2 or SR3, or the validation region VR. Which region they are assigned to depends on the DNN score of the large-$R$ jet and the presence of additional $b$-tagged jets. This assignment is detailed in Section~\ref{sec:categories_0l} and Figure~\ref{fig:regions}.
 
Events with a top-candidate jet and a $b$-candidate jet that is not $b$-tagged but otherwise satisfies all criteria outlined above are assigned to the template regions TR1--TR4. These events are used to estimate the shape of the reconstructed $m_{tb}$ distribution of the multi-jet background.
 
Finally, events with at least one top-proxy jet are separated similarly. Pairs consisting of a top-proxy jet and a $b$-candidate jet are assigned to different regions according to whether the $b$-candidate jet is $b$-tagged: if it is, then the pair is assigned to control regions~1 (CR1a or CR1b); if not, then the pair is assigned to control regions~2 (CR2a or CR2b). Each pair of top-proxy and $b$-candidate jets is categorised independently for events with more than one such configuration. By using all possible pairs, any bias that could arise by having to choose only one large-$R$ jet as the top-proxy jet is avoided without reducing the number of pairs available in the CR. The statistical correlations introduced by this choice are negligible.
 
The variable of interest in this analysis is the reconstructed mass of the top-quark--bottom-quark system, $m_{tb}$, which is defined in all cases as the invariant mass of the top-candidate (or top-proxy) jet and its associated $b$-candidate jet.

\subsection{Further categorisation}
\label{sec:categories_0l}
 
The regions defined in Section~\ref{sec:regions_0l} are further refined, taking advantage of the top-tagging properties of the large-$R$ jet and the presence of $b$-tagged jets in the event. In particular, the top-quark decay leads to a bottom quark, which can be reconstructed and identified with the $b$-tagging algorithm. Events are therefore separated into those where the top-candidate jet is $b$-tagged (1-$b$-tag-in-top category), and those where it is not (0-$b$-tag-in-top category). A top-candidate (or top-proxy) jet $J$ is considered $b$-tagged if at least one $b$-tagged small-$R$ jet ($j$) is close to it ($\Delta R(j,J) < 1.0$). This corresponds to the expected signature of a  $W' \rightarrow tb$ decay, characterised by the presence of one large-$R$ jet that is top-tagged and $b$-tagged and is back-to-back with a small-$R$ jet that is $b$-tagged. The remaining events are in the 0-$b$-tag-in-top category; they include some signal events where the $b$-quark from the top-quark decay is not identified. The background estimation procedure is performed separately for these two categories (see Figure~\ref{fig:regions}).

\begin{figure}[htbp]
\centering
\includegraphics[width=0.8\textwidth]{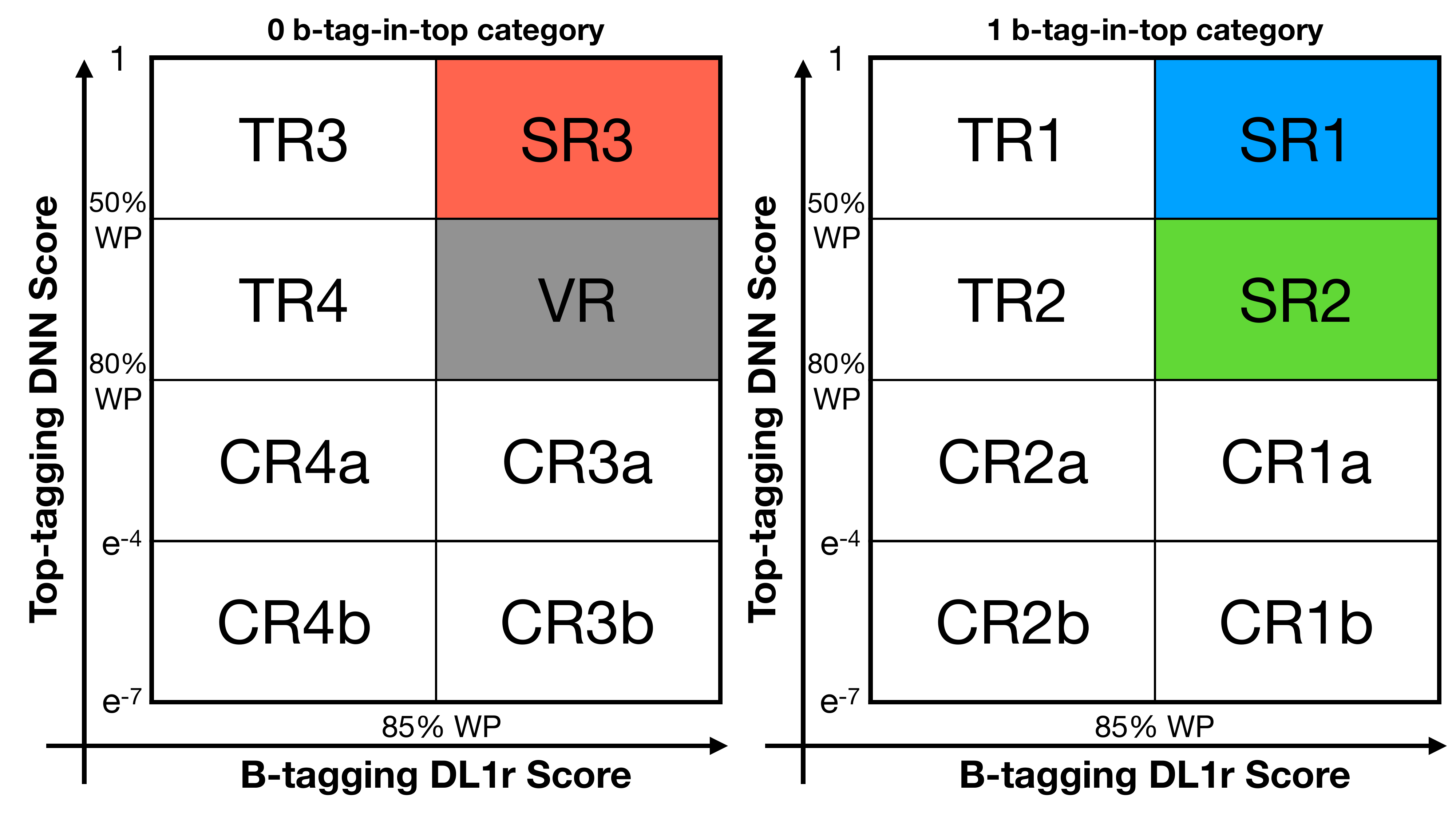}
\caption{ Events in the \Zerol channel are categorised according to the top-tagging and $b$-tagging status of the large-$R$ jets selected as top-candidate or top-proxy jets, and the $b$-tagging status of the small-$R$ jets selected as $b$-candidate jets.
Events are assigned to the right grid or the left grid, depending on whether the top-tagged or top-proxy jet is also $b$-tagged (1-$b$-tag-in-top category) or not (0-$b$-tag-in-top category).
Events with exactly one top-candidate jet are assigned to one of the top two rows depending on whether the top-candidate jet also fulfils the 50\% efficiency top-tagging working point. In the bottom two rows, each top-proxy jet from events without a top-candidate jet is considered. Events are further separated into columns based on the $b$-candidate jet: the right column if it passes the $b$-tag requirement, the left column if it does not.}
\label{fig:regions}
\end{figure}
 
In both the 0-$b$-tag-in-top and 1-$b$-tag-in-top categories, events are further categorised according to the top-tagging DNN score of the top-candidate (or top-proxy) jet and the $b$-tagging score of the $b$-candidate jet.
Signal-, validation- and template-region events in each category are assigned to one of the regions in the upper half of Figure~\ref{fig:regions}. Events with a top-candidate jet passing the 50\% efficiency working point are assigned to the upper row, while events with a top-candidate jet failing the 50\% efficiency working point but passing the 80\% efficiency working point are assigned to the second row from the top. Events in the signal regions and the validation region, with a $b$-tagged $b$-candidate jet, populate the top-right quadrant of each category, while events in the template regions populate the top-left quadrant.
 
Every pair consisting of a top-proxy jet and a $b$-candidate jet from events without a top-candidate jet is assigned to one of the control regions in the lower half of each category. Pairs in which the DNN score of the top-proxy jet is above $\text{e}^{-4}$ are assigned to the third row from the top, while the rest are assigned to the bottom row. Pairs with a $b$-tagged $b$-candidate jet are assigned to a region in the right column of each category, while events without one are assigned to a region in the left column.
 
The resulting regions are used in different ways:
 
\begin{itemize}
\item SR1, SR2 and SR3 are those where the signal-to-background ratio is the largest and the ones to be used in the statistical analysis described in Section~\ref{sec:results}.
\item VR is used to validate the data-driven multi-jet background estimation described in Section~\ref{sec:background_0l}.
\item TR1, TR2, TR3 and TR4 provide the initial template for the multi-jet background in SR1, SR2, SR3 and VR respectively.
\item CR1a and CR2a are used to obtain the multi-jet background in SR1 and SR2, while CR3a and CR4a are used to obtain the same background in SR3 and VR. They are also used to assess its uncertainty.
\item CR1b and CR2b are used to assess the uncertainty in the multi-jet background in regions SR1 and SR2, while CR3b and CR4b are used for the same uncertainty in regions SR3 and VR.
\end{itemize}

\begin{figure}[htb!]
\centering
\subfloat[]{\includegraphics[width=0.45\textwidth]{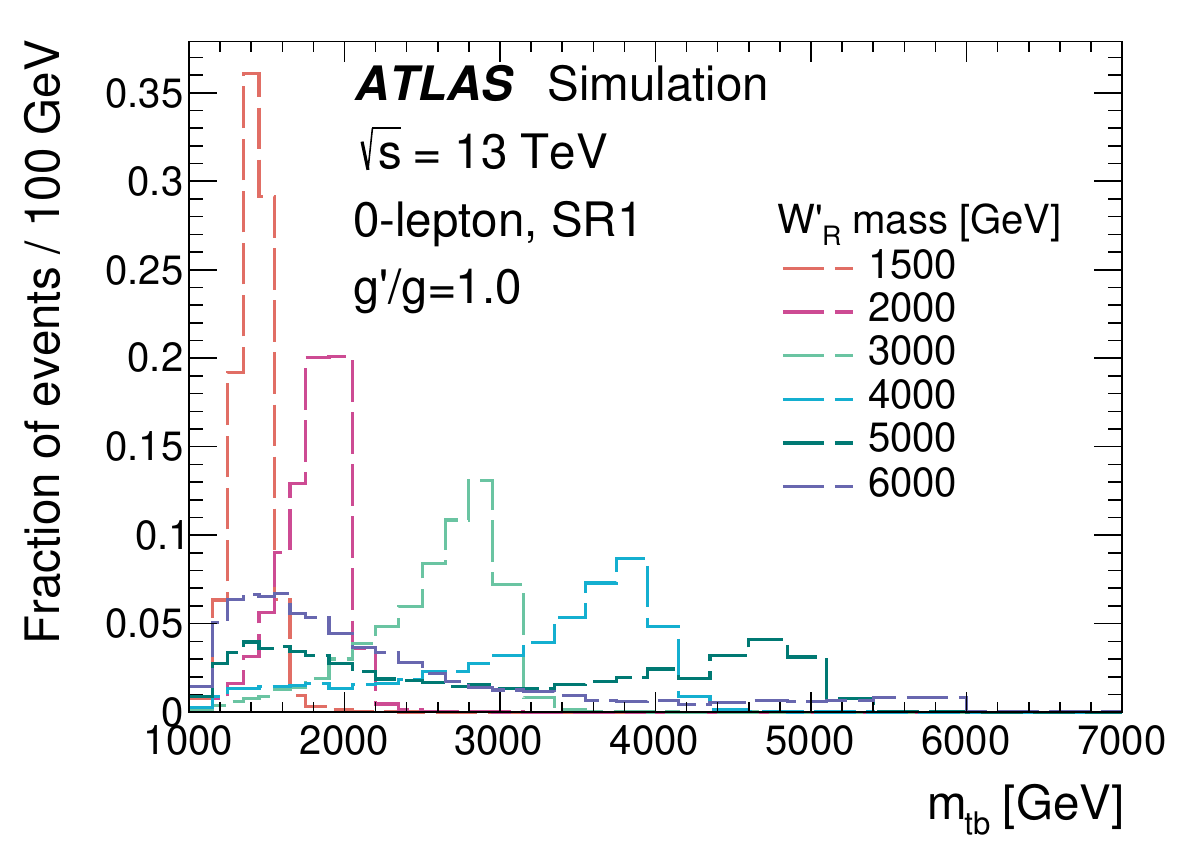}}
\subfloat[]{\includegraphics[width=0.45\textwidth]{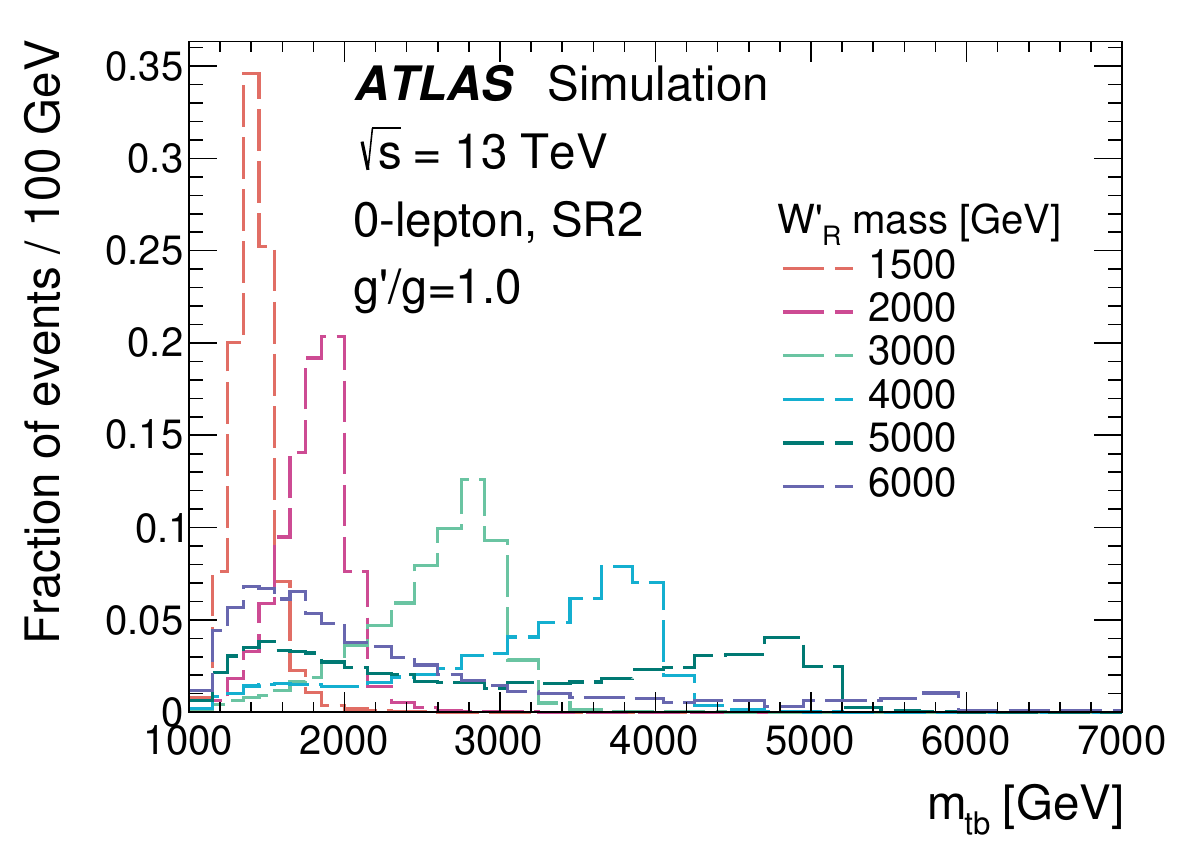}} \\
\subfloat[]{\includegraphics[width=0.45\textwidth]{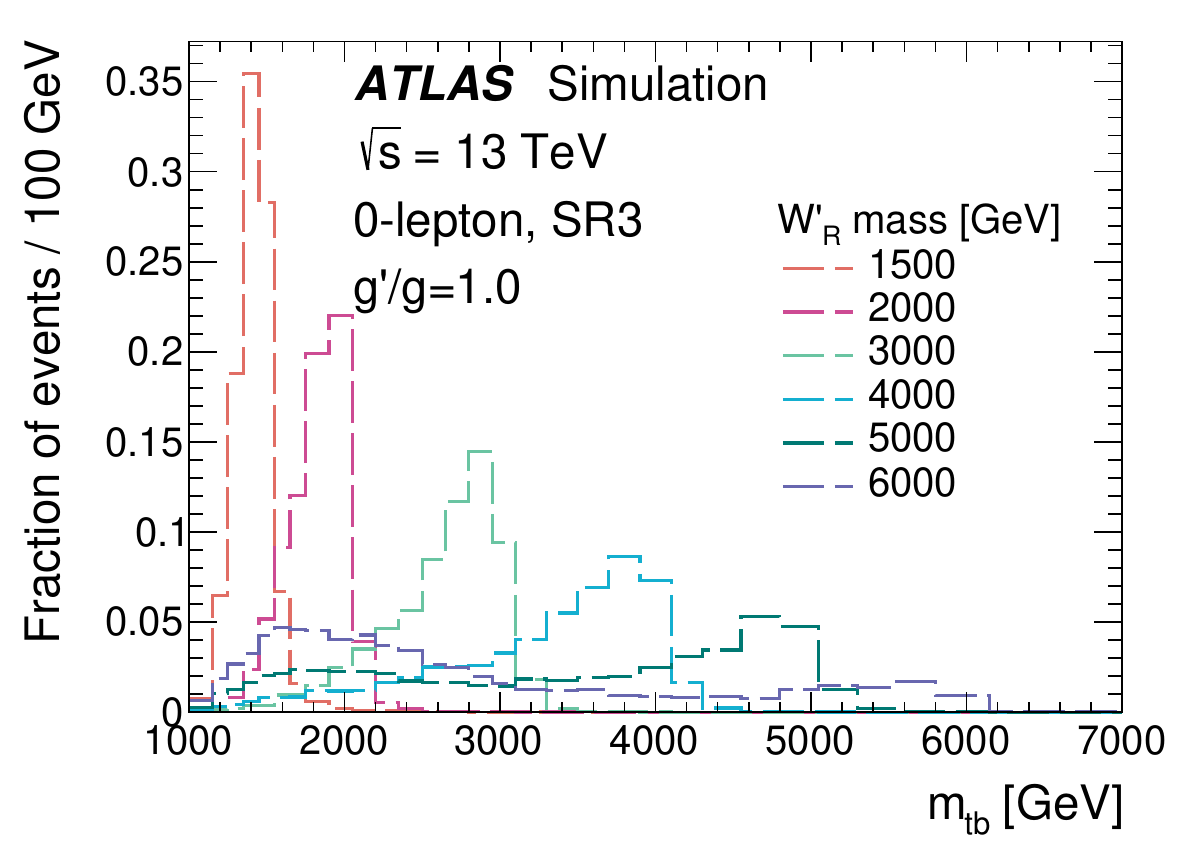}}
\caption{Reconstructed $m_{tb}$ distributions for the right-handed $W'$-boson signal with a coupling value of $g'/g=1$ in (a) signal region~1, (b) signal region~2 and (c) signal region~3 of the \Zerol channel. Distributions are normalised to unit area. The first and last bin in each distribution includes the underflow and overflow, respectively. }
\label{fig:signal_0l}
\end{figure}
 
The distribution of the reconstructed $m_{tb}$ in each of the three signal regions is shown in Figure~\ref{fig:signal_0l} for selected simulated signal samples with a right-handed $W'$~boson and a coupling value of $g'/g=1$. The distribution peaks at the $W'$-boson mass, but exhibits a tail to lower $m_{tb}$ that is more pronounced for higher $W'$-boson masses. The tail is due to the fact that, when the $W'$ pole-mass is high, the PDF values for producing on-shell $W'$~bosons are suppressed relative to the ones for producing low-mass off-shell $W'$~bosons. The product of fiducial acceptance and selection efficiency is shown in Figure~\ref{fig:efficiency_0l} for the three signal regions and the right- and left-handed chirality scenarios with a coupling value of $g'/g=1$. The fraction of $W'$-boson signal events in the template regions is small. Its impact on the background estimate is less than~3.5\% of the background for a $W'$-boson mass of 4~\TeV\ when the signal cross-section is normalised to the expected limit at 4~\TeV. It is negligible compared to the systematic uncertainties of the signal and background. The signal contamination in the control regions is negligible.
 
\begin{figure}[htb!]
\centering
\subfloat[]{\includegraphics[width=0.45\textwidth]{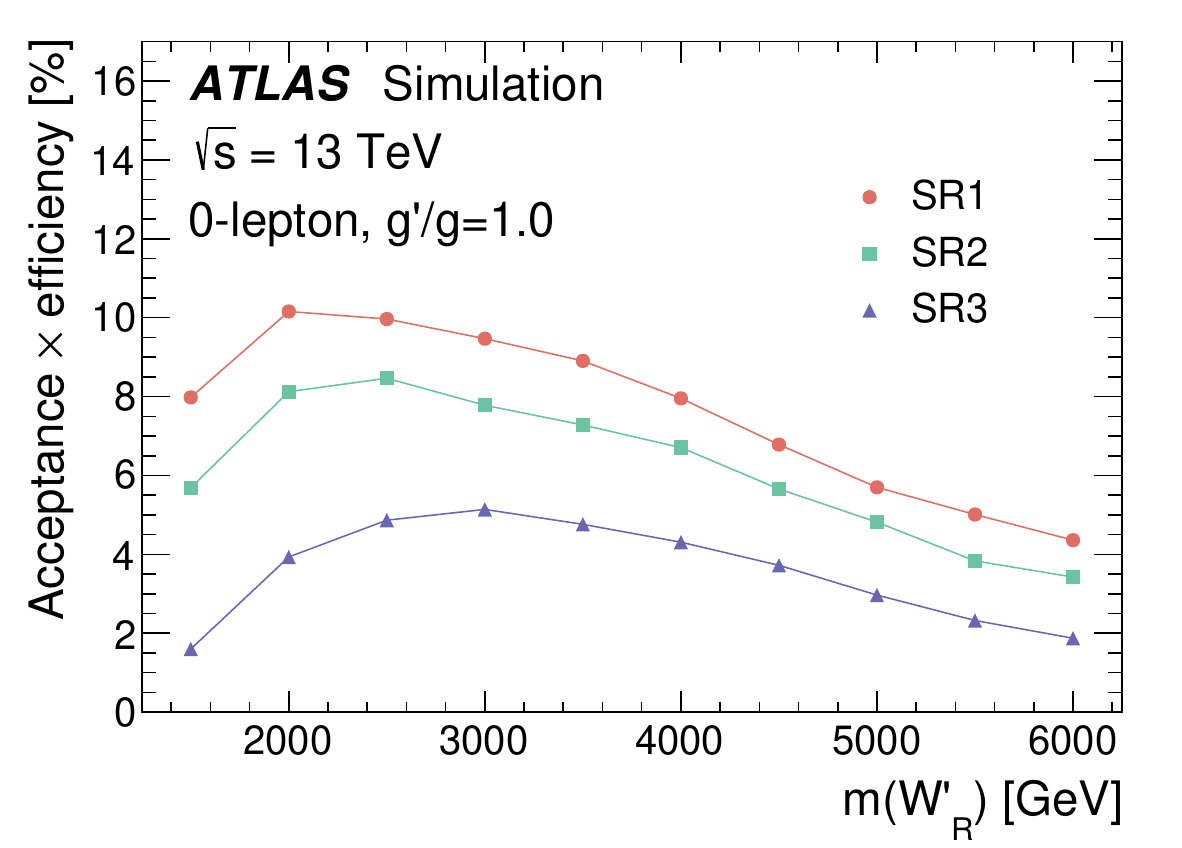}}
\subfloat[]{\includegraphics[width=0.45\textwidth]{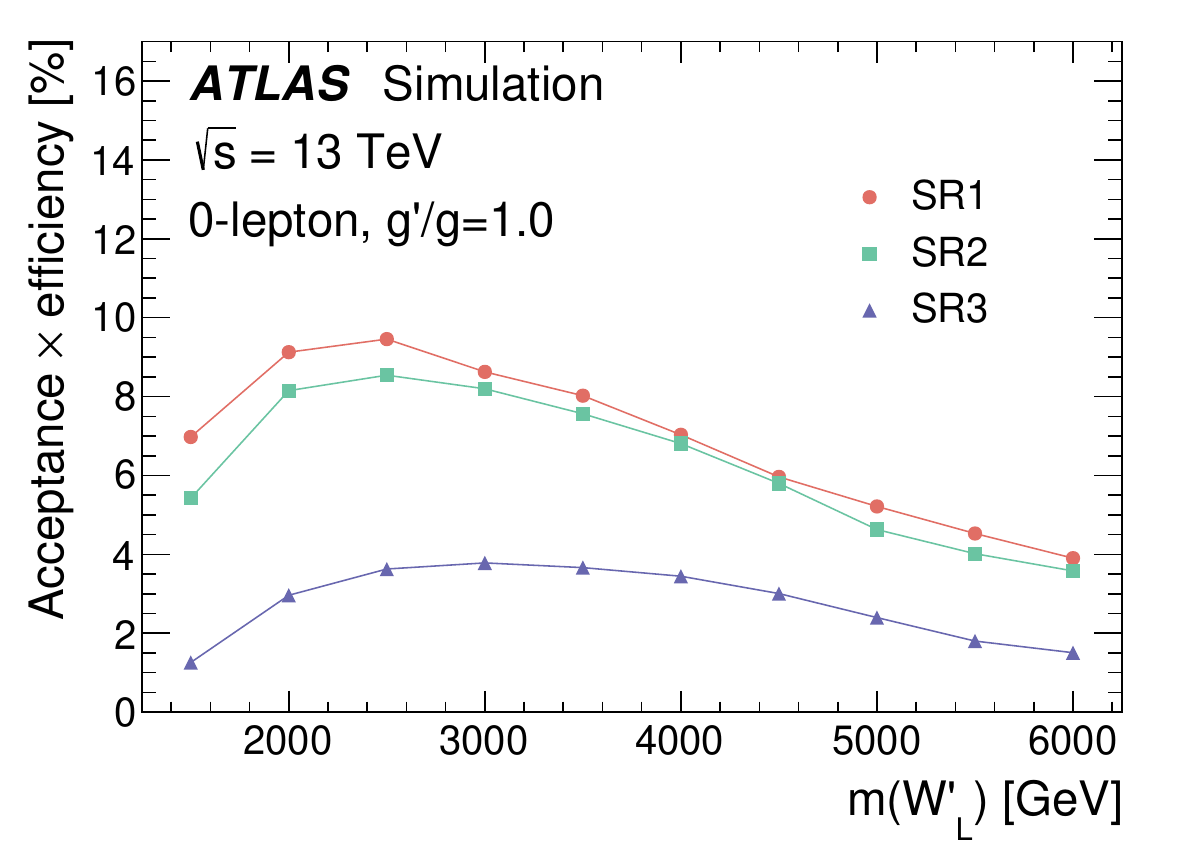}}
\caption{The product of fiducial acceptance and selection efficiency for the three signal regions of the \Zerol channel as a function of the mass of the $W'$~boson, for $W'$~bosons with (a) right-handed chirality and (b) left-handed chirality. The $W'$~boson's coupling strength is set to  $g'/g=1$.}
\label{fig:efficiency_0l}
\end{figure}
 
\section{Analysis strategy in the \Onel channel}
\label{sec:strategy_1l}
 
Events containing exactly one isolated lepton, two or more jets and a certain amount of \met are selected, based on the expected decay products of the $tb$ final state in the lepton+jets decay mode. Events passing these preselection requirements are categorised into different regions based on the number of jets, the number of $b$-tagged jets, and other kinematic variables.

\subsection{Event preselection}
\label{sec:preselection_1l}
 
Events are selected using a combination of single-lepton and \met triggers~\cite{TRIG-2016-01}. The \met triggers are only considered for events with a reconstructed $\met > 200$~\GeV\ to ensure 100\% efficiency of the trigger selection. The single-lepton triggers require the presence of a muon or an electron with \pt higher than a certain threshold and, in some cases, impose identification and lepton-isolation requirements. The lowest \pt threshold was 24 (20)~\GeV\ for electrons (muons) during the 2015 data-taking period and 26~\GeV\ for both the electrons and muons in the data-taking periods from 2016 to 2018. A trigger-matching requirement is applied to the reconstructed lepton, which must be within $\Delta R = 0.1$ of the corresponding object at the trigger~level~\cite{TRIG-2016-01}. The addition of $\met$ triggers offsets a small loss of signal efficiency that occurs for the muon trigger.
 
Events with noise bursts or coherent noise in the calorimeters are removed, as are events containing large energy deposits from non-collision or cosmic sources of background. Events without a reconstructed primary vertex are rejected.
 
Events are required to contain one lepton with $\pt > 50$~\GeV\ and $|\eta|<2.47$, and no additional lepton with $\pt > 30$~\GeV\ and $|\eta|<2.47$. Electrons in the transition region between the barrel and endcap EM calorimeters $(1.37 < |\eta| < 1.52)$ are not considered. Events are required to contain two or more jets with $\pt > 30$~\GeV\ and $|\eta|< 2.5$. Finally, events are required to have $\met > 100$~\GeV. These lepton selection criteria ensure that the trigger selection has a high efficiency for signal events, generally above 95\%.

\subsection{Event reconstruction}
\label{sec:reconstruction_1l}
 
Selected events contain exactly one lepton, missing transverse momentum, and at least two jets. These objects are used to reconstruct the $W'$~boson and the intermediate top quark and leptonically decaying $W$~boson from its decay, as shown in the right diagram of Figure~\ref{fig:fey}. The neutrino is reconstructed starting from the missing transverse momentum in the event. Assuming that all of the \met in an event is carried by the neutrino, $p_{x,\nu}$ and $p_{y,\nu}$ are given by the $x$- and $y$-component of the $\vec{p}_{\text{T}}^{\text{~miss}}$. The $p_{z,\nu}$ component is estimated by requiring that the squared sum of the lepton and neutrino four-momenta must yield the $W$-boson mass, which results in a quadratic equation. The possible solutions for $p_{z,\nu}$ are given by
 
\begin{equation*}\label{equ_NuRec3}
p_{z,\nu}^{\pm} = \frac{\mu\cdot p_{z,\ell}}{p_{\textrm T,\ell}^{2}} \pm \sqrt{\frac{\mu^{2}\cdot p_{z,\ell}^{2}}{p_{\textrm T,\ell}^{4}} - \frac{E_{\ell}^{2}\cdot (\met)^{2} - \mu^{2}}{p_{\textrm T,\ell}^{2}}} \; ,
\end{equation*}
 
with
 
\begin{equation*}\label{equ_NuRec2}
\mu = \frac{m_{W}^{2}}{2} + \cos\Delta\phi\cdot p_{\textrm T,\ell} \cdot p_{\textrm T,\nu}.
\end{equation*}
 
In these formulae, $m_{W}$ is set to 80.4~\GeV, $p_{\textrm T,\nu}$ is the transverse momentum of the neutrino and $\Delta\phi$ is the azimuthal angle between the charged lepton and the reconstructed $\vec{p}_{\text{T}}^{\text{~miss}}$. The $p_z$, transverse momentum, and energy of the charged lepton are given by $p_{z,\ell}$, $p_{\textrm T,\ell}$ and $E_{\ell}$, respectively.
 
If there are two real solutions for $p_{z,\nu}$, the one with the smaller absolute value is chosen. If the radicand is negative, the imaginary solution is avoided by multiplying $\vec{p}_{\text{T}}^{\text{~miss}}$ by a factor chosen to make the radicand exactly zero. This adjustment satisfies $m_{\textrm T}^{W} = m_{W}$, where $m _{\textrm T}^{W}$ is the transverse mass of the reconstructed $W$~boson, and results in a single real solution.
 
The $W$~boson is reconstructed as the sum of the four-vectors of the lepton and the neutrino. The top quark is then reconstructed by combining the $W$~boson with one of the jets, without considering $b$-tagging. The jet $j$ that provides the invariant mass of the $Wj$ system closest to the top-quark mass ($m_{\textrm{top}} = 172.5$~\GeV) is chosen and is referred to as \btop. Events with $p_{\textrm T}^{\textrm{top}} \le 200$~\GeV\ are rejected. Finally, the jet with the highest transverse momentum not selected as \btop is added to the top quark to obtain the reconstructed $W'$~boson and its mass $m_{tb}$. This jet is referred to as \bWP in the following. Events with $p_{\textrm T}^{\bWP} \le 200$~\GeV\ or $m_{tb} \le 500$~\GeV\ are rejected.
This simple method to reconstruct and identify the $tb$ candidate provides a $W'$-boson mass peak with good resolution without any efficiency reduction.

\subsection{Event categorisation}
\label{sec:categorisation_1l}
 
Events selected and reconstructed as described in the previous subsections are categorised into regions based on the reconstructed objects. Both lepton flavours, electron and muon, are kept together in the same region. Signal, validation and control regions are defined by selecting events with two or three jets, one or two  of which are required to be $b$-tagged, resulting in four possible combinations. These regions are referred to as 2j1b, 3j1b, 2j2b and 3j2b. Additional region-specific requirements are imposed to further suppress SM backgrounds:
\begin{itemize}
\item A requirement of $m_{\textrm T}^{W} > 20$~\GeV\ on the transverse mass of the reconstructed $W$~boson in regions with one $b$-tagged jet suppresses multi-jet events.
\item In regions with three jets, events in which the third jet (neither the \btop nor the \bWP) is $b$-tagged are rejected. This requirement reduces the $\ttbar$ background where the third jet is more likely to originate from a bottom quark.
\item In regions with three jets, events are rejected if they contain a reclustered jet with mass close to the top quark ($140~\GeV < m_{\text{vRC-jet}} < 200$~\GeV). This requirement removes both the $W$+jets and \ttbar background where the third jet is less likely to be $b$-tagged.
\item In regions with two jets and one $b$-tagged jet, only events in which the \bWP is $b$-tagged are kept. This requirement reduces the \ttbar background, which has a high fraction of events in which the \btop is $b$-tagged but the \bWP is not.
\end{itemize}
 
Each of the four initial regions is further divided into signal, validation and control regions. The signal regions are referred to as SR\,2j1b, SR\,3j1b, SR\,2j2b and SR\,3j2b and are defined by two additional requirements:
\begin{itemize}
\item The separation between the top-quark decay products in the $(\eta,\phi)$ plane is required to be small, as expected for a high-$\pt$ top quark. A requirement of $\Delta R(\ell,\btop) < 1.0$ is imposed.
\item The pseudorapidity difference between the top quark and the \bWP is required to satisfy $|\Delta \eta(\textrm{top},\bWP)| < 2.0$. This requirement rejects $\ttbar$ and multi-jet backgrounds, where the top quark and the \bWP are expected to be well separated.
\end{itemize}
 
Control regions for the $W$+jets background are defined using the same jet-multiplicity criteria as the signal regions (two or three jets) and requiring one of those jets to be $b$-tagged. Signal events are suppressed by requiring that the lepton and the \btop are separated in $\Delta R$. As the distance $\Delta R(\ell,\btop)$ increases, the amount of signal and $\ttbar$ background decreases, while the amount of $W$+jets background increases. Events are assigned to the control regions if they satisfy $1.5 < \Delta R(\ell,\btop) \leq 2.4$. These regions are referred to as CR\,2j1b and CR\,3j1b and are orthogonal to the signal regions.
 
Events with the same jet multiplicity and $b$-jet multiplicity as in CR\,2j1b or CR\,3j1b but which satisfy $1.0 < \Delta R(\ell,\btop) \le 1.5$ are assigned to validation regions for the $W$+jets background. These regions are referred to as VR\,2j1b and VR\,3j1b and are orthogonal to both the signal regions and the control regions.
 
Validation regions are also defined for the $\ttbar$ background by selecting events with two or three jets, two of which are required to be $b$-tagged. Orthogonality to the signal regions is maintained by requiring $1.0 < \Delta R(\ell,\btop) \leq 2.4$. These regions are referred to as VR\,2j2b and VR\,3j2b.
 
A summary of the region definitions is given in Table~\ref{tab:regionSummary}, while a schematic view is shown in Figure~\ref{fig:regions_1L}.
 
\begin{table}
\centering
\caption{Definition of the signal, control and validation regions in the \Onel channel.}
\begin{tabular}{c||c|c|c|c}
\hline
Regions &SR &$CR_{W+\textrm{jets}}$ &$VR_{W+\textrm{jets}}$ &$VR_{\ttbar}$\\
& \footnotesize{SR 2j1b, SR 2j2b,} & \footnotesize{CR 2j1b, CR 3j1b} & \footnotesize{VR 2j1b, VR 3j1b} & \footnotesize{VR 2j2b, VR 3j2b} \\
&  \footnotesize{SR3j1b, SR3j2b} & & & \\
\hline
Trigger &\multicolumn{4}{c}{\met OR one-lepton}\\
\hline
$N_{\textrm{jets}}$ &\multicolumn{4}{c}{$2,3$}\\
\hline
$N_{b\textrm{-jets}}$ &$1,2$&$1$& $1$&$2$\\
\hline
$p_{\textrm T}^{\textrm{lepton}}$ &\multicolumn{4}{c}{\phantom{0}$>50$~\GeV}\\
\met &\multicolumn{4}{c}{$>100$~\GeV}\\
$m_{\textrm T}^{W}$ (in 1-tag) &\multicolumn{4}{c}{\phantom{0}$>20$~\GeV}\\
$p_{\textrm T}^{\bWP}$ &\multicolumn{4}{c}{$>200$~\GeV}\\
$p_{\textrm T}^{\textrm{top}}$ &\multicolumn{4}{c}{$>200$~\GeV}\\
$m_{tb}$ &\multicolumn{4}{c}{$>500$~\GeV}\\
\hline
$|\Delta \eta(\textrm{top},\bWP)|$ &$<2.0$ & n/a & n/a & n/a \\
\hline
$\Delta R(\ell,\btop)$ &$<1.0$ &$>1.5$, $\leq2.4$ &$>1.0$, $\leq1.5$ &$>1.0$, $\leq 2.4$\\
\hline
$b$-tagging (2-jet regions) & \multicolumn{4}{c}{\bWP is $b$-tagged}\\
$b$-tagging (3-jet regions) & \multicolumn{4}{c}{third jet is not $b$-tagged}\\
\hline
vRC-jet (3-jet regions) & \multicolumn{4}{c}{veto events with $140~\GeV < m_{\textrm{vRC-jet}} < 200$~\GeV}\\
\hline
\end{tabular}
\label{tab:regionSummary}
\end{table}
 
\begin{figure}[htbp]
\centering
\includegraphics[width=0.9\textwidth]{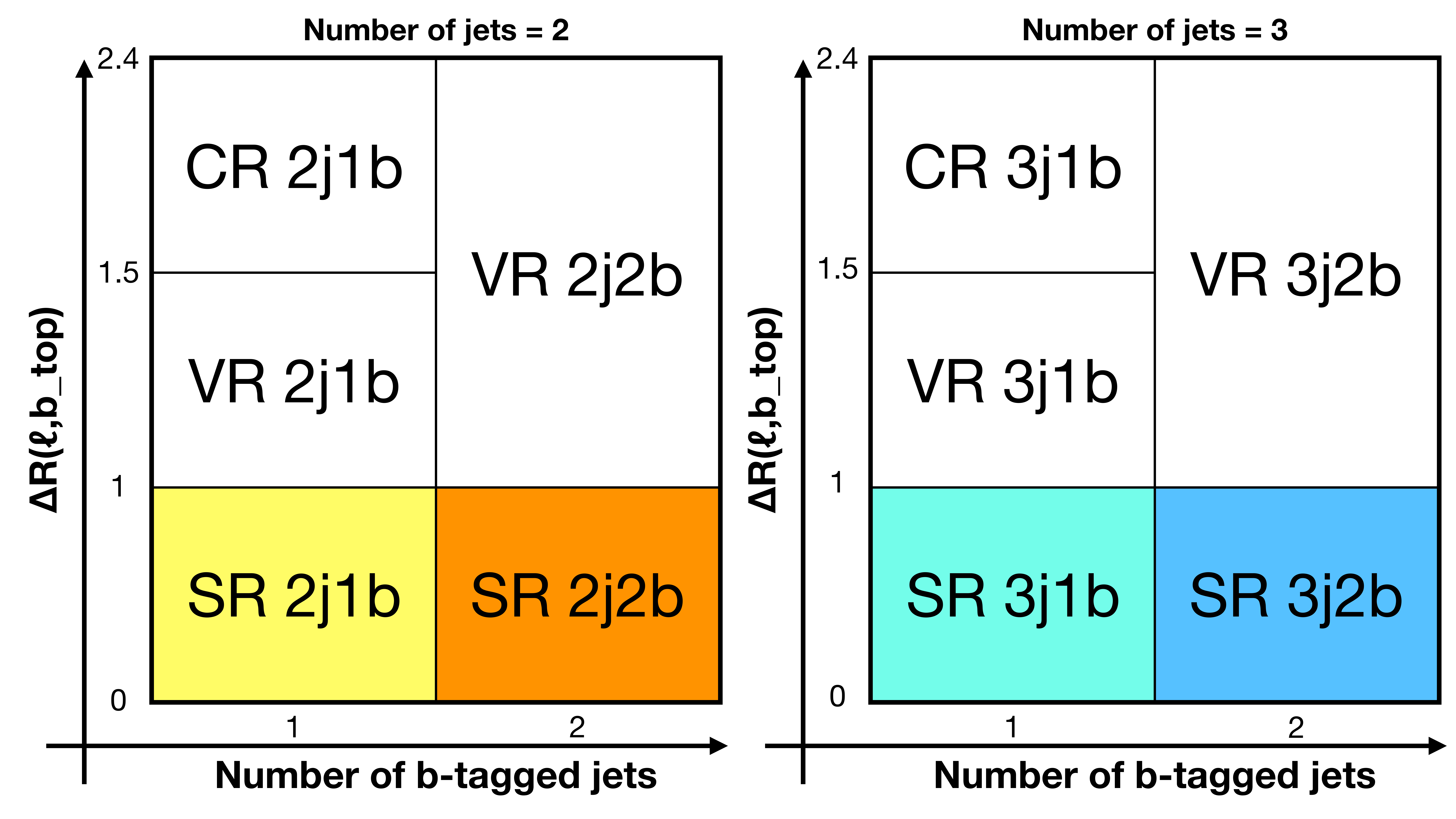}
\caption{ Events in the \Onel channel are categorised according to the number of jets and $b$-tagged jets in the event. Events are assigned to signal, control or validation regions depending on the angular separation between the lepton and jet used to reconstruct the top-quark candidate (\btop).}
\label{fig:regions_1L}
\end{figure}
 
The distribution of the reconstructed $m_{tb}$ in the four signal regions is shown in Figure~\ref{fig:signal_1l} for selected simulated signal samples with a right-handed $W'$~boson and a coupling value of $g'/g=1$. The behaviour is similar to the \Zerol case, with distributions peaking around the $W'$-boson mass and a long tail towards lower masses. The product of fiducial acceptance and selection efficiency for the same regions is shown in Figure~\ref{fig:efficiency_1l}. After an initial rise due to threshold effects for a $W'$-boson mass of 500~\GeV, the efficiency drops as the mass increases. This is mainly caused by $b$-tagging and lepton efficiency dropping as a function of \pt. As transverse momentum increases and the angular distance between the top-quark decay products is reduced, the efficiency to identify isolated leptons degrades accordingly. For high $W'$-boson masses the relative importance of the $W'$ peak becomes small due to reconstruction and PDF effects, as can be seen in Figure~\ref{fig:signal_1l}. When that happens the low mass tail dominates the efficiency calculation, causing it to increase slightly. The fraction of $W'$-boson signal events is negligible in both the validation and control regions.
 
\begin{figure}[htb!]
\centering
\subfloat[]{\includegraphics[width=0.45\textwidth]{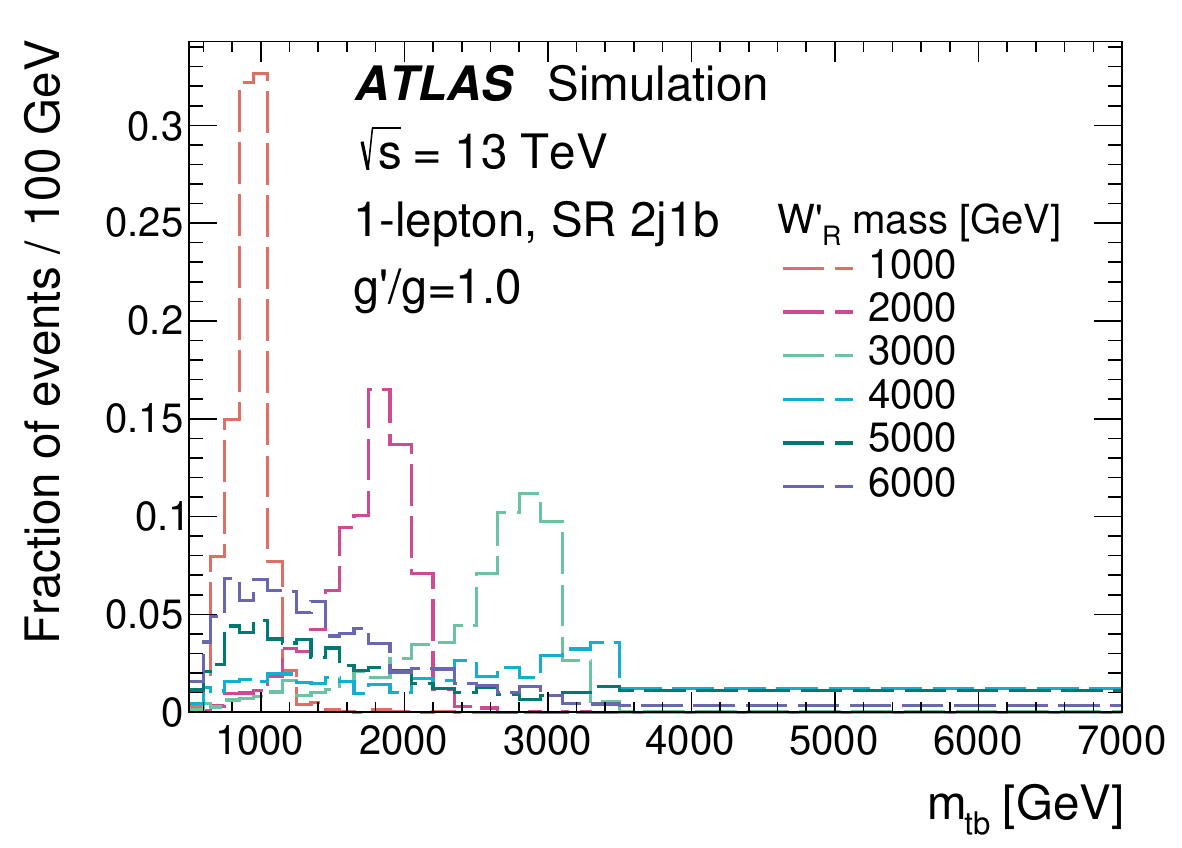}}
\subfloat[]{\includegraphics[width=0.45\textwidth]{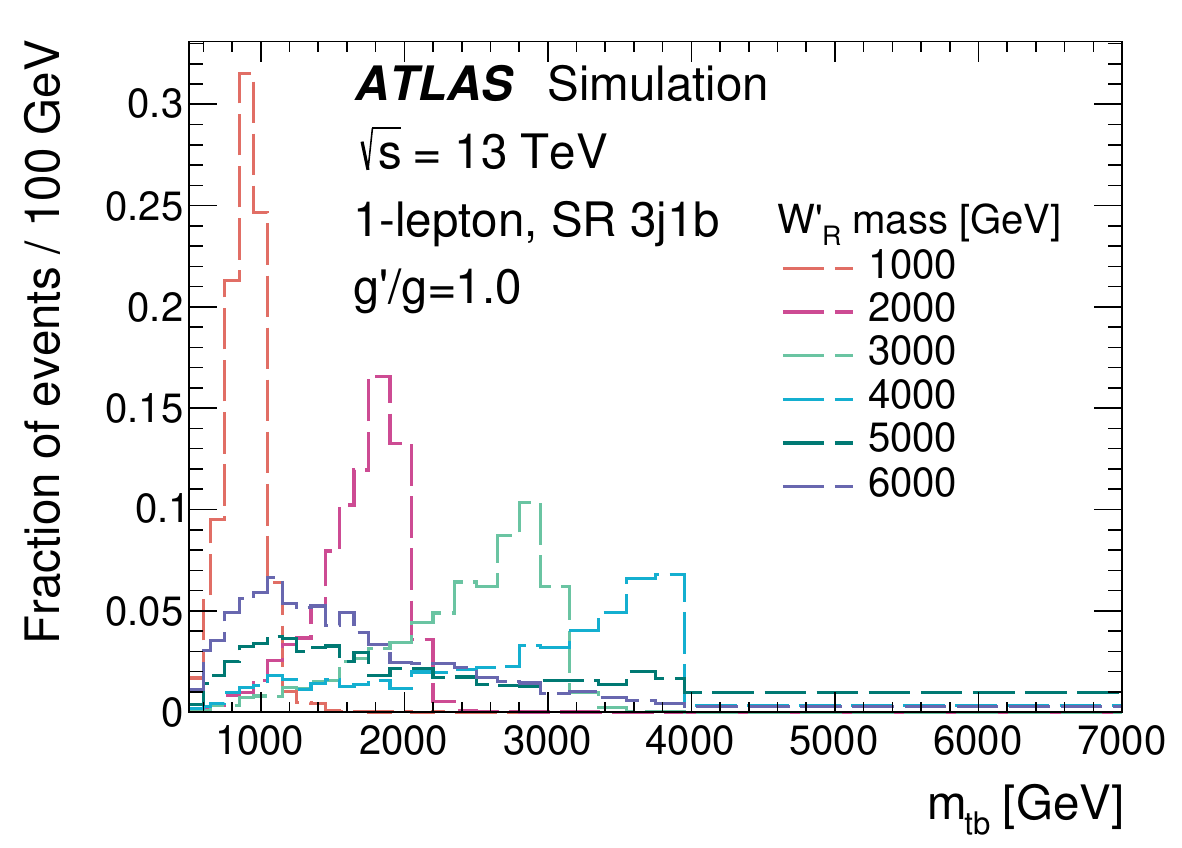}} \\
\subfloat[]{\includegraphics[width=0.45\textwidth]{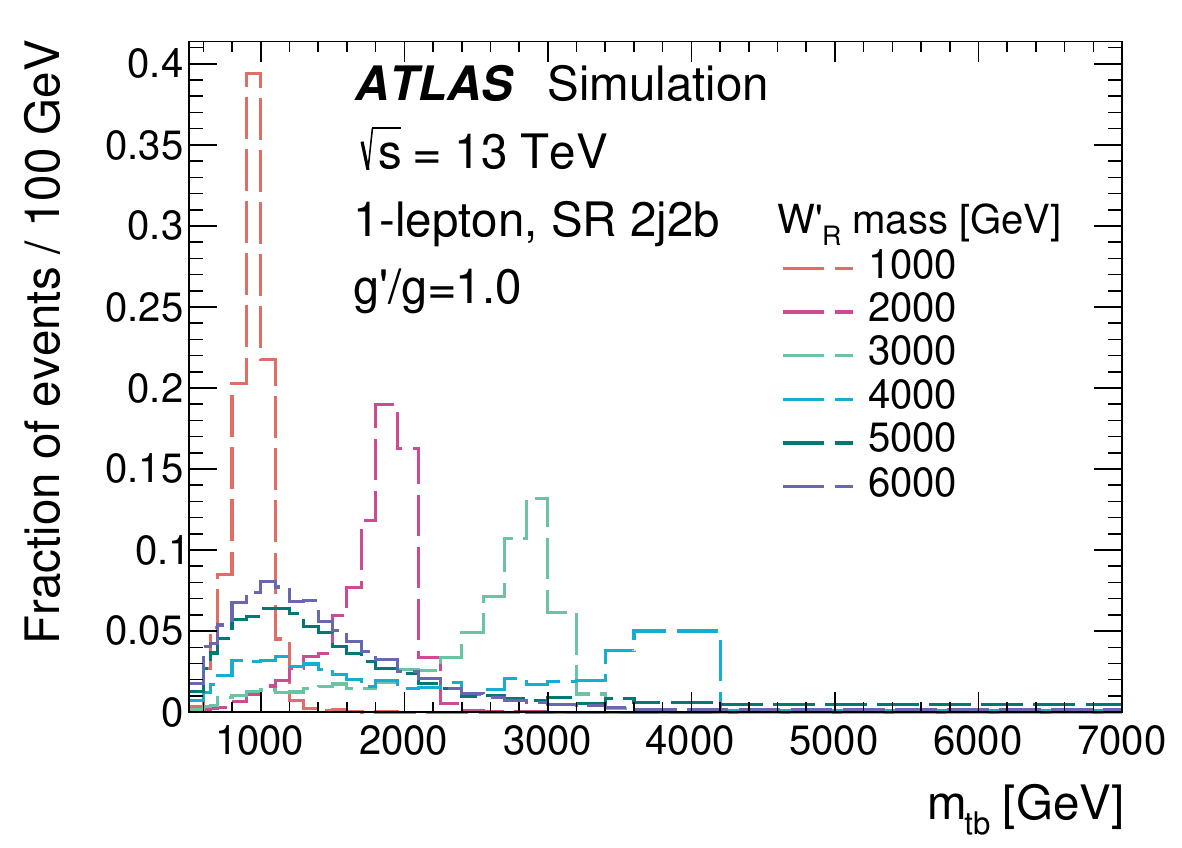}}
\subfloat[]{\includegraphics[width=0.45\textwidth]{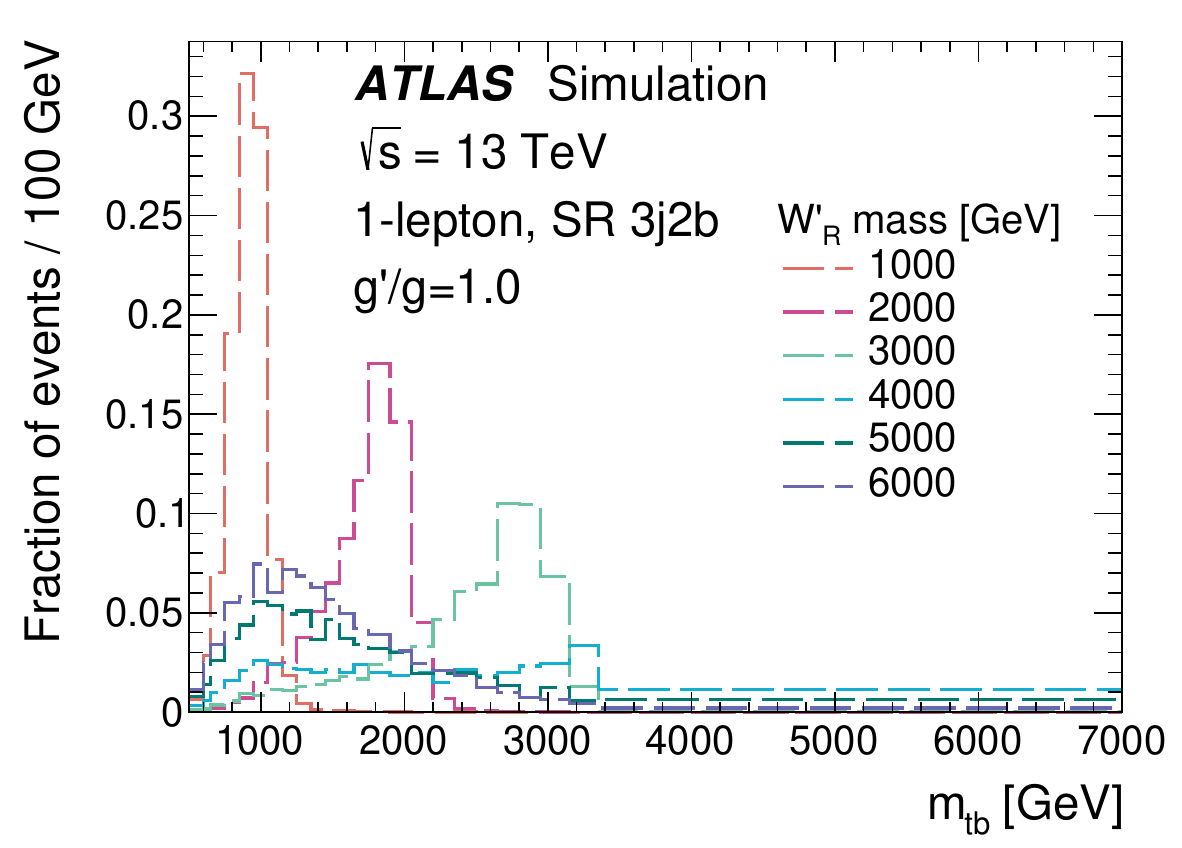}}
\caption{Reconstructed $m_{tb}$ distributions for the right-handed $W'$-boson signal with coupling value of $g'/g=1$ in (a) signal region 2j1b, (b) signal region 3j1b, (c) signal region 2j2b and (d) signal region 3j2b. Distributions are normalised to unit area. The first and last bin in each distribution includes the underflow and overflow, respectively.}
\label{fig:signal_1l}
\end{figure}
 
\begin{figure}[htb!]
\centering
\subfloat[]{\includegraphics[width=0.45\textwidth]{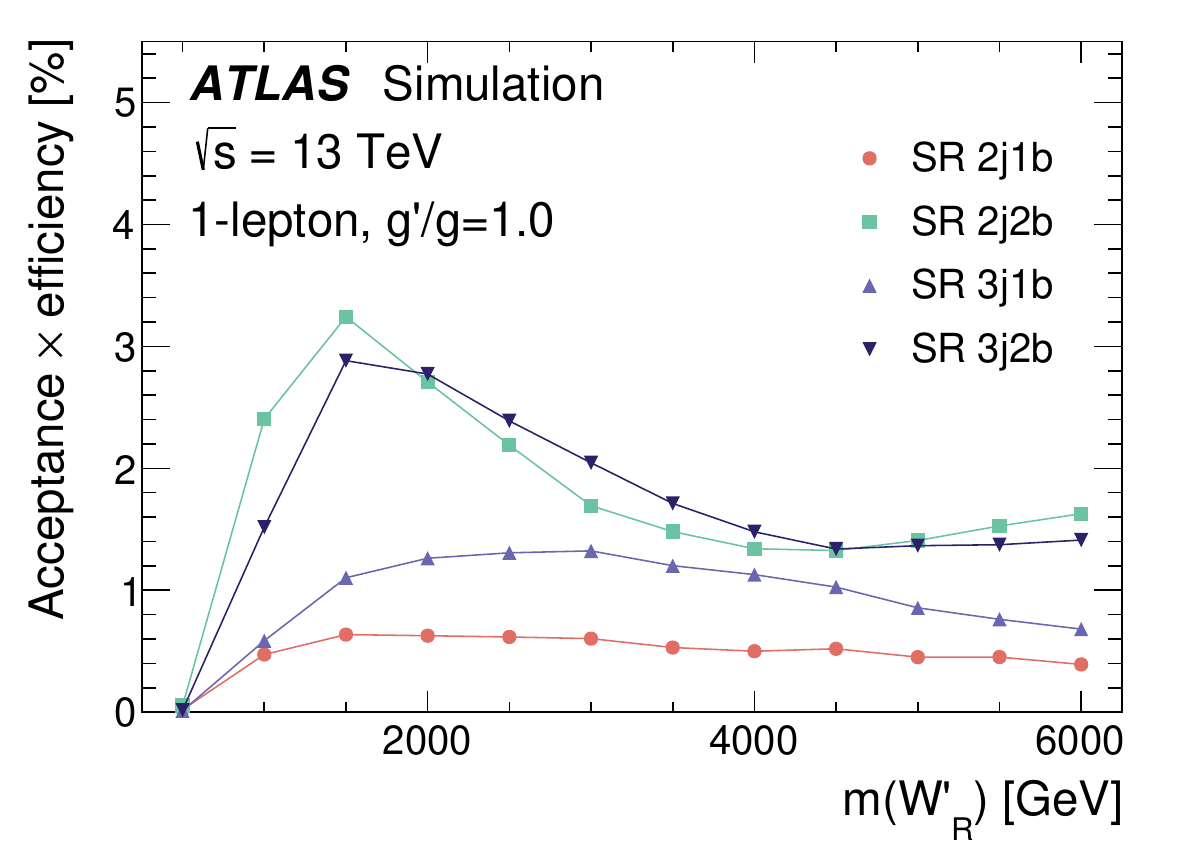}}
\subfloat[]{\includegraphics[width=0.45\textwidth]{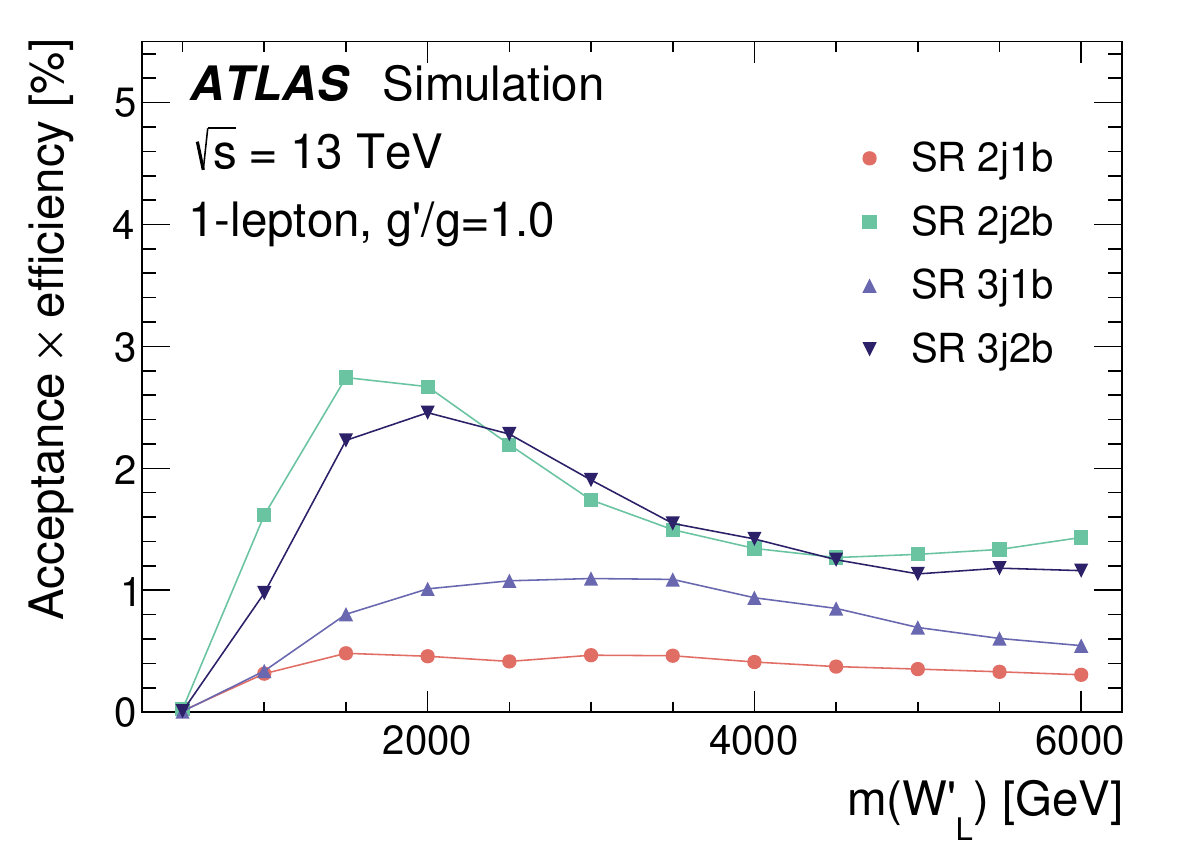}}
\caption{The product of fiducial acceptance and selection efficiency for the signal regions of the \Onel channel as a function of the mass of the $W'$~boson, for $W'$~bosons with (a) right-handed chirality and (b) left-handed chirality. The $W'$~boson coupling strength is set to  $g'/g=1.0$.}
\label{fig:efficiency_1l}
\end{figure}
 
\FloatBarrier

% End of text imported from the .//TeX/selection.tex input file

% The next lines are included from the .//TeX/background.tex input file
\section{Background estimation for the \Zerol channel}
\label{sec:background_0l}
 
The dominant background in the \Zerol channel, from multi-jet production, is estimated using a data-driven method that predicts both the shape and normalisation of the multi-jet $m_{tb}$ distribution in the signal and validation regions. The initial template for the multi-jet background in each signal region, SR$j$ in Figure~\ref{fig:regions} ($j = 1,2,3$), is the $m_{tb}$ histogram in the corresponding template region, TR$j$ in Figure~\ref{fig:regions} ($j = 1,2,3$). TR4 is used to obtain the template in the VR. The correct normalisation for each template in the target region is obtained by multiplying each bin in the template histograms by the ratio $N^{\text{obs}}_{\text{CR1a}}/N^{\text{obs}}_{\text{CR2a}}$ obtained in the same bin of regions CR1a and CR2a or the corresponding ratio $N^{\text{obs}}_{\text{CR3a}}/N^{\text{obs}}_{\text{CR4a}}$ obtained in CR3a and CR4a. This ratio represents the number of top-candidate jets for which the $b$-candidate jet is $b$-tagged divided by the  number of top-candidate jets for which the $b$-candidate jet is not $b$-tagged. It is expected to not depend strongly on the top-tagging criteria for the top-candidate. This ratio is obtained using the pairs composed of a top-proxy jet and a $b$-candidate jet in the control regions indicated with the letter \enquote{a} in Figure~\ref{fig:regions}, CR$j$a, which are completely dominated by multi-jet events. The pairs of jets in the control regions form the same kinematic relationship as the pairs of jets in the signal and validation regions. This equivalence allows the ratio to be used to scale the $m_{tb}$ distribution from the template to the signal and validation regions. The ratio varies between 0.14 (0.15) at low $m_{tb}$, around 1~\TeV, and 0.19 (0.21) at high $m_{tb}$, above 5~\TeV, for the 0-$b$-tag-in-top (1-$b$-tag-in-top) category.
 
The $\ttbar$ background is subdominant and is estimated using the simulated event samples described in Section~\ref{sec:samples}. It is non-negligible in the signal and template regions because it contains two $b$-hadrons. The predicted $\ttbar$ background ($N^{\ttbar}$) is subtracted from data ($N^{\text{obs}}$) in the template regions to obtain the multi-jet background template. The small background from $V+$jets is similar in flavour composition to the multi-jet background and is thus accounted for by the data-driven multi-jet background estimate.
 
The data-driven estimate of the multi-jet background in bin $i$ of $m_{tb}$ in each of the signal regions and the validation region is then given by
 
\begin{equation}
N^{\text{data-driven background}}_{\text{SR1,SR2}} (i)= R^{1}_{\text{corr}}(i) \times \left(N^{\text{obs}}_{\text{TR1,TR2}}(i)-N^{\ttbar}_{\text{TR1,TR2}}(i) \right) \times\frac{N^{\text{obs}}_{\text{CR1a}}(i)}{N^{\text{obs}}_{\text{CR2a}}(i)}
\label{eq:dd1}
\end{equation}
 
and
 
\begin{equation}
N^{\text{data-driven background}}_{\text{SR3,VR}} (i)= R^{0}_{\text{corr}}(i)\times \left( N^{\text{obs}}_{\text{TR3,TR4}}(i)-N^{\ttbar}_{\text{TR3,TR4}}(i) \right) \times\frac{N^{\text{obs}}_{\text{CR3a}}(i)}{N^{\text{obs}}_{\text{CR4a}}(i)}
\,.  \label{eq:dd2}
\end{equation}
 
The correction factors $R^{1,0}_{\text{corr}}$ take into account possible correlations between the top-tagging of the top-candidate jet and the $b$-tagging of the $b$-candidate jet in the 1-b-tag-in-top and 0-b-tag-in-top categories, respectively. The nominal value is $R_{\text{corr}}=1$ because the correlations are small. This is verified in simulated multi-jet samples by computing the corresponding ratio of yields, for example $(N_{\text{SR1}} N_{\text{CR2a}})/(N_{\text{TR1}} N_{\text{CR1a}})$ for SR1. Deviations from unity are considered as uncertainties of the method and are described in Section~\ref{sec:syst}.
 
In order to mitigate the impact of the smaller number of data events in the tails of the $m_{tb}$ distributions in the control regions, bins are merged from high to low $m_{tb}$ to ensure that the statistical uncertainty in each bin of the $b$-tagging ratio is less than 5\%. The same operation is done when calculating the correction factor $R_{\text{corr}}$.
 
The expected and observed event yields in the three signal regions and the validation region are shown in Table~\ref{tab:yields}, together with the predicted yields for a $W'$~boson with a mass of 3~\TeV, right-handed chirality and a coupling value of $g'/g=1.0$. The uncertainty in each estimate  is the sum in quadrature of the systematic uncertainties from all sources described in Section~\ref{sec:syst}. The uncertainty in the $\ttbar$ background is large, mostly due to the uncertainty in the theory modelling. The data-driven background uncertainty is small thanks to the very large event yields in data. It is larger in regions SR1 and SR2 than in SR3 and VR due to two main factors: the larger $\ttbar$ background propagated through the data-driven method and the presence of $b$-tagged jets inside the top-candidate (top-proxy) jet, which increases the impact of the flavour composition on the correlation factor.
 
\begin{table}[htbp]
\begin{center}
\caption{Predicted and observed event yields for the signal regions and the validation region of the \Zerol channel before the fit to data. The uncertainty in each estimate is the sum in quadrature of the systematic uncertainties from all sources described in Section~\ref{sec:syst}. Signal yields correspond to the theoretical prediction for a 3~\TeV\ right-handed $W'$ boson with a coupling strength of $g'/g=1.0$.}
\label{tab:yields}
\resizebox{\textwidth}{!}{
\begin{tabular}{l|cccc}
\hline
& SR1 & SR2 & SR3 & VR\\
\hline
Data-driven                         & 23190 $\pm$ 520~~~~ & 94000 $\pm$ 2000~~ & 76100 $\pm$ 500~~~~ & 297000 $\pm$ 1900~~~~ \\
All-hadronic $\ttbar$          & 4700 $\pm$ 830~~ & 6400 $\pm$ 1600 & 1090 $\pm$ 170~~ & 1400 $\pm$ 300~~~\\
Non-all-hadronic $\ttbar$   & 1220 $\pm$ 180~~ & 1730 $\pm$ 350~~ & 330 $\pm$ 46~~ & 445 $\pm$ 87~~~\\
\hline
Total background              & 29000 $\pm$ 1000~~ & 102400 $\pm$ 2400~~~~ & 77560 $\pm$ 550~~~~ & 299000 $\pm$ 2000~~~~ \\
\hline
$W'$ ($m = 3.0$~\TeV)   & 500 $\pm$ 100 & 390 $\pm$ 85~~ & 260 $\pm$ 58~~ & 190 $\pm$ 41~~ \\
 
\hline
Data   & 29220 & 100383 & 78407 & 298727\\
\hline
\end{tabular}
}
\end{center}
\end{table}

\section{Background estimation for the \Onel channel}
\label{sec:background_1l}
 
The dominant background components in the \Onel channel are those from $W$+jets and \ttbar production. They are estimated using the simulated event samples described in Section~\ref{sec:samples}. The subdominant diboson, $Z$+jets and single-top-quark processes are also estimated using simulated event samples. The small multi-jet contamination from jets misreconstructed as isolated leptons is estimated using a data-driven method known as the template method.
 
The initial template for the multi-jet distribution is obtained by defining \enquote{loose} regions with exactly the same selection requirements as those described in Section~\ref{sec:categorisation_1l} except for a looser lepton selection. The lepton selection is modified as follows to obtain regions enriched in multi-jet events:
 
\begin{itemize}
\item Electrons must pass the \enquote{medium} but not the \enquote{tight} likelihood identification requirements~\cite{PERF-2017-01}.
\item Muons must pass the \enquote{loose} but not the \enquote{medium} identification requirements~\cite{PERF-2015-10}.
\item Electrons and muons must fail the isolation requirements described in Section~\ref{sec:objects}.
\end{itemize}
 
Templates are obtained in each loose region for two variables, $m_{tb}$ and $m_{\textrm T}^{W}$, by subtracting the background components described above from data.
 
The distribution of $m_{\textrm T}^{W}$ is used to obtain correction factors from a binned maximum-likelihood fit performed independently in each signal, control and validation region. In these $m_{\textrm T}^{W}$ fits, the contributions from \ttbar and $W$+jets, as well as the initial multi-jet template from the corresponding loose region, are allowed to float freely. The  resulting multi-jet normalisation factors, one per region, are used to scale the corresponding loose multi-jet $m_{tb}$ template and obtain the $m_{tb}$ multi-jet distribution to be used in the statistical analysis. The $m_{\textrm T}^{W}$ distribution is chosen for this method because multi-jet contribution's shape is different from that of other backgrounds and the bin-by-bin signal significance is extremely small, even for signal regions.
 
The expected and observed event yields in the signal, control and validation regions are shown in Tables~\ref{tab:yields_1l_sr}, \ref{tab:yields_1l_cr} and~\ref{tab:yields_1l_vr} respectively, together with the predicted yields for a $W'$~boson with a mass of 3~\TeV, right-handed chirality and a coupling value of $g'/g=1.0$. The uncertainty in each estimate is the sum in quadrature of the systematic uncertainties from all sources described in Section~\ref{sec:syst}. The amount of multi-jet background is small in all regions, particularly in those with two $b$-tagged jets, where it is compatible with zero.
 
\begin{table}[htbp]
\begin{center}
\caption{Predicted and observed event yields for the signal regions of the \Onel channel before the fit to data. The uncertainty in each estimate is the sum in quadrature of the systematic uncertainties from all sources described in Section~\ref{sec:syst}. Signal yields correspond to the theoretical prediction for a 3~\TeV\ right-handed $W'$ boson with a coupling strength of $g'/g=1.0$.}
\label{tab:yields_1l_sr}
\begin{tabular}{l|cccc}
\hline
& {SR 2j1b} & {SR 2j2b} & {SR 3j1b} & {SR 3j2b}\\
\hline
\ttbar    & 856 $\pm$ 51~~ & 3910 $\pm$ 220~~ & 8150 $\pm$ 210~~ & 7480 $\pm$ 250~~ \\
$W$+jets   & 3140 $\pm$ 170~~ & 329 $\pm$ 32~~ & 3600 $\pm$ 230~~ & 204 $\pm$ 22~~ \\
$Z$+jets    & 205 $\pm$ 95~~ & 100 $\pm$ 44~~ & 380 $\pm$ 160 & 64 $\pm$ 28 \\
Single-top-quark & 300 $\pm$ 40~~ & 1130 $\pm$ 110~~ & 1660 $\pm$ 270~~ & 990 $\pm$ 140 \\
Diboson  & 69 $\pm$ 28 & 13 $\pm$ 6~~ & 190 $\pm$ 77~~ & 15.7 $\pm$ 7.2~~ \\
Multi-jet  & 89 $\pm$ 11 & 82 $\pm$ 37 & 179 $\pm$ 24~~ & 11 $\pm$ 11 \\
\hline
Total background  & 4670 $\pm$ 220~~ & 5560 $\pm$ 290~~ & 14160 $\pm$ 490~~~~ & 8760 $\pm$ 310~~ \\
$W'$ ($m = 3.0$~\TeV)   & 15.2 $\pm$ 1.2~~  & 42.8  $\pm$ 4.9~~ & 33.4 $\pm$ 3.6~~  & 51.7 $\pm$ 5.3~~ \\
\hline
Data   & 5081 & 5150 & 14496 & 8060 \\
\hline
\end{tabular}
\end{center}
\end{table}
 
\begin{table}[htbp]
\begin{center}
\caption{Predicted and observed event yields for the control regions of the \Onel channel before the fit to data. The uncertainty in each estimate is the sum in quadrature of the systematic uncertainties from all sources described in Section~\ref{sec:syst}.}
\label{tab:yields_1l_cr}
\begin{tabular}{l|cc}
\hline
& {CR 2j1b} & {CR 3j1b} \\
\hline
\ttbar    & 1386 $\pm$ 58~~~~ & 4940 $\pm$ 160~~ \\
$W$+jets   & 7720 $\pm$ 470~~ & 6780 $\pm$ 530~~ \\
$Z$+jets   & 150 $\pm$ 60~~ & 160 $\pm$ 66~~ \\
Single-top-quark   & 640 $\pm$ 160 & 1380 $\pm$ 360~~ \\
Diboson   & 168 $\pm$ 68~~ & 300 $\pm$ 120 \\
Multi-jet  & 236 $\pm$ 26~~ & 273 $\pm$ 38~~  \\
\hline
Total background  & 10300 $\pm$ 520~~~~ & 13700 $\pm$ 800~~~~ \\
\hline
Data  & 11553 & 14431 \\
\hline
\end{tabular}
\end{center}
\end{table}
\begin{table}[htbp]
\begin{center}
\caption{Predicted and observed event yields for the validation regions of the \Onel channel before the fit to data. The uncertainty in each estimate is the sum in quadrature of the systematic uncertainties from all sources described in Section~\ref{sec:syst}.}
\label{tab:yields_1l_vr}
\begin{tabular}{l|cccc}
\hline
& {VR 2j1b} & {VR 3j1b} & {VR 2j2b} & {VR 3j2b}\\
\hline
\ttbar    & 677 $\pm$ 36~~ & 6420 $\pm$ 160~~ & 3010 $\pm$ 180~~ & 6500 $\pm$ 240~~ \\
$W$+jets   & 3730 $\pm$ 220~~ & 4160 $\pm$ 300~~ & 1340 $\pm$ 110~~ & 620 $\pm$ 68~~ \\
$Z$+jets   &  100 $\pm$ 40~~ & 153 $\pm$ 64~~ & 82 $\pm$ 34 & 40 $\pm$ 17 \\
Single-top-quark   & 330 $\pm$ 80~~ & 1640 $\pm$ 240~~ & 1660 $\pm$ 340~~ & 1370 $\pm$ 390~~ \\
Diboson   & 83 $\pm$ 35 & 203 $\pm$ 83~~ & 31 $\pm$ 13 & 33 $\pm$ 14 \\
Multi-jet  & 83 $\pm$ 9~~ & 195 $\pm$ 27~~ & 283 $\pm$ 55~~ & 92 $\pm$ 66  \\
\hline
Total background  & 5000 $\pm$ 260~~ & 12770 $\pm$ 540~~~~ & 6410 $\pm$ 440~~ & 8650 $\pm$ 490~~ \\
\hline
Data  & 5398 & 13091 & 6413 & 8310 \\
\hline
\end{tabular}
\end{center}
\end{table}
 
\FloatBarrier

% End of text imported from the .//TeX/background.tex input file

% The next lines are included from the .//TeX/systematics.tex input file
\section{Systematic uncertainties}
\label{sec:syst}
 
The modelling of signal, $\ttbar$, $V$+jets, single-top-quark, and diboson events described in Section~\ref{sec:samples} is affected by experimental uncertainties related to the reconstruction and calibration of the physics objects. In addition, uncertainties in the theoretical modelling of the $\ttbar$, single-top-quark, and $V$+jets backgrounds are also taken into account.
 
In the \Zerol channel, these uncertainties affecting the simulated backgrounds also affect the data-driven background estimate because they are propagated through the $\ttbar$ subtraction in the template regions. Additional sources of uncertainty affecting the data-driven background in the \Zerol channel are considered in order to account for possible deviations from the core assumptions of the method described in Section~\ref{sec:background_0l}. The multi-jet background in the \Onel channel is small, so all uncertainties in this background are expected to be covered by a single normalisation uncertainty.

\subsection{Experimental uncertainties}
 
Uncertainties related to the energy scale and resolution of small- and large-$R$ jets are evaluated by combining information about detector reconstruction performance in simulated events with \textit{in situ} methods using data collected with ATLAS during LHC \RunTwo~\cite{JETM-2018-02,JETM-2018-05}. Uncertainties related to the mass scale of large-$R$ jets are evaluated by using a forward-folding technique combining fits to the $W$-boson and top-quark mass peaks in order to extract both the mass scale and resolution differences between data and simulation~\cite{ATLAS-CONF-2020-022}. This approach is complemented by the $R_{\text{trk}}$ method~\cite{JETM-2018-02}. A constant jet mass resolution uncertainty of 20\% is assigned to the mass of large-$R$ jets~\cite{JETM-2018-02}.
 
Uncertainties in the correction factors for the $b$-tagging identification response are derived from dedicated flavour-enriched samples in data. An additional term is included to extrapolate the measured uncertainties to the high-\pt region with jet $\pt > 400$~\GeV. This term is calculated from simulated events by considering variations of the quantities affecting the $b$-tagging performance, such as the impact parameter resolution, percentage of poorly measured tracks, description of the detector material, and track multiplicity per jet. The dominant uncertainty affecting the extrapolation to high-\pt is related to the interactions of high-\pt $b$-hadrons in the innermost pixel layer, which were not considered in the simulation of the samples used for this analysis~\cite{FTAG-2018-01}.
 
Uncertainties in the correction factors for the top-tagging identification are considered~\cite{JETM-2018-06}, taking into account effects on the selection and reconstruction of jets involved in the scale factor estimation. These uncertainties are obtained by taking into account uncertainties related to the jet energy scale and $b$-tagging, as well as MC generator uncertainties and statistical uncertainties. Additional uncertainties related to the modelling of the samples used for the scale factor estimation are also taken into account. Uncertainties are also considered for jets with $\pt>800$~\GeV\ in the extrapolation of the measured uncertainties to the high-\pt region.
 
Uncertainties are considered on the electrons energy scale and energy resolution, the muons momentum scale and resolution and on the data-to-MC correction factors applied to the trigger, reconstruction, identification and isolation efficiencies~\cite{EGAM-2018-01,MUON-2018-03}. Uncertainties are also considered on the soft term used in the \met calculation and on the \met energy scale and resolution~\cite{ATLAS-CONF-2018-023}.
 
Variations in the reweighting applied to simulated event samples to match the mean number of $pp$ interactions observed in each bunch crossing in data are included. They cover the uncertainty in the ratio of the predicted and measured inelastic cross-sections. A constant 1.7\%~\cite{ATLAS-CONF-2019-021} normalisation uncertainty is applied to all simulated event samples to account for uncertainty in the combined 2015--2018 integrated luminosity, obtained using the LUCID-2 detector \cite{LUCID2} for the primary luminosity measurements, complemented by measurements using the inner detector and calorimeters.

\subsection{Modelling uncertainties in background simulations}
\label{sec:model_sys}
 
For the \Zerol channel, uncertainties in modelling the \ttbar background are included. Other possible backgrounds are only included as part of the data-driven estimation. For the \Onel channel, uncertainties in modelling the \ttbar, single-top-quark, and $W$+jets backgrounds are included.
 
Several uncertainties in the theoretical modelling of the $\ttbar$ background samples are considered. Systematic uncertainties due to the choice of parton shower and hadronisation model are evaluated by comparing the nominal \ttbar sample with a sample produced with the \POWHEGBOX[v2] generator using the \NNPDF[3.0nlo] PDF set. Events in the latter sample were passed to \HERWIG[7.04]~\cite{Bahr:2008pv,Bellm:2015jjp}, which used the H7UE set of tuned parameters~\cite{Bellm:2015jjp} and the \MMHT[lo] PDF set~\cite{Harland-Lang:2014zoa}. To assess the uncertainty in the matching of NLO matrix elements to the parton shower, the nominal \POWHEG sample is compared with a sample of events generated with \MGNLO[2.6.0]~\cite{Alwall:2014hca} interfaced with \PYTHIA[8.230]. The \MGNLO calculation used the \NNPDF[3.0nlo] PDF set, and \PYTHIA[8] used the A14 tune and the \NNPDF[2.3lo] PDF set. Before the comparisons, these samples were corrected to match the NNLO predictions of the top-quark $\pT$ distribution using the procedure outlined in Section~\ref{sec:samples-0lep}.  Systematic uncertainties associated with alternative choices of renormalisation and factorisation scales, in which their nominal values are varied by factors of 0.5 and 2.0, and with the choice of PDF set, which is changed to \textsc{LUXqed+PDF4LHC15}~\cite{Butterworth:2015oua,Manohar:2016nzj}, are included by correcting the nominal sample to dedicated alternative NNLO calculations~\cite{Czakon:2017wor}. Additional uncertainties are calculated using internal weights associated with each event for alternative MC tune choices~\cite{ATL-PHYS-PUB-2014-021} corresponding to changes in the amount of initial-state and final-state radiation and in the modelling of multiple parton interactions.
 
For the three production modes contributing to the single-top-quark background, the uncertainty due to the parton shower and hadronisation model is evaluated by comparing the nominal sample of events with a sample where the events generated with the \POWHEGBOX[v2] generator are interfaced to \HERWIG[7.04], which used the H7UE tune and the \MMHT[lo] PDF set. To estimate the uncertainty in the matching of NLO matrix elements to the parton shower, the nominal samples are compared with samples generated with the \MGNLO[2.6.2] generator at NLO in QCD using the five-flavour scheme and the \NNPDF[2.3nlo] PDF set. The events are interfaced with \PYTHIA[8.230], which used the A14 tune and the \NNPDF[2.3lo] PDF set. Additional uncertainties are considered for the choice of PDF set, analogous to the $\ttbar$ uncertainties. These are included by using weights related to variations of the \NNPDF[3.0nlo] set and alternative baseline PDF sets, namely the \MMHT[nlo] set~\cite{Harland-Lang:2014zoa} and the \CT[14nlo] set~\cite{Dulat:2015mca}. Finally, the nominal $tW$-channel \POWPY[8] sample is also compared with an alternative sample generated using the diagram subtraction scheme~\cite{Frixione:2008yi,ATL-PHYS-PUB-2016-020} to estimate the uncertainty arising from the interference with \ttbar production.
 
For the $W$+jets background in the \Onel channel, internal weights are used to consider alternative renormalisation and factorisation scale choices. Uncertainties related to the choice of PDF set are estimated using internal weights corresponding to the \NNPDF[3.0nnlo] set and two alternative baseline PDF sets, \CT[18nnlo]~\cite{Hou:2019efy} and \MSHT[nnlo]~\cite{Bailey:2020ooq}, as well as two variations of the \NNPDF[3.0nnlo] set with different values of \alphas. Uncertainties associated with the inclusion of approximate NLO electroweak corrections are included by using internal weights corresponding to different ways of combining the QCD and electroweak contributions: additive, multiplicative, or exponentiated~\cite{Kallweit:2014xda}. Each option is compared with the nominal prediction independently.
 
A 6\% uncertainty is assigned to the $\ttbar$ normalisation in the \Zerol channel, in accord with the inclusive cross-section calculation described in Section~\ref{sec:samples}. No overall normalisation uncertainty is assigned to the $\ttbar$ or $W$+jet inclusive cross-section for the \Onel channel as the normalisation of each of these background components is controlled by free-floating parameters in the final likelihood fit described in Section~\ref{sec:results}. An overall normalisation uncertainty of 5\% is assigned to the single-top-quark backgrounds to account for the inclusive cross-section uncertainty. A conservative 40\% normalisation uncertainty is assigned to both $Z$+jets and diboson production to take into account any possible mismodelling in the production of additional jets~\cite{Alwall:2007fs} and heavy-flavour jets~\cite{HIGG-2017-03,} in these minor backgrounds.
 
\subsection{Uncertainties related to the data-driven background estimation}
\label{sec:corr}
 
Dedicated uncertainties in the data-driven background estimation in the \Zerol channel are obtained by measuring the correlation between the top-tagging DNN score of the top-candidate jet and the $b$-tagging score of the $b$-candidate jet directly in data, using the control regions defined in Section~\ref{sec:background_0l}.
 
The assumption of $R_{\text{corr}}=1$ in Eqs.~(\ref{eq:dd1}) and (\ref{eq:dd2}) is replaced by the ratio of diagonal products,
 
\begin{equation*}
R^{1}_{\text{corr}}(i) = \frac{N^{\text{obs}}_{\text{CR1a}}(i)\times N^{\text{obs}}_{\text{CR2b}}(i)}{N^{\text{obs}}_{\text{CR2a}}(i)\times N^{\text{obs}}_{\text{CR1b}}(i)}
\end{equation*}
 
and
 
\begin{equation*}
R^{0}_{\text{corr}}(i) = \frac{N^{\text{obs}}_{\text{CR3a}}(i)\times N^{\text{obs}}_{\text{CR4b}}(i)}{N^{\text{obs}}_{\text{CR4a}}(i)\times N^{\text{obs}}_{\text{CR3b}}(i)}\,,
\end{equation*}
 
for the 1-$b$-tag-in-top and 0-$b$-tag-in-top categories, respectively, to obtain the varied background estimates used to define the uncertainty. The counts $N_{\text{CR}j}(i)$ represent the number of events in bin $i$ of the $m_{tb}$ distribution in region CR$j$. These $R_{\text{corr}}$ values deviate from unity as the correlation between top-tagging and $b$-tagging increases.
 
The background rejection factors for top-tagging and $b$-tagging decrease for high-\pT\ large-$R$ and small-$R$ jets, resulting in a dependence on $m_{tb}$. This can cause $R_{\text{corr}}$ to deviate from unity. In addition, the flavour composition of the partons in multi-jet events changes as a function of $m_{tb}$. The uncertainty in  $R_{\text{corr}}$ is separated into two uncorrelated components, one for low masses ($m_{tb} \le 2$~\TeV) and the other for high masses ($m_{tb} > 2$~\TeV), to account for these two effects. This uncertainty ranges from 1\% at low $m_{tb}$ values to approximately 5\% for large $m_{tb}$ in SR3. It takes slightly larger values for SR2 and SR1, with values up to 13\% in the high $m_{tb}$ region.
 
In the \Onel channel, the uncertainty in the normalisation factor for the multi-jet background described in Section~\ref{sec:background_1l} is taken into account.

\subsection{Uncertainty impact}
\label{sec:uncert_impact}
 
In order to estimate the importance of the different categories of systematic uncertainties, their post-fit impact is calculated for right-handed $W'$~bosons with various masses and $g'/g=1.0$. The result of such an estimation in the combined fit of the two channels is shown in Table~\ref{tab:syst} as the fractional contribution of each category to the total uncertainty in the observed signal strength. For each category, the fit to data is repeated with the corresponding group of nuisance parameters fixed to their best-fit values. Each category's contribution is evaluated from the difference of the squares of the uncertainty of the original fit and the modified fit, by dividing the square root of this difference by the uncertainty of the original fit. The sum in quadrature is different from one due to correlations among nuisance parameters in the fit.
 
The relative importance of systematic uncertainties falls with increasing $W'$-boson mass, and the measurement becomes very statistically dominated at large masses. The MC and data-driven background statistics category from Table~\ref{tab:syst} is dominated by the statistical uncertainty of the data-driven background in the 0-lepton channel. Among the systematic uncertainties, background modelling uncertainties dominate mostly at large masses, while for lower masses the top- and flavour-tagging uncertainty components become more important.
 
\begin{table}[htbp]
\begin{center}
\caption{Post-fit fractional contributions of different uncertainty categories to the total uncertainty in the observed signal strength, as determined in the combined fit of the \Zerol and \Onel channels. Different masses of a right-handed $W'$ boson with $g'/g=1.0$ are considered. For each category, the fit to data is repeated with the corresponding group of nuisance parameters fixed to their best-fit values. Each category's contribution is evaluated from the difference of the squares of the uncertainty of the original fit and the modified fit, by dividing the square root of this difference by the uncertainty of the original fit. The sum in quadrature is different from unity due to correlations among nuisance parameters in the fit.}
\label{tab:syst}
\begin{tabular}{l|cc}
\hline
Uncertainty & {$m(W')=2$~\TeV} & {$m(W')=4$~\TeV} \\\hline
\textbf{Background modelling}                           & 0.36 & 0.42  \\
$\ttbar$                                                & 0.21 & 0.11  \\
$W$+jets                                                & 0.19 & 0.11  \\
Multi-jet and data-driven background         & 0.16 & 0.37  \\
Single-top-quark                                        & 0.20 & 0.21  \\
Other processes                                         & 0.03 & 0.00  \\\hline
\textbf{Instrumental}                                   & 0.40 & 0.21  \\
Top-tagging                                             & 0.14 & 0.16  \\
Flavour-tagging                                         & 0.35 & 0.11  \\
Large-$R$ jets                                            & 0.04 & 0.00  \\
Small-$R$ jets                                            & 0.11 & 0.05  \\
Other                                                   & 0.01 & 0.00  \\\hline
\textbf{Total systematic uncertainty}                   & 0.61 & 0.47  \\
\textbf{MC and data-driven bkg. statistics}             & 0.25 & 0.42  \\
\textbf{Statistical uncertainty}                        & 0.75 & 0.79  \\
\hline
\end{tabular}
\end{center}
\end{table}

% End of text imported from the .//TeX/systematics.tex input file

% The next lines are included from the .//TeX/results.tex input file
\section{Statistical analysis and results}
\label{sec:results}
 
In order to test for the presence of a massive resonance, templates in the variable $m_{tb}$ obtained from the simulated signal samples, background samples, and data-driven predictions are fitted to the data. The fit uses a binned maximum-likelihood approach based on the \textsc{RooStats} framework~\cite{Baak:2014wma}. Separate fits are performed for each signal mass and chirality hypothesis. Each fit includes the three signal regions defined in Section~\ref{sec:strategy_0l} for the 0-lepton channel and the six signal and control regions defined in Section~\ref{sec:strategy_1l} for the 1-lepton channel, making a total of nine regions in which the $m_{tb}$ distribution is fitted simultaneously.
 
The systematic uncertainties described in Section~\ref{sec:syst} can change the acceptance, normalisation and shape of the $m_{tb}$ distribution for the signal and the background processes. They are incorporated into the fit as nuisance parameters with a log-normal or Gaussian constraint. The signal and background expectations in each bin are functions of these nuisance parameters.
 
As already indicated in Section~\ref{sec:syst}, the normalisations of the $\ttbar$ and $W$+jets background components in the \Onel channel are allowed to float freely in the fit. Two independent normalisation factors are used for each of the $W$+jets and $\ttbar$ background components, one for the 2-jet regions and one for the 3-jet regions, making a total of four normalisation factors. This choice is motivated by the previous search~\cite{EXOT-2016-18}, where significant differences in the modelling as a function of the jet multiplicity were observed. For the same reason, all modelling uncertainties in the \Onel channel are kept uncorrelated between regions with different jet multiplicities. Both the normalisation and modelling uncertainties are assumed to be correlated between regions with different $b$-tagged jet multiplicities but the same jet multiplicity. In the \Zerol channel, all modelling uncertainties are kept fully uncorrelated between the three signal regions.
 
Given the different treatment of the modelling uncertainties in the different channels, these uncertainties are considered uncorrelated between the \Zerol and \Onel channels. The same is true for the normalisation uncertainties present in the \Zerol channel and the floating normalisations present in the \Onel channel. Experimental uncertainties, when present in both channels, are considered correlated.
 
The probability that the data are compatible with the background-only hypothesis is estimated by integrating the distribution of the log-likelihood ratio, approximated using the asymptotic formulae described in Ref.~\cite{Cowan:2010js-witherr}. In the absence of any significant excess above the expected background, upper limits at the $95\%$ confidence level (CL) on the signal production cross-section times the $W' \rightarrow tb$ decay branching ratio are derived using the $\text{CL}_\text{s}$ method~\cite{Read:2002hq}.
 
For left-handed $W'$ hypotheses, interference with $s$-channel single-top-quark production is included in the fit by changing the signal template shape. If the signal is scaled by a factor $\mu_{s}$, the interference contribution is scaled by $\sqrt{\mu_{s}}$ and the signal template is modified correspondingly by the interference contribution~\cite{EXOT-2016-04}.
 
The $m_{tb}$ distribution in the validation region of the \Zerol channel before the likelihood fit is shown in Figure~\ref{fig:distributions_0l}(a). There is good agreement between data and prediction, and the uncertainty becomes significant only for $m_{tb}$ above 3~\TeV. Figures~\ref{fig:distributions_0l}(b)--\ref{fig:distributions_0l}(d) show the $m_{tb}$ distributions in the three signal regions of the 0-lepton channel after a background-only fit to data. The pre-fit background prediction is also shown; it is very close to the post-fit background in all three regions. The maximum value of $m_{tb}$ observed in data is 7.8~\TeV, for an event in SR3.
 
\begin{figure}[htb!]
\centering
\subfloat[]{\includegraphics[width=0.45\textwidth]{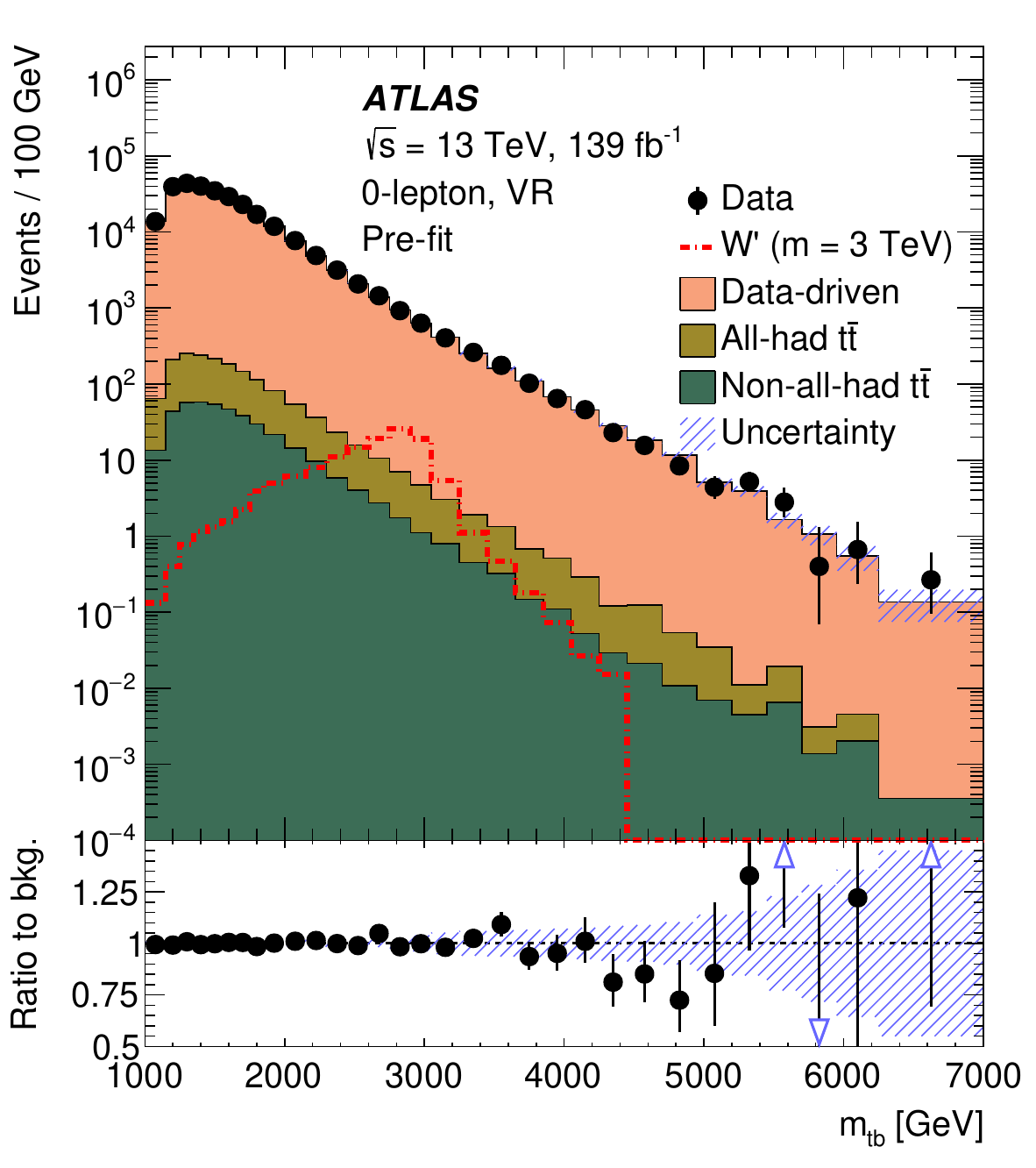}}
\subfloat[]{\includegraphics[width=0.45\textwidth]{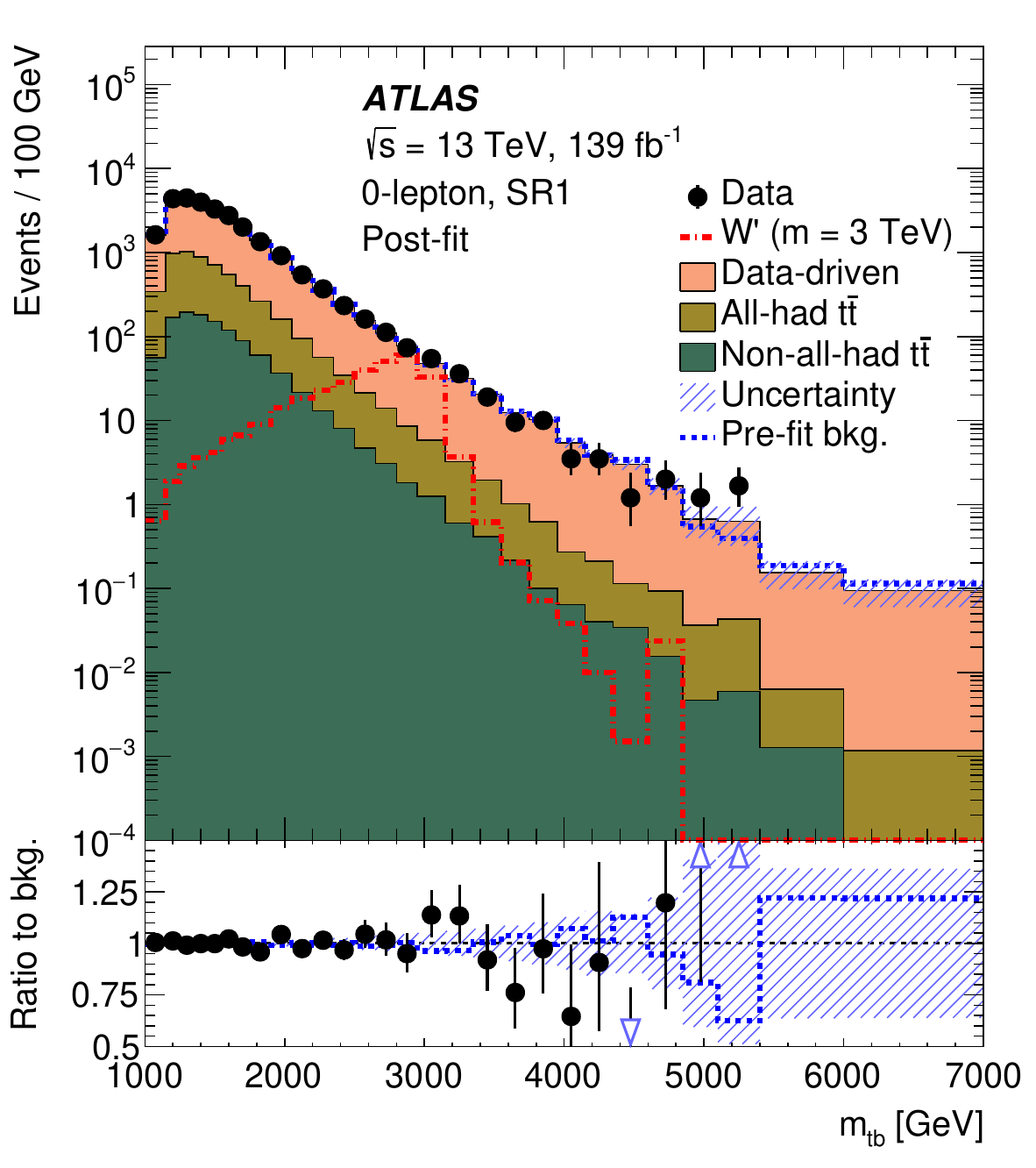}} \\
\subfloat[]{\includegraphics[width=0.45\textwidth]{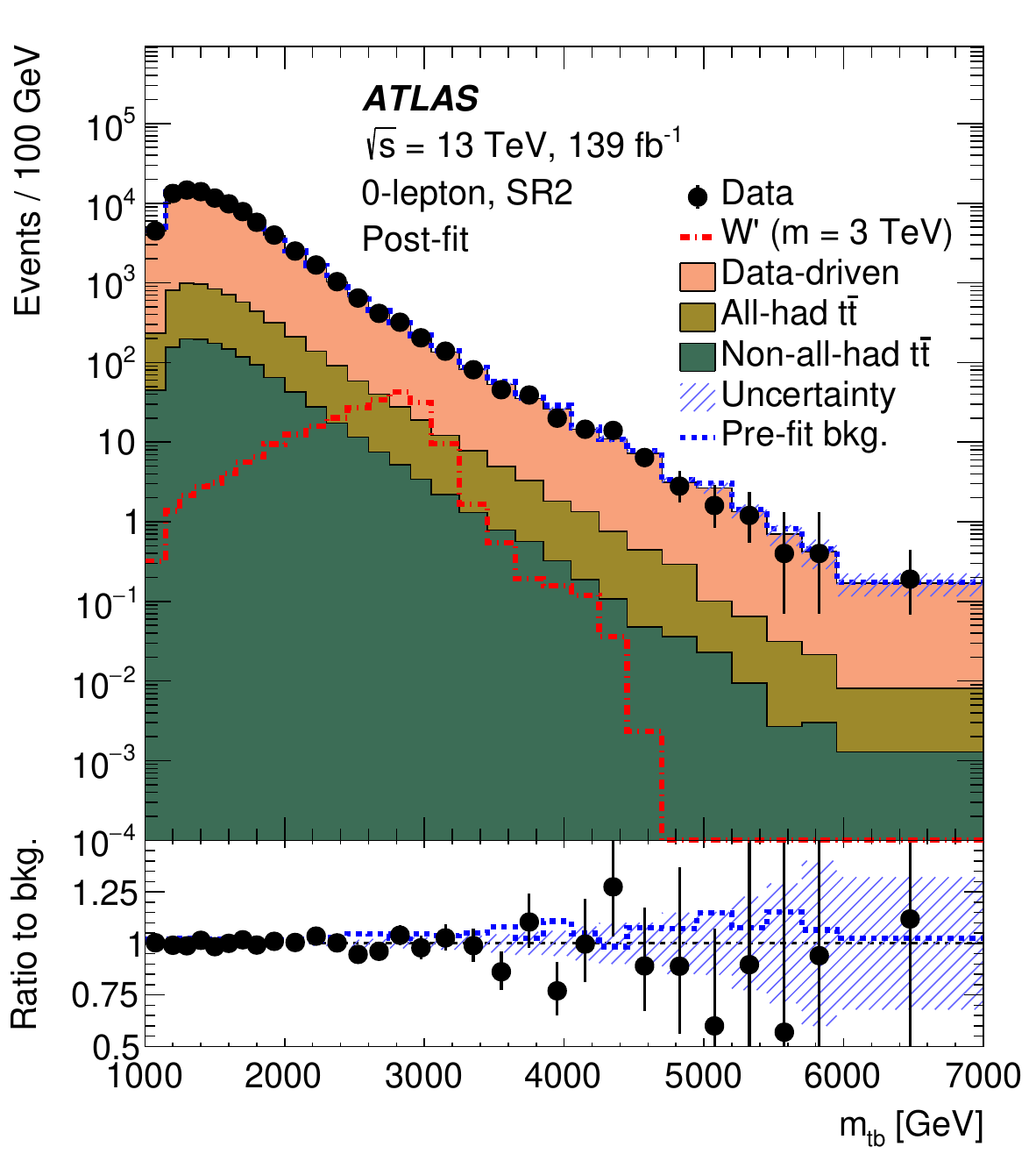}}
\subfloat[]{\includegraphics[width=0.45\textwidth]{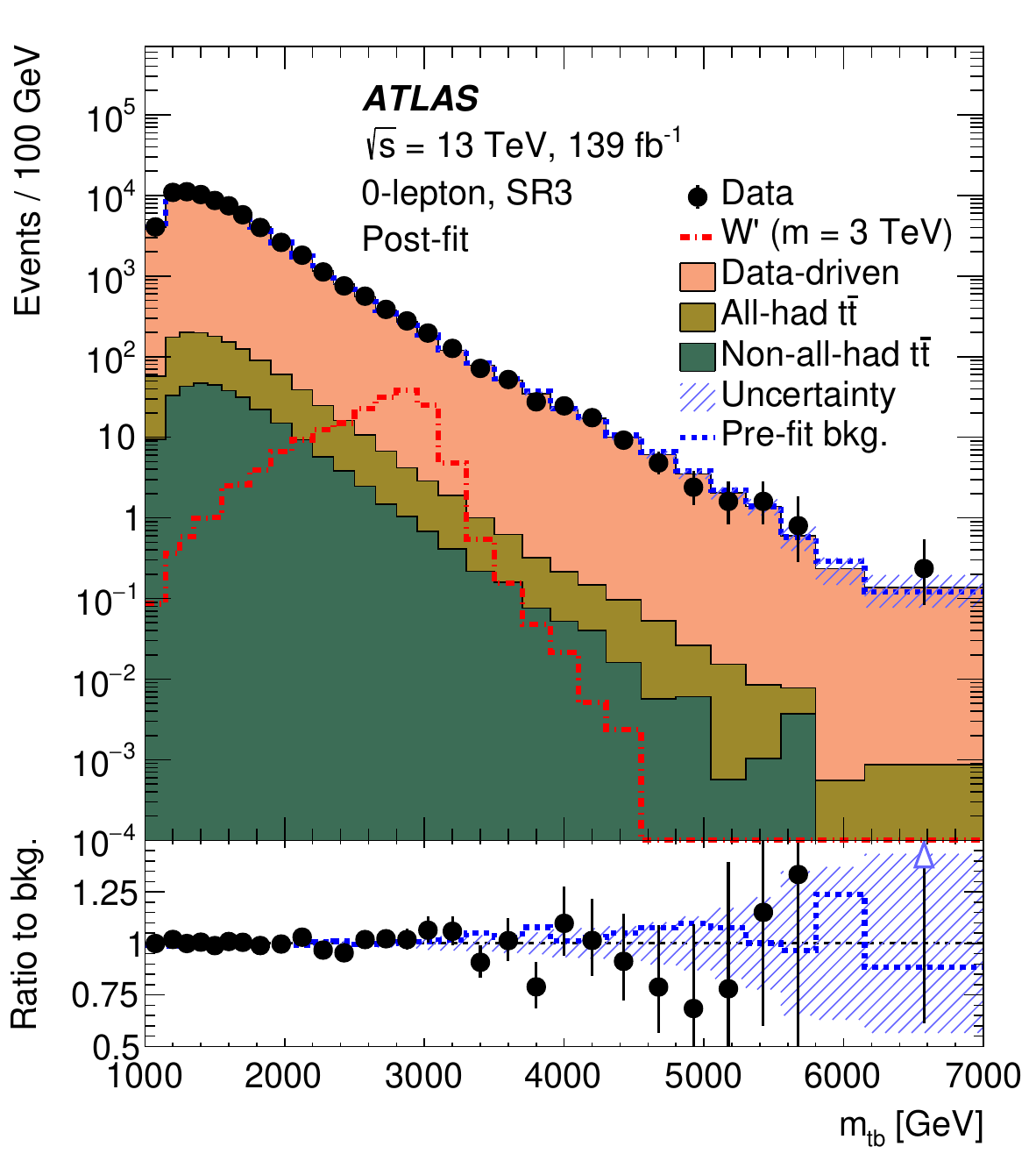}}
\caption{Distribution of the reconstructed $m_{tb}$ for data and backgrounds in the \Zerol channel in (a) the validation region before the fit to data, and the three signal regions after the background-only fit to data: (b) signal region 1, (c) signal region 2 and (d) signal region 3. The bottom panel in each plot shows the ratio of data to the background sum. For the signal regions, the dashed blue line shows the pre-fit background sum, and in the bottom panel the ratio of pre-fit to post-fit background sum. The hatched band includes all of the systematic uncertainties (a) before and (b, c, d) after the fit to data. The dashed red line shows the distribution of the $W'$-boson signal for a mass of 3~\TeV and a coupling strength of $g'/g=1.0$, normalised to the predicted cross-section. The last bin in each distribution includes overflow. The blue arrows in the ratio panel indicate that the data point is outside the range shown. }
\label{fig:distributions_0l}
\end{figure}
 
The distributions of  $m_{tb}$ in the \Onel channel are shown in Figure~\ref{fig:distributions_1l_cr} for the two control regions and the two $W$+jets validation regions after a background-only fit to data. Agreement is good in all \Onel-channel regions, with the uncertainty remaining relatively small until values of $m_{tb}$ are higher than 3.5~\TeV. Figure~\ref{fig:distributions_1l_sr} shows the distributions for the four signal regions. The post-fit normalisation factors for the $\ttbar$ background component have values of $0.89\pm 0.07$ and $0.92\pm 0.04$ for the 2-jet and 3-jet regions respectively. The corresponding factors for the $W$+jets background component are $1.19\pm0.07$ and $1.21\pm0.11$ for the 2-jet and 3-jet regions respectively.
 
\begin{figure}[htb!]
\centering
\subfloat[]{\includegraphics[width=0.45\textwidth]{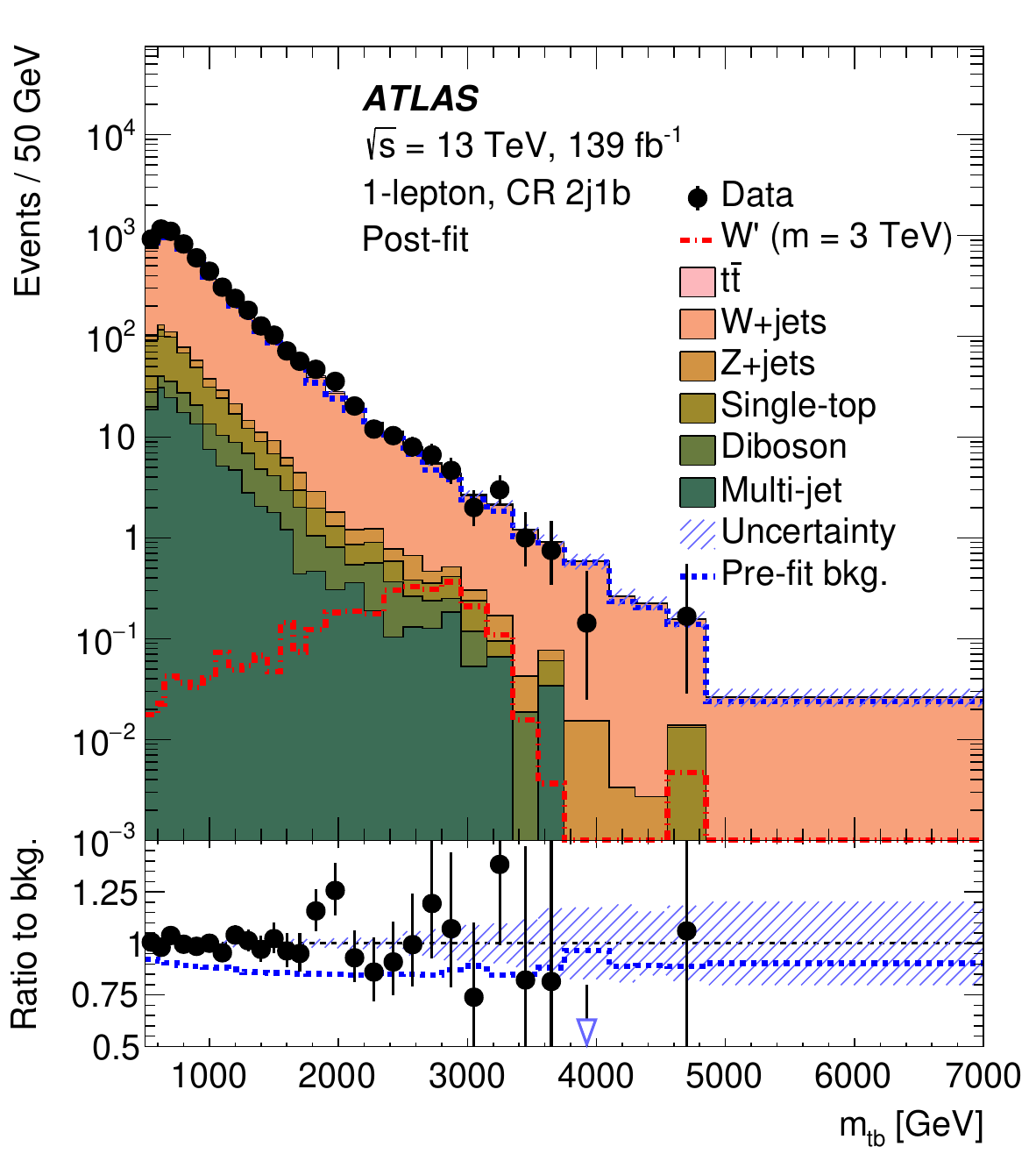}}
\subfloat[]{\includegraphics[width=0.45\textwidth]{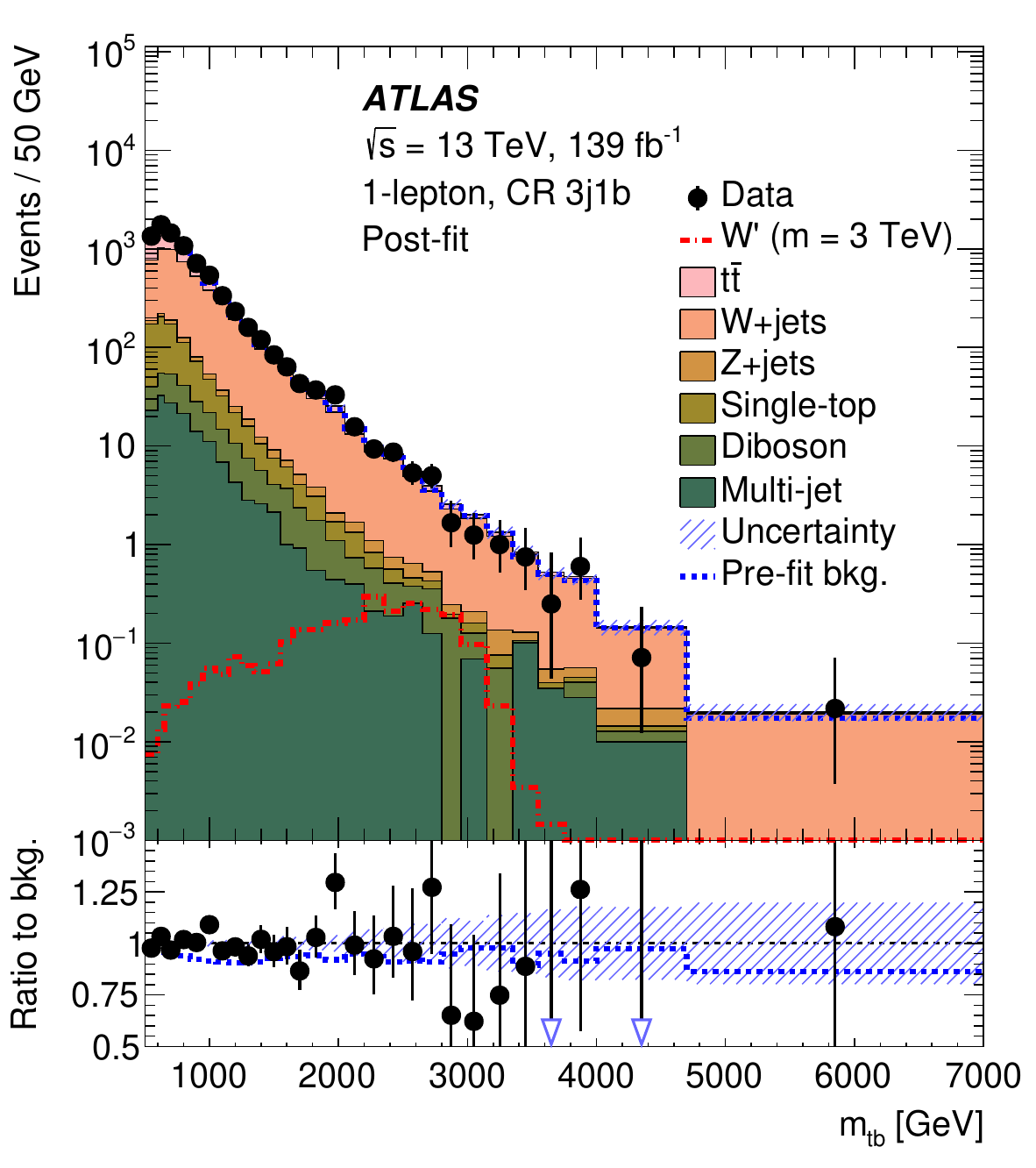}} \\
\subfloat[]{\includegraphics[width=0.45\textwidth]{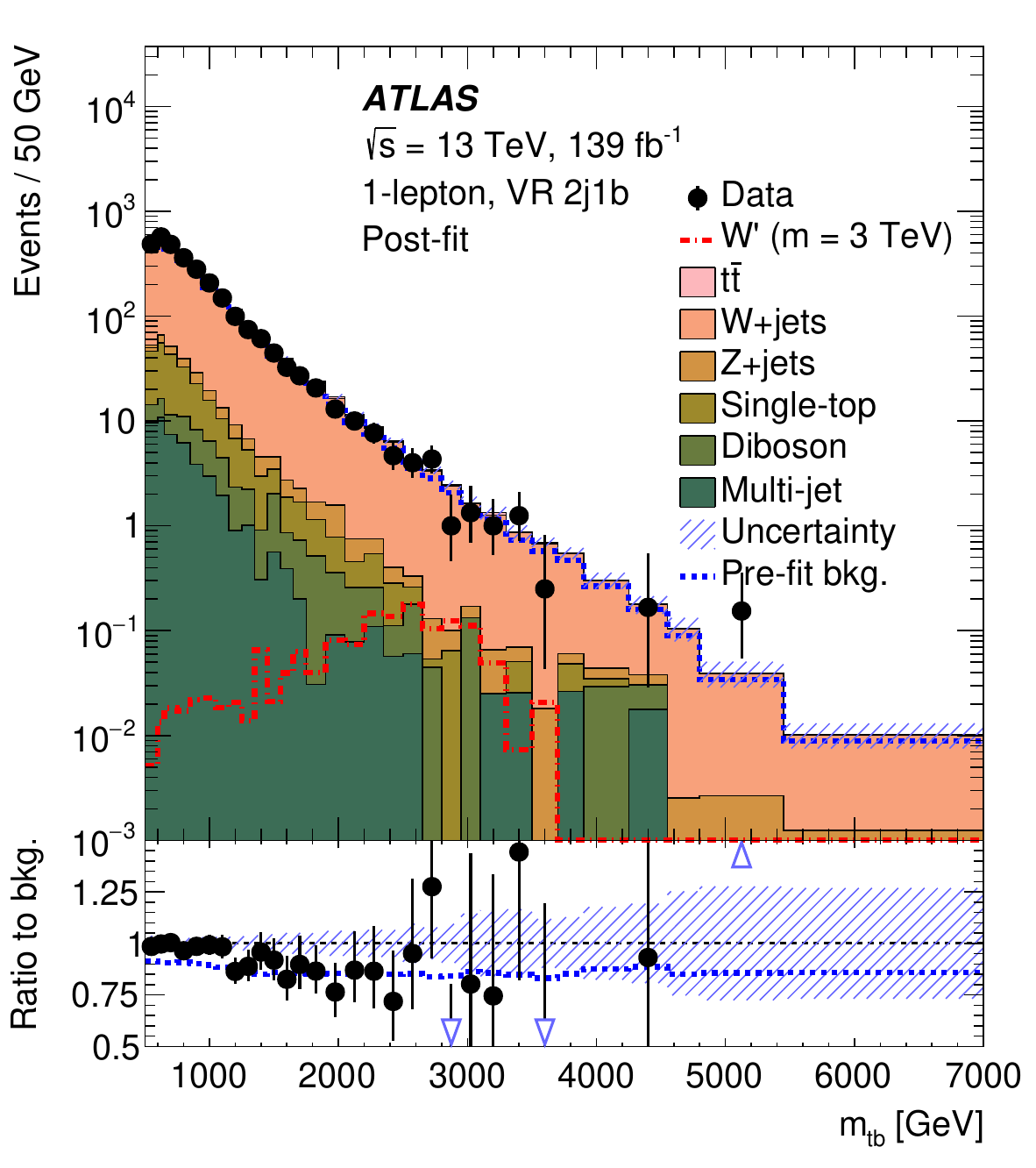}}
\subfloat[]{\includegraphics[width=0.45\textwidth]{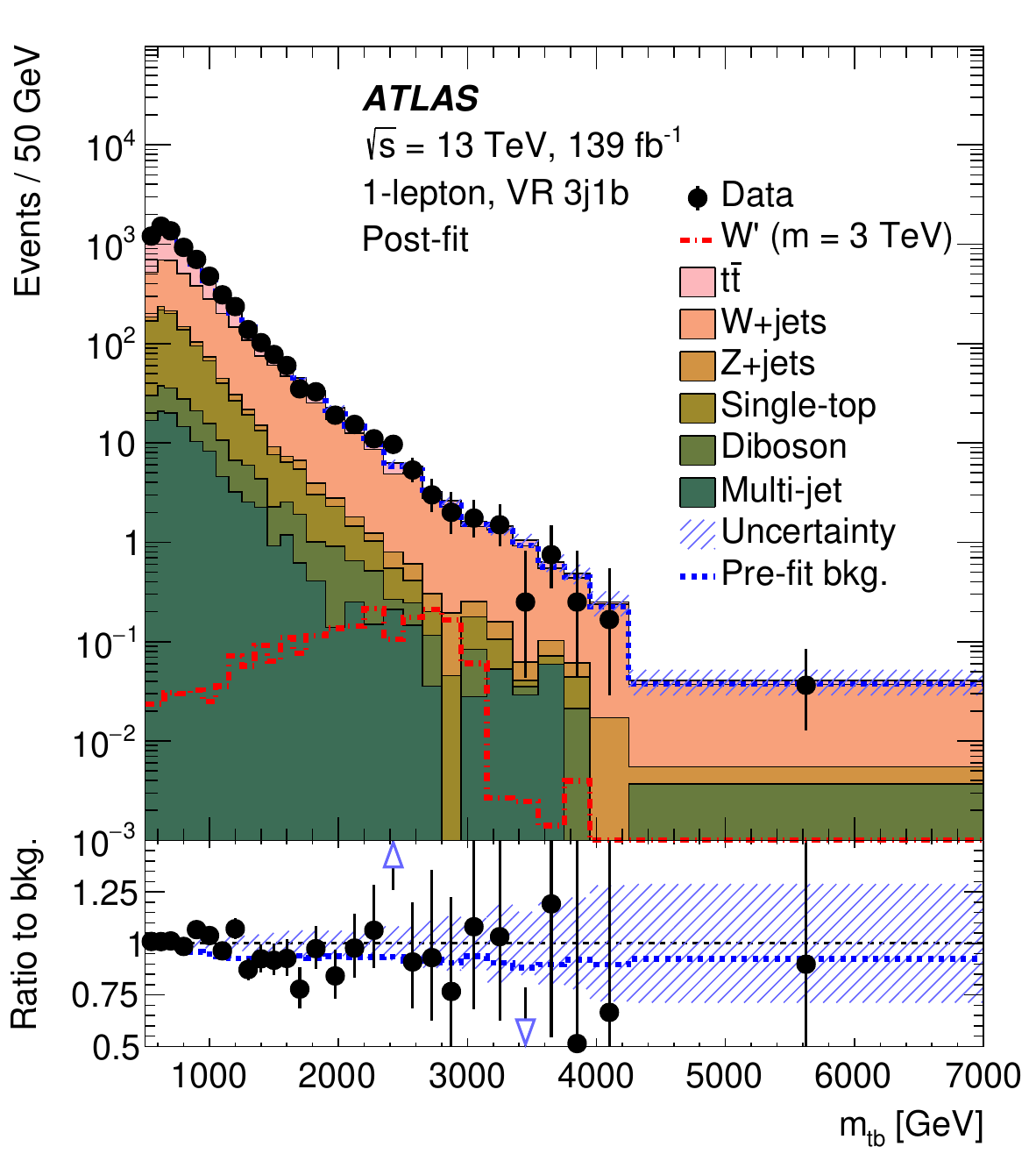}}
\caption{Reconstructed $m_{tb}$ distributions for data and backgrounds in the \Onel channel, (a) and (b) in the control regions, and (c) and (d) in the $W$+jets validation regions. They are shown after the background-only fit to data. Each bottom panel shows the ratio of data to the background sum. The dashed blue line shows the pre-fit background sum, and in the bottom panel the ratio of pre-fit to post-fit background sum. The hatched band includes all of the systematic uncertainties after the fit to data. The dashed red line shows the distribution of the $W'$-boson signal for a mass of 3~\TeV and a coupling strength of $g'/g=1.0$, normalised to the predicted cross-section. The last bin in each distribution includes overflow. The blue arrows in the ratio panel indicate that the data point is outside the range shown. }
\label{fig:distributions_1l_cr}
\end{figure}

\begin{figure}[htb!]
\centering
\subfloat[]{\includegraphics[width=0.45\textwidth]{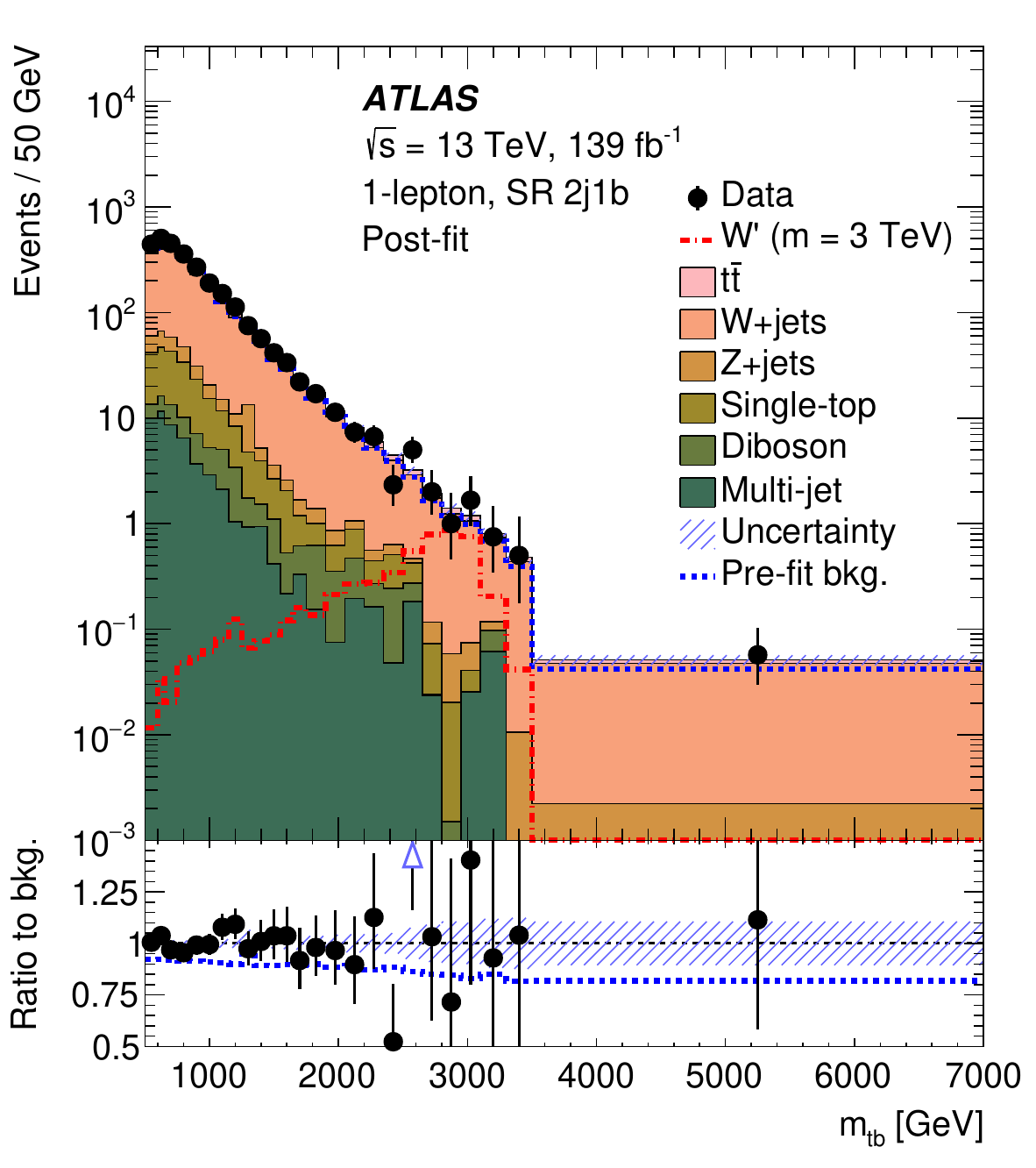}}
\subfloat[]{\includegraphics[width=0.45\textwidth]{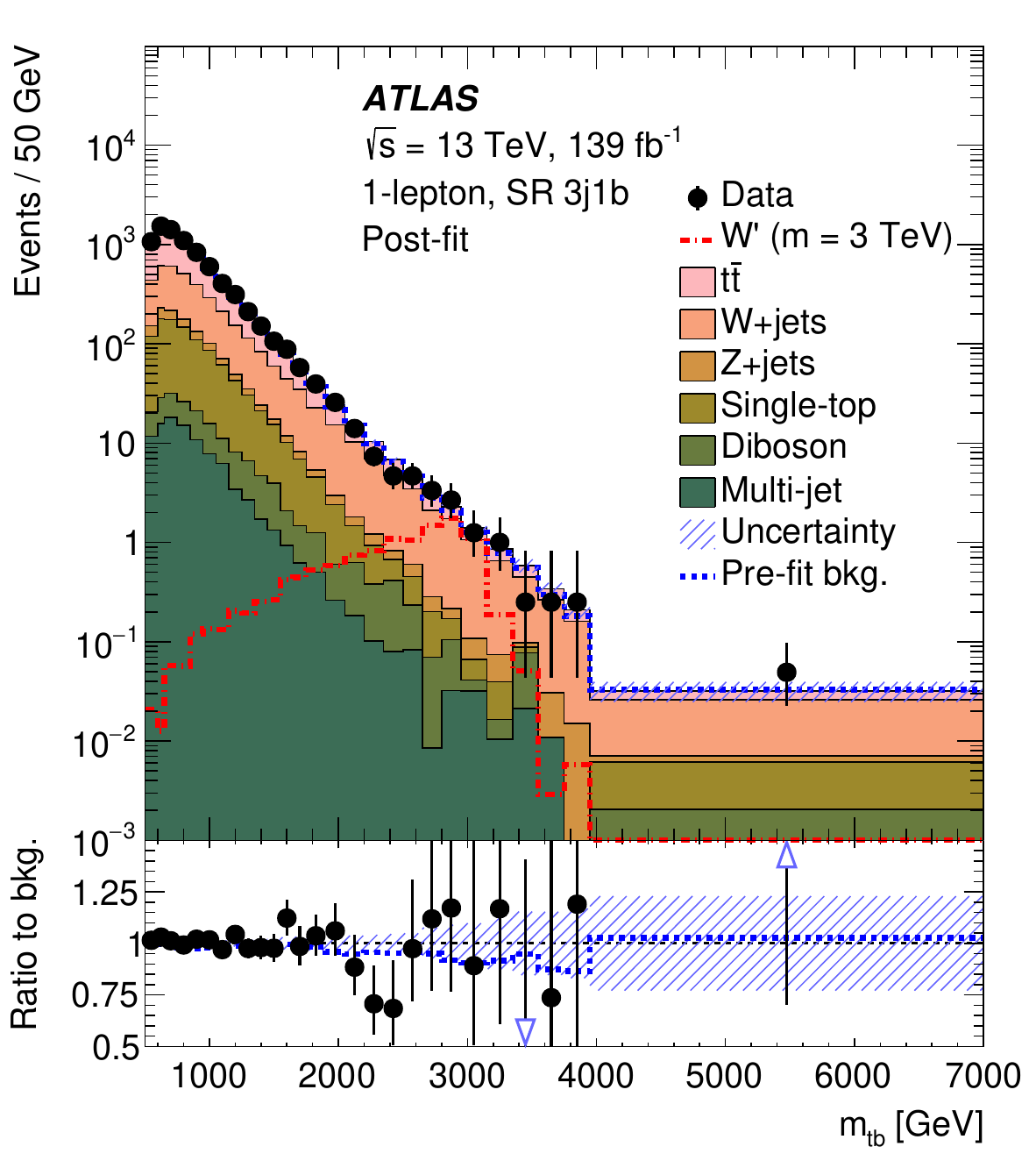}} \\
\subfloat[]{\includegraphics[width=0.45\textwidth]{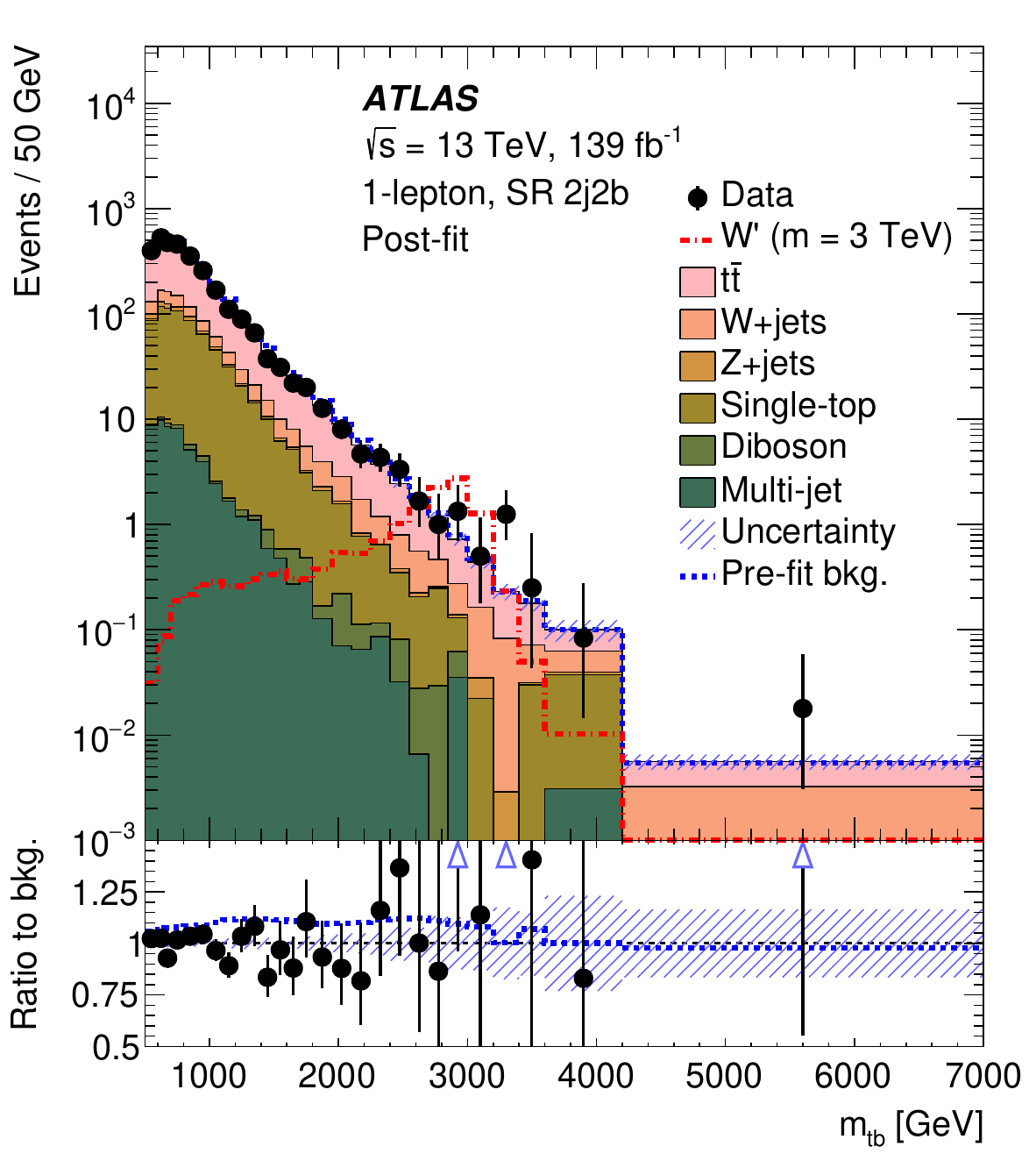}}
\subfloat[]{\includegraphics[width=0.45\textwidth]{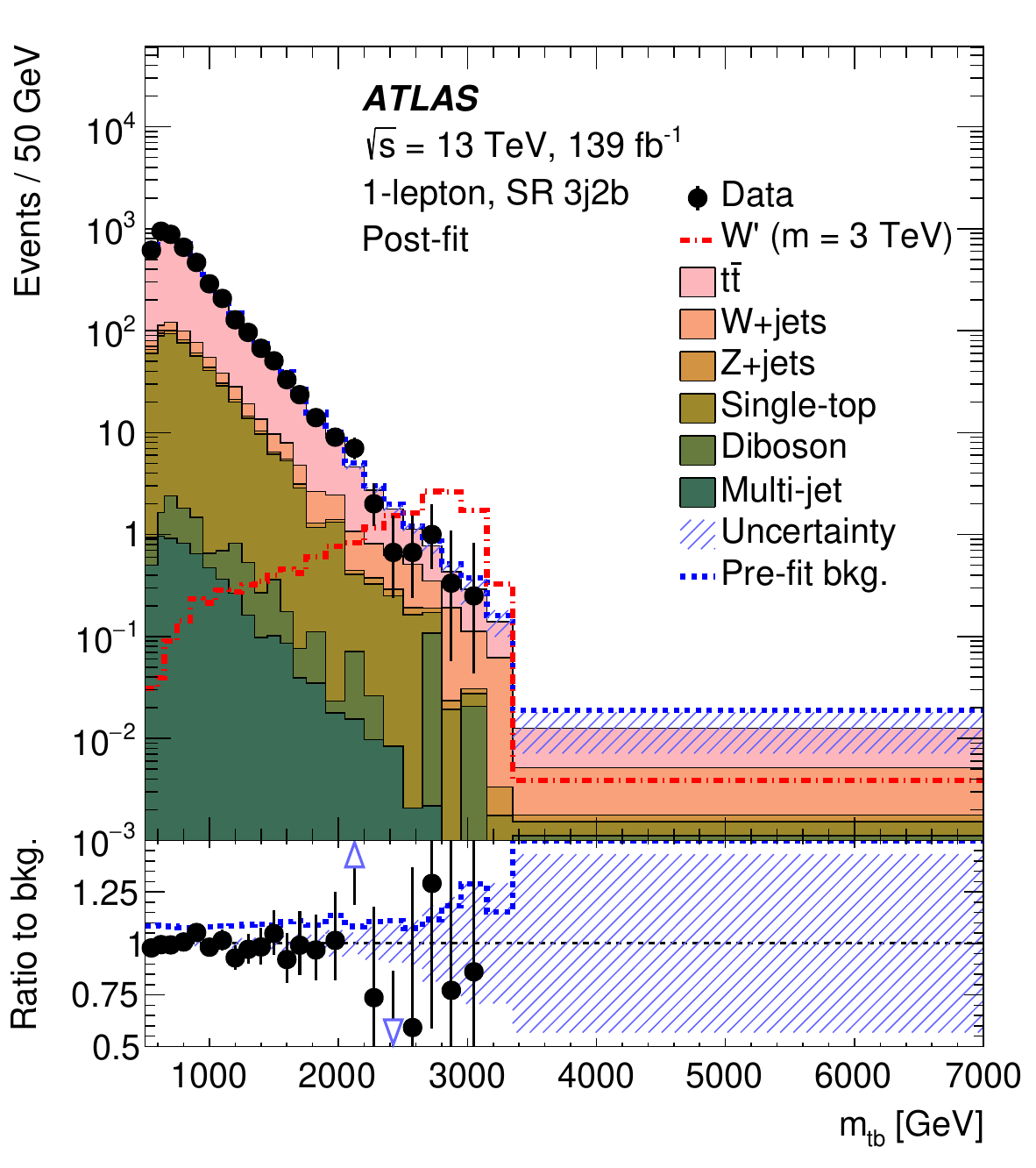}}
\caption{Reconstructed $m_{tb}$ distributions for data and backgrounds in the four \Onel signal regions after the background-only fit to data. Each bottom panel shows the ratio of data to the background sum. For the signal regions, the dashed blue line shows the pre-fit background sum, and in the bottom panel the ratio of pre-fit to post-fit background sum. The hatched band includes all of the systematic uncertainties after the fit to data. The dashed red line shows the distribution of the $W'$-boson signal for a mass of 3~\TeV and a coupling strength of $g'/g=1.0$, normalised to the predicted cross-section. The last bin in each distribution includes overflow. The blue arrows in the ratio panel indicate that the data point is outside the range shown. }
\label{fig:distributions_1l_sr}
\end{figure}

Good agreement between the background prediction and data is observed in all regions. Upper limits on the production cross-section times decay branching ratio as a function of the $W'$-boson mass are therefore derived and are shown in Figure~\ref{fig:limits_rh} for a right-handed $W'$~boson and in Figure~\ref{fig:limits_lh} for a left-handed $W'$~boson. For each chirality, three different values of $g'/g$ are used to generate upper limits.  In all cases, the expected limit in each channel is shown in addition to the combination. The observed limits and expected limits are derived by linear interpolation between those obtained for several different signal mass hypotheses. Mass-limit values are obtained from the intersection of the limit curves with the theory curve, obtained at NLO using \textsc{ZTOP}~\cite{Duffty_2012}. For a right-handed $W'$~boson, masses below 4.6~\TeV\ (4.2~\TeV) are observed (expected) to be excluded for a $g'/g$ value of 1.0, while for a left-handed $W'$~boson, masses below 4.2~\TeV\ (4.1~\TeV) are observed (expected) to be excluded for the same coupling value. The observed limits are higher than expected because of statistical fluctuations of the data around $m_{tb}=4$~\TeV\ in the signal regions of the \Zerol channel. The lower exclusion limits obtained for a left-handed $W'$~boson are partially explained by the higher branching ratio to $tb$ in the right-handed scenario, where the $W'$~boson cannot decay leptonically. The sensitivity to high $W'$-boson masses is limited by statistical uncertainties.  For a right-handed $W'$~boson with $g'/g=1.0$, the expected mass limit is more than 1~\TeV\ higher than in the previous combination of the two channels~\cite{EXOT-2016-18}. The mass limit for a left-handed $W'$~boson is also a more than 1~\TeV\ improvement on the previous 0-lepton-channel-only results~\cite{EXOT-2017-02}.
 
Figures~\ref{fig:2d_limits}(a) and \ref{fig:2d_limits}(b) show the observed and expected exclusion contours as functions of the $W'$-boson mass and coupling strength for the right-handed and left-handed hypotheses respectively. The interpolation between coupling values is performed using a quadratic function. In both figures, the expected limit in each channel is shown in addition to the combination. For low $W'$-boson masses, the \Onel channel dominates the sensitivity because the large multi-jet background reduces the sensitivity in the \Zerol channel. For high $W'$-boson masses, the efficiency of the signal selection in the \Onel channel decreases due to the lepton isolation requirement, while the \Zerol channel remains highly efficient. For very high coupling strengths, the width of the $W'$~boson increases and the reconstructed signal peaks become very wide or disappear completely. In this scenario, the signal distributions shift towards lower $m_{tb}$ values, making the \Onel channel competitive even at high $W'$ masses.
 
\begin{figure}[htb!]
\centering
\subfloat[]{\includegraphics[width=0.45\textwidth]{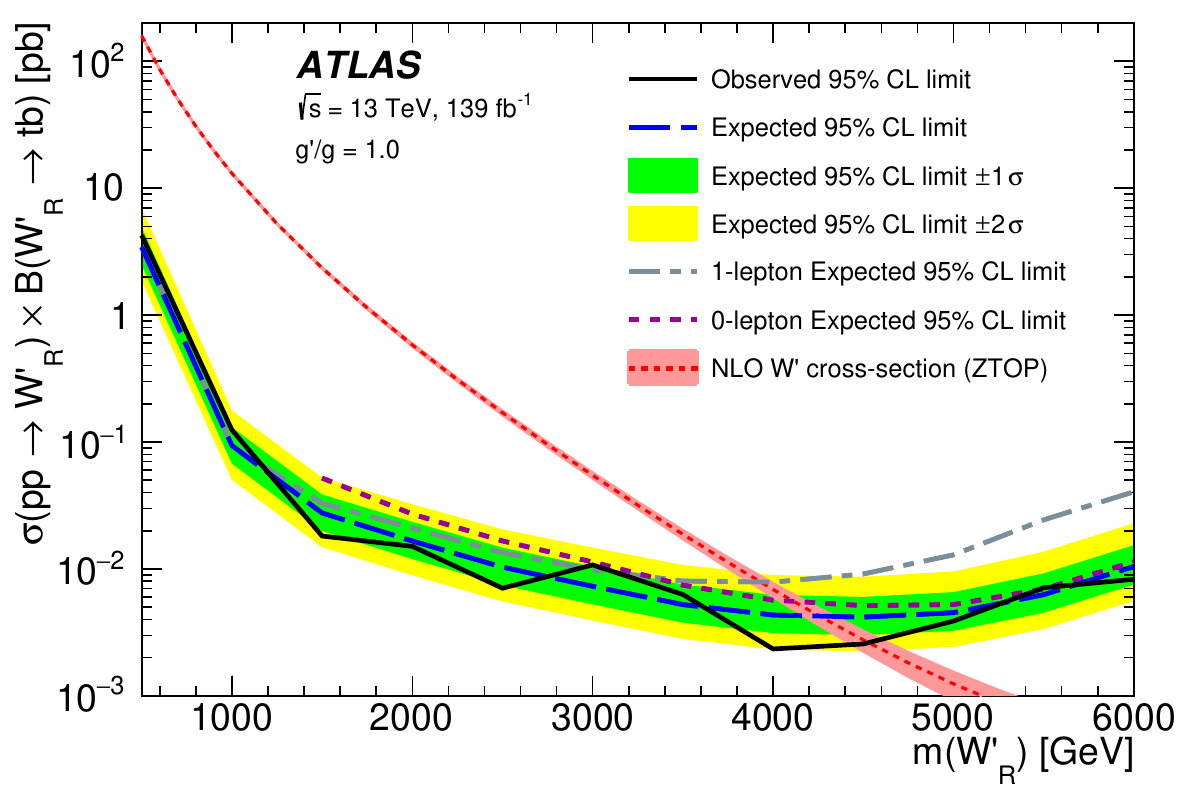}}\\
\subfloat[]{\includegraphics[width=0.45\textwidth]{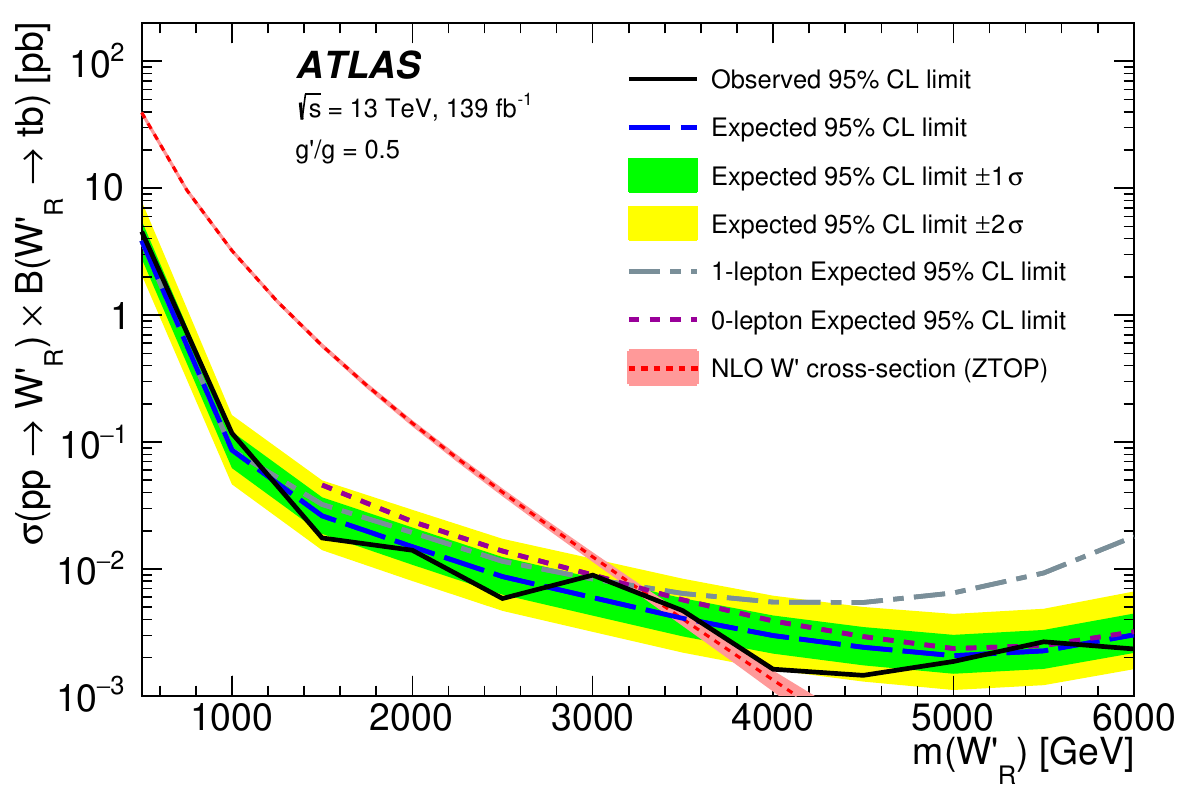}}
\subfloat[]{\includegraphics[width=0.45\textwidth]{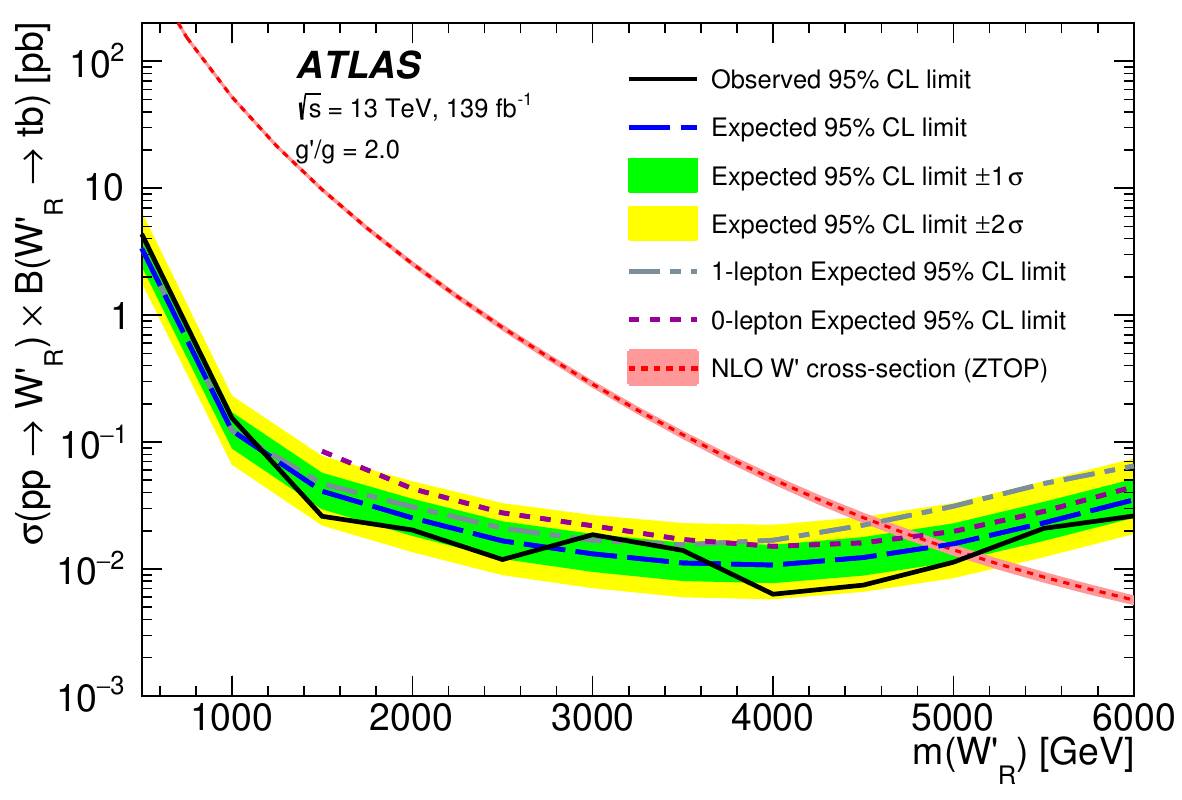}}
\caption{Observed and expected 95\% CL limits on the cross-section times branching ratio for the production of a $W'$~boson with decay into $tb$ and right-handed couplings as a function of the mass of the $W'$~boson and a coupling value of (a) $g'/g=1.0$, (b) $g'/g=0.5$ and (c) $g'/g=2.0$. They are obtained from the combination of the \Zerol and \Onel channels. The expected limits for the individual channels are also shown. The observed limits and expected limits are derived by linear interpolation between those obtained for several different signal mass hypotheses. The uncertainty in the theory prediction includes components from the factorisation and renormalisation scales, PDFs, strong coupling constant, and top-quark mass.}
\label{fig:limits_rh}
\end{figure}
 
\begin{figure}[htb!]
\centering
\subfloat[]{\includegraphics[width=0.45\textwidth]{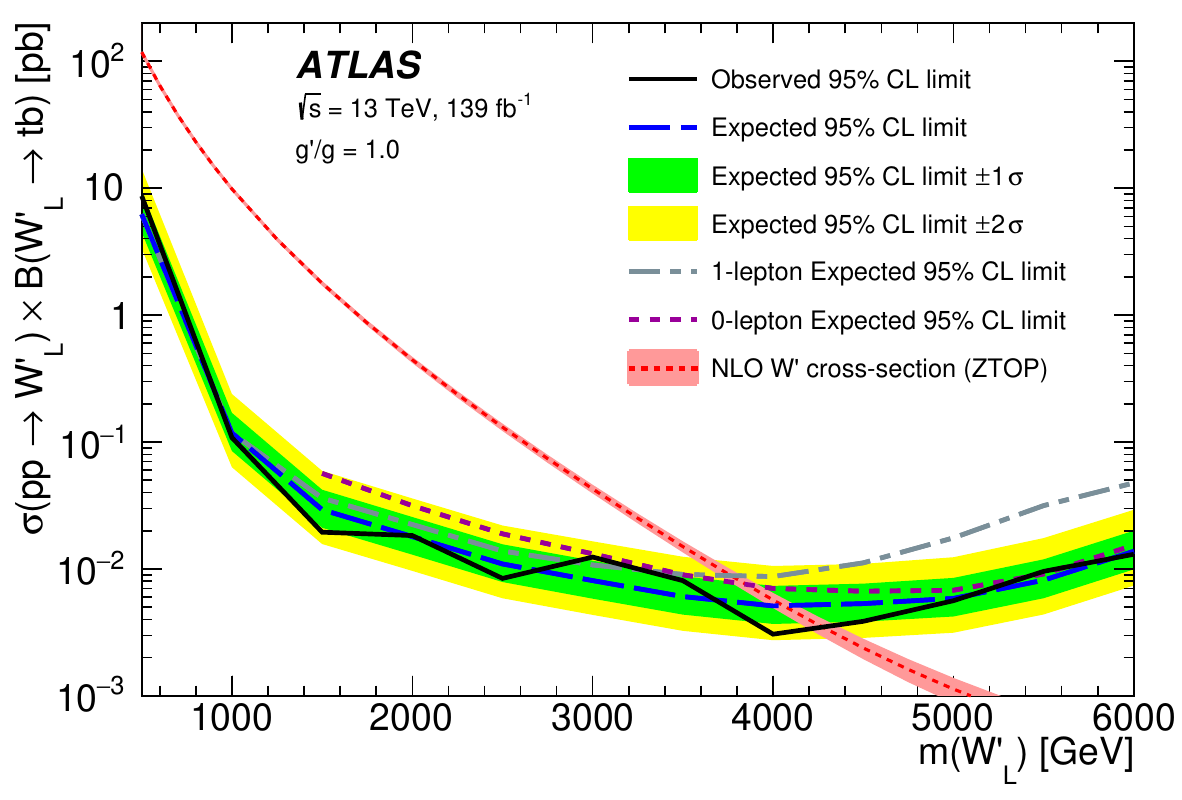}}\\
\subfloat[]{\includegraphics[width=0.45\textwidth]{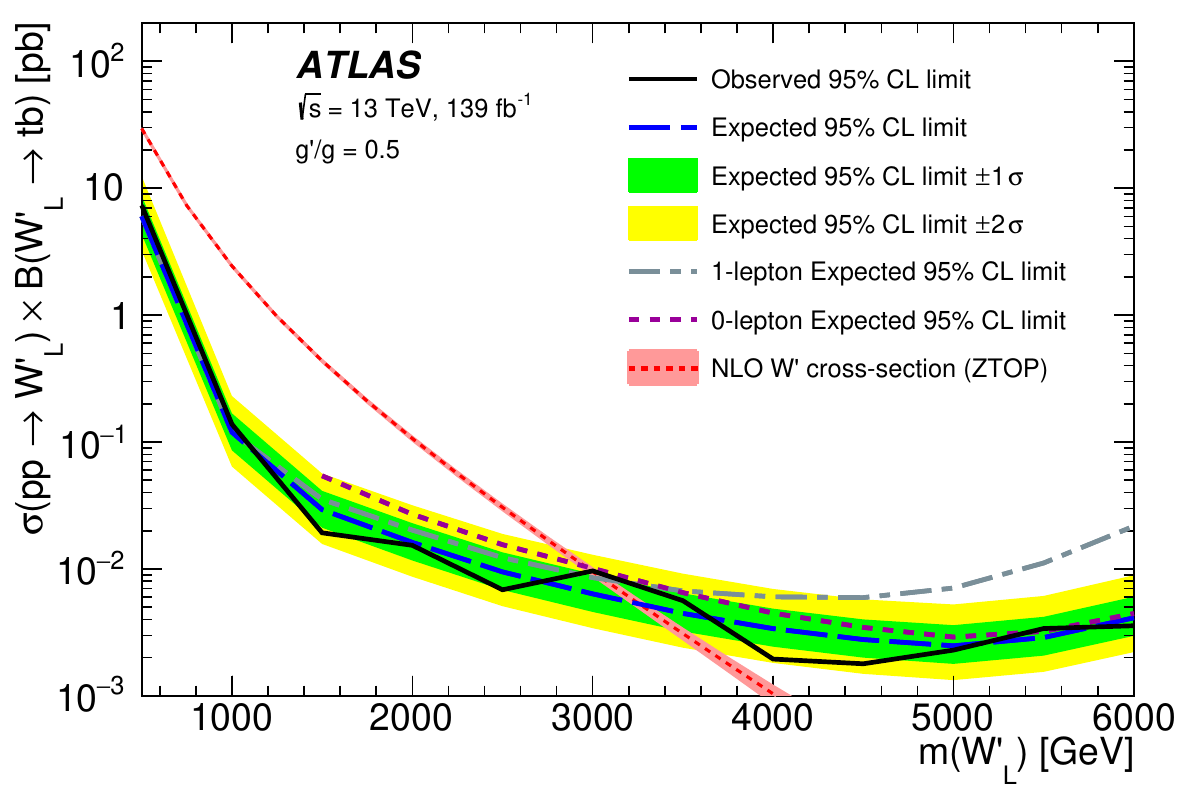}}
\subfloat[]{\includegraphics[width=0.45\textwidth]{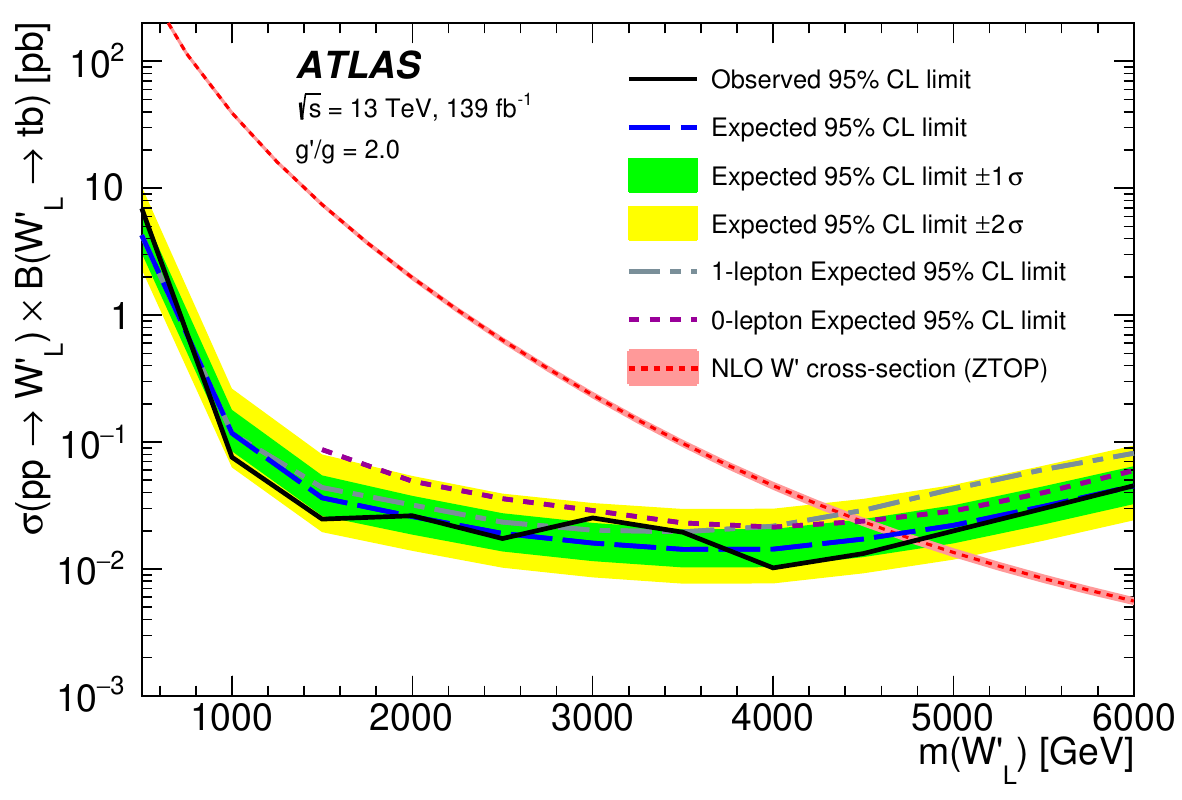}}
\caption{Observed and expected 95\% CL limits on the cross-section times branching ratio for the production of a $W'$~boson with decay into $tb$ and left-handed couplings as a function of the mass of the $W'$~boson and a coupling value of (a) $g'/g=1.0$, (b) $g'/g=0.5$ and (c) $g'/g=2.0$. They are obtained from the combination of the \Zerol and \Onel channels. The expected limits for the individual channels are also shown. The observed limits and expected limits are derived by linear interpolation between those obtained for several different signal mass hypotheses. The uncertainty in the theory prediction includes components from the factorisation and renormalisation scales, PDFs, strong coupling constant, and top-quark mass.}
\label{fig:limits_lh}
\end{figure}

\begin{figure}[htb!]
\centering
\subfloat[]{\includegraphics[width=0.45\textwidth]{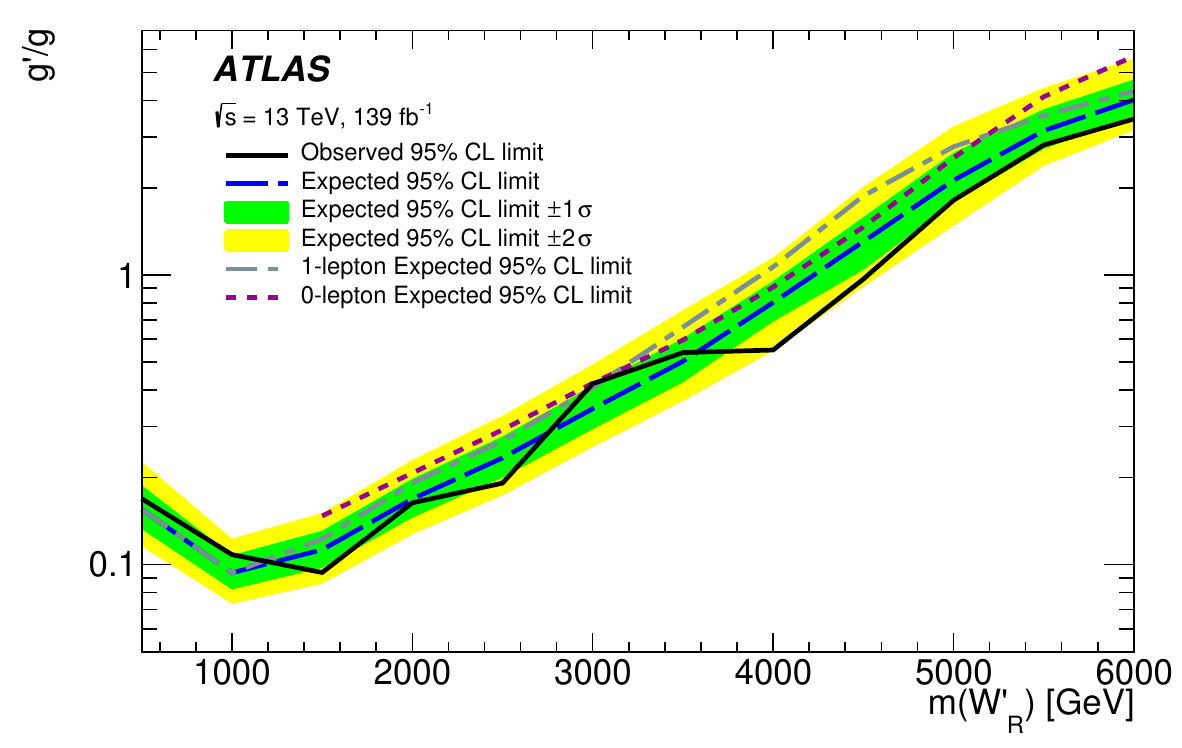}}
\subfloat[]{\includegraphics[width=0.45\textwidth]{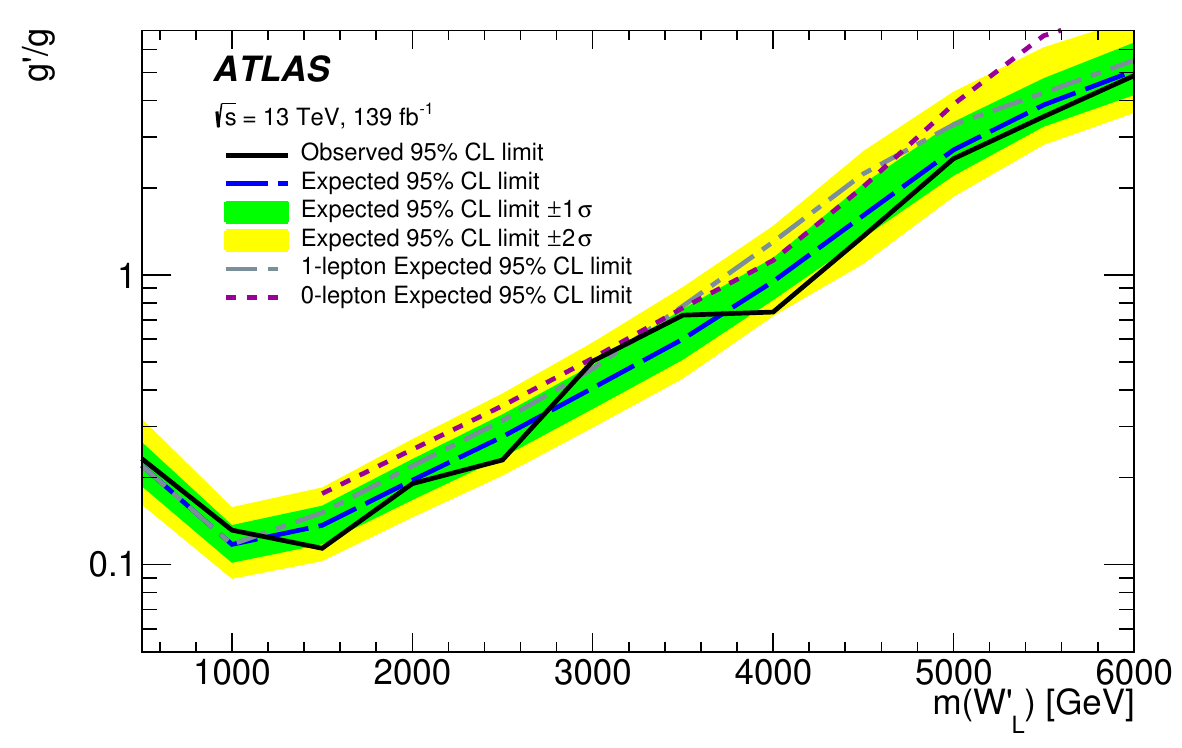}}
\caption{Observed and expected limits as a function of the coupling value and the $W'$-boson mass for (a) right-handed and (b) left-handed $W'$-boson couplings. They are obtained from the combination of the \Zerol and \Onel channels. The expected limits for the individual channels are also shown. The area above the line is excluded.}
\label{fig:2d_limits}
\end{figure}

% End of text imported from the .//TeX/results.tex input file

% The next lines are included from the .//TeX/summary.tex input file
\FloatBarrier
\section{Conclusions}
\label{sec:summary}
 
A search for $W'\rightarrow tb$ using 139~$\ifb$ of $\sqrt{s}=13$~\TeV\ $pp$ collision data collected with the ATLAS detector at the LHC is presented. The search combines two channels, named according to the targeted decay of the top quark. The \Zerol channel employs a DNN-based algorithm to identify large-radius jets originating from hadronically decaying top quarks. They are combined with small-radius jets selected with a $b$-tagging algorithm to reconstruct the $W'$~boson. The dominant background from multi-jet production is estimated using a data-driven method. The \Onel channel selects events with one lepton (electron or muon), a certain amount of \met, and two or more jets. These objects are combined using top-quark and $W$-boson mass constraints to reconstruct the $W'$~boson. The dominant backgrounds come from $\ttbar$ and $W$+jets production.
 
The observed distributions of the reconstructed $W'$-boson mass in various analysis regions are consistent with the background-only prediction, and exclusion limits at 95\% CL are set on the production cross-section times branching ratio for $W'\rightarrow tb$. Several signal hypotheses are considered: $W'$-boson masses in the range \numrange[range-phrase = --]{0.5}{6}~\TeV, right-handed and left-handed couplings, and different coupling strengths relative to the coupling of the $W$~boson to fermions in the SM. Effects of interference between the left-handed $W'$~boson and the SM $W$~boson are taken into account.
 
Right-handed $W'$~bosons with masses below 4.6~\TeV\ (4.2~\TeV) are observed (expected) to be excluded for a coupling value of $g'/g = 1.0$. For the same coupling value, left-handed $W'$~bosons with masses below 4.2~\TeV\ (4.1~\TeV) are observed (expected) to be excluded.
The observed mass limits for right-handed $W'$~bosons with $g'/g=1.0$ are more than 1~\TeV\ higher than in the previous 0-lepton-channel CMS search and the previous ATLAS combination of the two channels. The observed mass limits for left-handed $W'$~bosons with the same coupling strength are approximately 0.8~\TeV\ higher than in the previous 0-lepton-channel CMS search. The obtained limits are the most stringent to date.

% End of text imported from the .//TeX/summary.tex input file

\section*{Acknowledgements}

% The next lines are included from the .//acknowledgements/Acknowledgements.tex input file

We thank CERN for the very successful operation of the LHC, as well as the
support staff from our institutions without whom ATLAS could not be
operated efficiently.
 
We acknowledge the support of
ANPCyT, Argentina;
YerPhI, Armenia;
ARC, Australia;
BMWFW and FWF, Austria;
ANAS, Azerbaijan;
CNPq and FAPESP, Brazil;
NSERC, NRC and CFI, Canada;
CERN;
ANID, Chile;
CAS, MOST and NSFC, China;
Minciencias, Colombia;
MEYS CR, Czech Republic;
DNRF and DNSRC, Denmark;
IN2P3-CNRS and CEA-DRF/IRFU, France;
SRNSFG, Georgia;
BMBF, HGF and MPG, Germany;
GSRI, Greece;
RGC and Hong Kong SAR, China;
ISF and Benoziyo Center, Israel;
INFN, Italy;
MEXT and JSPS, Japan;
CNRST, Morocco;
NWO, Netherlands;
RCN, Norway;
MEiN, Poland;
FCT, Portugal;
MNE/IFA, Romania;
MESTD, Serbia;
MSSR, Slovakia;
ARRS and MIZ\v{S}, Slovenia;
DSI/NRF, South Africa;
MICINN, Spain;
SRC and Wallenberg Foundation, Sweden;
SERI, SNSF and Cantons of Bern and Geneva, Switzerland;
MOST, Taiwan;
TENMAK, T\"urkiye;
STFC, United Kingdom;
DOE and NSF, United States of America.
In addition, individual groups and members have received support from
BCKDF, CANARIE, Compute Canada and CRC, Canada;
PRIMUS 21/SCI/017 and UNCE SCI/013, Czech Republic;
COST, ERC, ERDF, Horizon 2020, ICSC-NextGenerationEU and Marie Sk{\l}odowska-Curie Actions, European Union;
Investissements d'Avenir Labex, Investissements d'Avenir Idex and ANR, France;
DFG and AvH Foundation, Germany;
Herakleitos, Thales and Aristeia programmes co-financed by EU-ESF and the Greek NSRF, Greece;
BSF-NSF and MINERVA, Israel;
Norwegian Financial Mechanism 2014-2021, Norway;
NCN and NAWA, Poland;
La Caixa Banking Foundation, CERCA Programme Generalitat de Catalunya and PROMETEO and GenT Programmes Generalitat Valenciana, Spain;
G\"{o}ran Gustafssons Stiftelse, Sweden;
The Royal Society and Leverhulme Trust, United Kingdom.
 
The crucial computing support from all WLCG partners is acknowledged gratefully, in particular from CERN, the ATLAS Tier-1 facilities at TRIUMF (Canada), NDGF (Denmark, Norway, Sweden), CC-IN2P3 (France), KIT/GridKA (Germany), INFN-CNAF (Italy), NL-T1 (Netherlands), PIC (Spain), ASGC (Taiwan), RAL (UK) and BNL (USA), the Tier-2 facilities worldwide and large non-WLCG resource providers. Major contributors of computing resources are listed in Ref.~\cite{ATL-SOFT-PUB-2023-001}.

% End of text imported from the .//acknowledgements/Acknowledgements.tex input file

\clearpage
\printbibliography
 
\clearpage
\input{atlas_authlist}

\end{document}

%% file: atlas_authlist.tex
% ATLAS Collaboration author list
% Reference date of EXOT-2021-36 is 2023-04-28
% Author list last updated on date 28-NOV-23
% Data extracted on 28-Nov-2023 for paper reference EXOT-2021-36
% at 10:10am
 
\begin{flushleft}
\hypersetup{urlcolor=black}
{\Large The ATLAS Collaboration}

\bigskip

\AtlasOrcid[0000-0002-6665-4934]{G.~Aad}$^\textrm{\scriptsize 102}$,
\AtlasOrcid[0000-0002-5888-2734]{B.~Abbott}$^\textrm{\scriptsize 120}$,
\AtlasOrcid[0000-0002-1002-1652]{K.~Abeling}$^\textrm{\scriptsize 55}$,
\AtlasOrcid[0000-0001-5763-2760]{N.J.~Abicht}$^\textrm{\scriptsize 49}$,
\AtlasOrcid[0000-0002-8496-9294]{S.H.~Abidi}$^\textrm{\scriptsize 29}$,
\AtlasOrcid[0000-0002-9987-2292]{A.~Aboulhorma}$^\textrm{\scriptsize 35e}$,
\AtlasOrcid[0000-0001-5329-6640]{H.~Abramowicz}$^\textrm{\scriptsize 151}$,
\AtlasOrcid[0000-0002-1599-2896]{H.~Abreu}$^\textrm{\scriptsize 150}$,
\AtlasOrcid[0000-0003-0403-3697]{Y.~Abulaiti}$^\textrm{\scriptsize 117}$,
\AtlasOrcid[0000-0003-0762-7204]{A.C.~Abusleme~Hoffman}$^\textrm{\scriptsize 137a}$
% ...
\bigskip
\\

% ...
$^{29}$Physics Department, Brookhaven National Laboratory, Upton NY; United States of America.\\
% ...
$^{35}$$^{(a)}$Facult\'e des Sciences Ain Chock, R\'eseau Universitaire de Physique des Hautes Energies - Universit\'e Hassan II, Casablanca;$^{(b)}$Facult\'{e} des Sciences, Universit\'{e} Ibn-Tofail, K\'{e}nitra;$^{(c)}$Facult\'e des Sciences Semlalia, Universit\'e Cadi Ayyad, LPHEA-Marrakech;$^{(d)}$LPMR, Facult\'e des Sciences, Universit\'e Mohamed Premier, Oujda;$^{(e)}$Facult\'e des sciences, Universit\'e Mohammed V, Rabat;$^{(f)}$Institute of Applied Physics, Mohammed VI Polytechnic University, Ben Guerir; Morocco.\\
% ...
$^{49}$Fakult\"{a}t Physik , Technische Universit{\"a}t Dortmund, Dortmund; Germany.\\
% ...
$^{55}$II. Physikalisches Institut, Georg-August-Universit\"{a}t G\"ottingen, G\"ottingen; Germany.\\
% ...
$^{102}$CPPM, Aix-Marseille Universit\'e, CNRS/IN2P3, Marseille; France.\\
% ...
$^{117}$Department of Physics, New York University, New York NY; United States of America.\\
% ...
$^{120}$Homer L. Dodge Department of Physics and Astronomy, University of Oklahoma, Norman OK; United States of America.\\
% ...
$^{137}$$^{(a)}$Departamento de F\'isica, Pontificia Universidad Cat\'olica de Chile, Santiago;$^{(b)}$Millennium Institute for Subatomic physics at high energy frontier (SAPHIR), Santiago;$^{(c)}$Instituto de Investigaci\'on Multidisciplinario en Ciencia y Tecnolog\'ia, y Departamento de F\'isica, Universidad de La Serena;$^{(d)}$Universidad Andres Bello, Department of Physics, Santiago;$^{(e)}$Instituto de Alta Investigaci\'on, Universidad de Tarapac\'a, Arica;$^{(f)}$Departamento de F\'isica, Universidad T\'ecnica Federico Santa Mar\'ia, Valpara\'iso; Chile.\\
% ...
$^{150}$Department of Physics, Technion, Israel Institute of Technology, Haifa; Israel.\\
$^{151}$Raymond and Beverly Sackler School of Physics and Astronomy, Tel Aviv University, Tel Aviv; Israel.\\
% ...
$^{a}$ Also Affiliated with an institute covered by a cooperation agreement with CERN.\\
$^{b}$ Also at An-Najah National University, Nablus; Palestine.\\
$^{c}$ Also at Borough of Manhattan Community College, City University of New York, New York NY; United States of America.\\
$^{d}$ Also at Center for High Energy Physics, Peking University; China.\\
$^{e}$ Also at Center for Interdisciplinary Research and Innovation (CIRI-AUTH), Thessaloniki; Greece.\\
$^{f}$ Also at Centro Studi e Ricerche Enrico Fermi; Italy.\\
% ...

\end{flushleft}

% Created with Glance <Atlas.Glance@cern.ch>

%% file: ANA-EXOT-2021-36-PAPER.bib
@article{Dienes_1999,
   title={Grand unification at intermediate mass scales through extra dimensions},
   volume={537},
   DOI={10.1016/s0550-3213(98)00669-5},
   number={1-3},
   eprint = "hep-ph/9806292",
   archivePrefix = "arXiv",
   journal={Nucl. Phys. B},
   publisher={Elsevier BV},
   author={Dienes, Keith R. and Dudas, Emilian and Gherghetta, Tony},
   year={1999},
   month={01},
   pages={47–108}
}

@article{PhysRevD.19.1277,
  title = {Implications of dynamical symmetry breaking: An addendum},
  author = {Weinberg, S.},
  journal = {Phys. Rev. D},
  volume = {19},
  pages = {1277--1280},
  numpages = {0},
  year = {1979},
  month = {02},
  publisher = {American Physical Society},
  doi = {10.1103/PhysRevD.19.1277},
}

@article{PhysRevD.20.2619,
  title = {Dynamics of spontaneous symmetry breaking in the Weinberg-Salam theory},
  author = {Susskind, Leonard},
  journal = {Phys. Rev. D},
  volume = {20},
  pages = {2619--2625},
  numpages = {0},
  year = {1979},
  month = {11},
  publisher = {American Physical Society},
  doi = {10.1103/PhysRevD.20.2619},
}

@article{DIMOPOULOS1979237,
title = {Mass without scalars},
journal = {Nucl. Phys. B},
volume = {155},
number = {1},
pages = {237-252},
year = {1979},
doi = {https://doi.org/10.1016/0550-3213(79)90364-X},
author = {Savas Dimopoulos and Leonard Susskind},
}

@article{EICHTEN1980125,
title = {Dynamical breaking of weak interaction symmetries},
journal = {Phys. Lett. B},
volume = {90},
number = {1},
pages = {125-130},
year = {1980},
doi = {https://doi.org/10.1016/0370-2693(80)90065-9},
author = {Estia Eichten and Kenneth Lane},
}

@article{Muller_1996,
   title={Topflavor: a separate SU(2) for the third family},
   volume={383},
   DOI={10.1016/0370-2693(96)00745-9},
   eprint = "hep-ph/9602390",
   archivePrefix = "arXiv",
   number={3},
   journal={Phys. Lett. B},
   publisher={Elsevier BV},
   author={Muller, David J and Nandi, Satyanarayan},
   year={1996},
   month={09},
   pages={345–350}
}

@article{Burdman_2006,
   title={Resonances from two universal extra dimensions},
   volume={74},
   DOI={10.1103/physrevd.74.075008},
   number={7},
   journal={Phys. Rev. D},
   eprint = "hep-ph/0601186",
   rchivePrefix = "arXiv",
   publisher={American Physical Society (APS)},
   author={Burdman, Gustavo and Dobrescu, Bogdan A. and Pontón, Eduardo},
   year={2006},
   pages={075008}
}

@article{Malkawi_1996,
   title={A model of strong flavor dynamics for the top quark},
   volume={385},
   DOI={10.1016/0370-2693(96)00859-3},
   number={1-4},
   journal={Phys. Lett. B},
   eprint = "hep-ph/9603349",
   archivePrefix = "arXiv",
   publisher={Elsevier BV},
   author={Malkawi, Ehab and Tait, Tim and Yuan, C.-P.},
   year={1996},
   month={09},
   pages={304–310}
}

@article{PhysRevD.10.275,
  title = {Lepton number as the fourth "color"},
  author = {Pati, Jogesh C. and Salam, Abdus},
  journal = {Phys. Rev. D},
  volume = {10},
  pages = {275--289},
  numpages = {0},
  year = {1974},
  month = {07},
  publisher = {American Physical Society},
  doi = {10.1103/PhysRevD.10.275},
  addendum = "[Erratum: \href{https://doi.org/10.1103/PhysRevD.11.703.2}{Phys. Rev. D \textbf{11} (1975) 703}]",
}

@article{Hill_1995,
   title={Topcolor assisted technicolor},
   volume={345},
   DOI={10.1016/0370-2693(94)01660-5},
   number={4},
   journal={Phys. Lett. B},
   eprint = "hep-ph/9411426",
   archivePrefix = "arXiv",
   publisher={Elsevier BV},
   author={Hill, Christopher T.},
   year={1995},
   month={02},
   pages={483–489}
}

@article{Altarelli:1989ff,
    author = "Altarelli, Guido and Mele, B. and Ruiz-Altaba, M.",
    title = "{Searching for New Heavy Vector Bosons in $p \bar{p}$ Colliders}",
    reportNumber = "CERN-TH-5323-89",
    doi = "10.1007/BF01556677",
    journal = "Z. Phys. C",
    volume = "45",
    pages = "109",
    year = "1989",
    addendum       = "[Erratum: \href{https://doi.org/10.1007/BF01552335}{Z. Phys. C \textbf{47} (1990) 676}]",
}

@article{Sullivan_2002,
   title={Fully differential W' production and decay at next-to-leading order in QCD},
   volume={66},
   DOI={10.1103/physrevd.66.075011},
   eprint = "hep-ph/0207290",
   archivePrefix = "arXiv",
   number={7},
   journal={Phys. Rev. D},
   publisher={American Physical Society (APS)},
   author={Sullivan, Zack},
   year={2002},
   pages={075011},
}

@article{Duffty_2012,
   title={Model independent reach for W' bosons at the LHC},
    eprint = "1208.4858",
    archivePrefix = "arXiv",
    primaryClass = "hep-ph",
   volume={86},
   DOI={10.1103/physrevd.86.075018},
   number={7},
   journal={Phys. Rev. D},
   publisher={American Physical Society (APS)},
   author={Duffty, Daniel and Sullivan, Zack},
   year={2012},
   pages={075018}
}

@article{Abazov:2011xs,
    author = "{D0 Collaboration}",
    title = "{Search for $W{^\prime}\rightarrow tb$ resonances with left- and right-handed couplings to fermions}",
    eprint = "1101.0806",
    archivePrefix = "arXiv",
    primaryClass = "hep-ex",
    reportNumber = "FERMILAB-PUB-11-004-E",
    doi = "10.1016/j.physletb.2011.03.066",
    journal = "Phys. Lett. B",
    volume = "699",
    pages = "145--150",
    year = "2011"
}

@article{CDF:2015aej,
    author = "{CDF Collaboration}",
    title = "{Search for Resonances Decaying to Top and Bottom Quarks with the CDF Experiment}",
    eprint = "1504.01536",
    archivePrefix = "arXiv",
    primaryClass = "hep-ex",
    reportNumber = "FERMILAB-PUB-15-123-PPD",
    doi = "10.1103/PhysRevLett.115.061801",
    journal = "Phys. Rev. Lett.",
    volume = "115",
    number = "6",
    pages = "061801",
    year = "2015"
}

@article{LUCID2,
  author={G. Avoni and others},
  title={The new LUCID-2 detector for luminosity measurement and monitoring in ATLAS},
  journal={JINST},
  volume={13},
  number={07},
  pages={P07017},
  doi="10.1088/1748-0221/13/07/P07017",
  year={2018},
}

@article{Thaler:2010tr,
    author = "Thaler, Jesse and Van Tilburg, Ken",
    title = "{Identifying Boosted Objects with N-subjettiness}",
    eprint = "1011.2268",
    archivePrefix = "arXiv",
    primaryClass = "hep-ph",
    reportNumber = "MIT-CTP-4191",
    doi = "10.1007/JHEP03(2011)015",
    journal = "JHEP",
    volume = "03",
    pages = "015",
    year = "2011"
}

@article{Thaler:2008ju,
    author = "Thaler, Jesse and Wang, Lian-Tao",
    title = "{Strategies to identify boosted tops}",
    eprint = "0806.0023",
    archivePrefix = "arXiv",
    primaryClass = "hep-ph",
    doi = "10.1088/1126-6708/2008/07/092",
    journal = "JHEP",
    volume = "07",
    pages = "092",
    year = "2008"
}

@article{Larkoski:2013eya,
    author = "Larkoski, Andrew J. and Salam, Gavin P. and Thaler, Jesse",
    title = "{Energy correlation functions for jet substructure}",
    eprint = "1305.0007",
    archivePrefix = "arXiv",
    primaryClass = "hep-ph",
    reportNumber = "MIT-CTP-4446, CERN-PH-TH-2013-066, LPN13-026",
    doi = "10.1007/JHEP06(2013)108",
    journal = "JHEP",
    volume = "06",
    pages = "108",
    year = "2013"
}

@article{Krohn_2010,
   title={Jet trimming},
   volume={2010},
   DOI={10.1007/jhep02(2010)084},
   volume={02},
   journal={JHEP},
   publisher={Springer Science and Business Media LLC},
   author={Krohn, David and Thaler, Jesse and Wang, Lian-Tao},
   year={2010},
   pages={084}
}

@article{CATANI1993187,
title = {Longitudinally-invariant $k_\bot$-clustering algorithms for hadron-hadron collisions},
journal = {Nucl. Phys. B},
volume = {406},
number = {1},
pages = {187-224},
year = {1993},
doi = {https://doi.org/10.1016/0550-3213(93)90166-M},
author = {S. Catani and Yu.L. Dokshitzer and M.H. Seymour and B.R. Webber},
}

@article{Ellis_1993,
   title={Successive combination jet algorithm for hadron collisions},
   volume={48},
   DOI={10.1103/physrevd.48.3160},
   number={7},
   journal={Phys. Rev. D},
   publisher={American Physical Society (APS)},
   author={Ellis, Stephen D. and Soper, Davison E.},
   year={1993},
   month={10},
   pages={3160–3166}
}

@article{Cacciari_2008,
   title={The catchment area of jets},
   volume={2008},
   DOI={10.1088/1126-6708/2008/04/005},
   volume={04},
   journal={JHEP},
   publisher={Springer Science and Business Media LLC},
   author={Cacciari, Matteo and Salam, Gavin P and Soyez, Gregory},
   year={2008},
   month={04},
   pages={005–005}
}

@article{Boos:2006xe,
    author = "Boos, E. and Bunichev, V. and Dudko, L. and Perfilov, M.",
    title = "{Interference between $W^\prime$ and $W$ in single-top quark production processes}",
    eprint = "hep-ph/0610080",
    archivePrefix = "arXiv",
    doi = "10.1016/j.physletb.2007.03.064",
    journal = "Phys. Lett. B",
    volume = "655",
    pages = "245--250",
    year = "2007"
}

@article{Nachman:2014kla,
      author         = "Nachman, Benjamin and Nef, Pascal and Schwartzman, Ariel
                        and Swiatlowski, Maximilian and Wanotayaroj, Chaowaroj",
      title          = "{Jets from Jets: Re-clustering as a tool for large radius
                        jet reconstruction and grooming at the LHC}",
      journal        = "JHEP",
      volume         = "02",
      year           = "2015",
      pages          = "075",
      doi            = "10.1007/JHEP02(2015)075",
      eprint         = "1407.2922",
      archivePrefix  = "arXiv",
      primaryClass   = "hep-ph",
      SLACcitation   = "%%CITATION = ARXIV:1407.2922;%%"
}

@article{Krohn:2009zg,
      author         = "Krohn, David and Thaler, Jesse and Wang, Lian-Tao",
      title          = "{Jets with variable R}",
      journal        = "JHEP",
      volume         = "06",
      year           = "2009",
      pages          = "059",
      doi            = "10.1088/1126-6708/2009/06/059",
      eprint         = "0903.0392",
      archivePrefix  = "arXiv",
      primaryClass   = "hep-ph",
      SLACcitation   = "%%CITATION = ARXIV:0903.0392;%%"
}

@article{Hou:2019efy,
    author = "Hou, Tie-Jiun and others",
    title = "{New CTEQ global analysis of quantum chromodynamics with high-precision data from the LHC}",
    eprint = "1912.10053",
    archivePrefix = "arXiv",
    primaryClass = "hep-ph",
    reportNumber = "MSUHEP-19-025, PITT-PACC-1911, SMU-HEP-19-03",
    doi = "10.1103/PhysRevD.103.014013",
    journal = "Phys. Rev. D",
    volume = "103",
    number = "1",
    pages = "014013",
    year = "2021"
}

@article{Bailey:2020ooq,
    author = "Bailey, S. and Cridge, T. and Harland-Lang, L. A. and Martin, A. D. and Thorne, R. S.",
    title = "{Parton distributions from LHC, HERA, Tevatron and fixed target data: MSHT20 PDFs}",
    eprint = "2012.04684",
    archivePrefix = "arXiv",
    primaryClass = "hep-ph",
    reportNumber = "IPPP/20/58",
    doi = "10.1140/epjc/s10052-021-09057-0",
    journal = "Eur. Phys. J. C",
    volume = "81",
    number = "4",
    pages = "341",
    year = "2021"
}

@article{Czakon:2017wor,
    author = "Czakon, Michal and Heymes, David and Mitov, Alexander and Pagani, Davide and Tsinikos, Ioannis and Zaro, Marco",
    title = "{Top-pair production at the LHC through NNLO QCD and NLO EW}",
    eprint = "1705.04105",
    archivePrefix = "arXiv",
    primaryClass = "hep-ph",
    reportNumber = "CAVENDISH-HEP-17-07, CP3-17-12, TUM-HEP-1084-17, TTK-17-15",
    doi = "10.1007/JHEP10(2017)186",
    journal = "JHEP",
    volume = "10",
    pages = "186",
    year = "2017"
}

@article{Alwall:2007fs,
    author = "Alwall, Johan and others",
    title = "{Comparative study of various algorithms for the merging of parton showers and matrix elements in hadronic collisions}",
    eprint = "0706.2569",
    archivePrefix = "arXiv",
    primaryClass = "hep-ph",
    reportNumber = "SLAC-PUB-12604, CERN-PH-TH-2007-066, LU-TP-07-13, KA-TP-06-2007, DCPT-07-62, IPPP-07-31",
    doi = "10.1140/epjc/s10052-007-0490-5",
    journal = "Eur. Phys. J. C",
    volume = "53",
    pages = "473--500",
    year = "2008"
}

@article{Manohar:2016nzj,
    author = "Manohar, Aneesh and Nason, Paolo and Salam, Gavin P. and Zanderighi, Giulia",
    title = "{How bright is the proton? A precise determination of the photon parton distribution function}",
    eprint = "1607.04266",
    archivePrefix = "arXiv",
    primaryClass = "hep-ph",
    reportNumber = "CERN-TH-2016-155",
    doi = "10.1103/PhysRevLett.117.242002",
    journal = "Phys. Rev. Lett.",
    volume = "117",
    number = "24",
    pages = "242002",
    year = "2016"
}

@Article{Cowan:2010js-witherr,
   Author = {Cowan, Glen and Cranmer, Kyle and Gross, Eilam and Vitells, Ofer},
   Title = {{Asymptotic formulae for likelihood-based tests of new physics}},
   Journal = {Eur. Phys. J. C},
   Volume = {71},
   Year = {2011},
   Pages = {1554},
   doi = "10.1140/epjc/s10052-011-1554-0",
   Eprint = "1007.1727",
   Archiveprefix = {arXiv},
   Primaryclass = {physics.data-an},
   addendum = "[Erratum: \href{https://doi.org/10.1140/epjc/s10052-013-2501-z}{Eur. Phys. J. C \textbf{73} (2013) 2501}]",
}


%% file: ATLAS-SUSY.bib
@article{Baak:2014wma,
      author         = "Baak, M. and Besjes, G.J. and C{\^o}t{\'e}, D. and Koutsman, A. and Lorenz, J. and Short, D.",
      title          = "{HistFitter software framework for statistical data analysis}",
      journal        = "Eur. Phys. J. C",
      volume         = "75",
      year           = "2015",
      pages          = "153",
      doi            = "10.1140/epjc/s10052-015-3327-7",
      eprint         = "1410.1280",
      archivePrefix  = "arXiv",
      primaryClass   = "hep-ex",
      SLACcitation   = "%%CITATION = ARXIV:1410.1280;%%",
}


%% file: ATLAS-useful.bib
@Article{Cacciari:2008gp,
     author    = "Cacciari, Matteo and Salam, Gavin P. and Soyez, Gregory",
     title     = "{The anti-\(k_{t}\) jet clustering algorithm}",
     journal   = "JHEP",
     volume    = "04",
     year      = "2008",
     pages     = "063",
     eprint    = "0802.1189",
     archivePrefix = "arXiv",
     primaryClass  =  "hep-ph",
     doi       = "10.1088/1126-6708/2008/04/063",
     SLACcitation  = "%%CITATION = 0802.1189;%%"
}

@Article{Fastjet,
      author         = "Cacciari, Matteo and Salam, Gavin P. and Soyez, Gregory",
      title          = "{FastJet user manual}",
      journal        = "Eur. Phys. J. C",
      volume         = "72",
      year           = "2012",
      pages          = "1896",
      doi            = "10.1140/epjc/s10052-012-1896-2",
      eprint         = "1111.6097",
      archivePrefix  = "arXiv",
      primaryClass   = "hep-ph",
      reportNumber   = "CERN-PH-TH-2011-297",
      SLACcitation   = "%%CITATION = ARXIV:1111.6097;%%"
}

@Article{Butterworth:2015oua,
      author         = "Butterworth, Jon and others",
      title          = "{PDF4LHC recommendations for LHC Run II}",
      journal        = "J. Phys. G",
      volume         = "43",
      year           = "2016",
      pages          = "023001",
      doi            = "10.1088/0954-3899/43/2/023001",
      eprint         = "1510.03865",
      archivePrefix  = "arXiv",
      primaryClass   = "hep-ph",
      reportNumber   = "OUTP-15-17P, SMU-HEP-15-12, TIF-UNIMI-2015-14,
                        LCTS-2015-27, CERN-PH-TH-2015-249",
      SLACcitation   = "%%CITATION = ARXIV:1510.03865;%%"
}

@article{Dulat:2015mca,
      author         = "Dulat, Sayipjamal and Hou, Tie-Jiun and Gao, Jun and
                        Guzzi, Marco and Huston, Joey and Nadolsky, Pavel and
                        Pumplin, Jon and Schmidt, Carl and Stump, Daniel and Yuan,
                        C. P.",
      title          = "{New parton distribution functions from a global analysis
                        of quantum chromodynamics}",
      journal        = "Phys. Rev. D",
      volume         = "93",
      year           = "2016",
      number         = "3",
      pages          = "033006",
      doi            = "10.1103/PhysRevD.93.033006",
      eprint         = "1506.07443",
      archivePrefix  = "arXiv",
      primaryClass   = "hep-ph",
      SLACcitation   = "%%CITATION = ARXIV:1506.07443;%%"
}

@article{Harland-Lang:2014zoa,
      author         = "Harland-Lang, L. A. and Martin, A. D. and Motylinski, P.
                        and Thorne, R. S.",
      title          = "{Parton distributions in the LHC era: MMHT 2014 PDFs}",
      journal        = "Eur. Phys. J. C",
      number         = "5",
      volume         = "75",
      pages          = "204",
      doi            = "10.1140/epjc/s10052-015-3397-6",
      year           = "2015",
      eprint         = "1412.3989",
      archivePrefix  = "arXiv",
      primaryClass   = "hep-ph",
      reportNumber   = "LCTS-2014-47, IPPP-14-97, DCPT-14-194",
      SLACcitation   = "%%CITATION = ARXIV:1412.3989;%%",
}

@article{Ball:2012cx,
      author         = "{NNPDF Collaboration} and Ball, Richard D. and others",
      title          = "{Parton distributions with LHC data}",
      journal        = "Nucl. Phys. B",
      volume         = "867",
      year           = "2013",
      pages          = "244",
      doi            = "10.1016/j.nuclphysb.2012.10.003",
      eprint         = "1207.1303",
      archivePrefix  = "arXiv",
      primaryClass   = "hep-ph",
      reportNumber   = "EDINBURGH-2012-08, IFUM-FT-997, FR-PHENO-2012-014,
                        RWTH-TTK-12-25, CERN-PH-TH-2012-037, SFB-CPP-12-47\,
                        --CERN-PH-TH-2012-037",
      SLACcitation   = "%%CITATION = ARXIV:1207.1303;%%"
}

@article{Ball:2014uwa,
      author         = "{The NNPDF Collaboration} and Ball, Richard D. and others",
      title          = "{Parton distributions for the LHC run II}",
      journal        = "JHEP",
      volume         = "04",
      year           = "2015",
      pages          = "040",
      doi            = "10.1007/JHEP04(2015)040",
      eprint         = "1410.8849",
      archivePrefix  = "arXiv",
      primaryClass   = "hep-ph",
      reportNumber   = "EDINBURGH-2014-15, IFUM-1034-FT, CERN-PH-TH-2013-253,
                        OUTP-14-11P, CAVENDISH-HEP-14-11",
      SLACcitation   = "%%CITATION = ARXIV:1410.8849;%%"
}

@article{Czakon:2012pz,
      author         = "Czakon, Michal and Mitov, Alexander",
      title          = "{NNLO corrections to top pair production at hadron
                        colliders: the quark-gluon reaction}",
      journal        = "JHEP",
      volume         = "01",
      pages          = "080",
      doi            = "10.1007/JHEP01(2013)080",
      year           = "2013",
      eprint         = "1210.6832",
      archivePrefix  = "arXiv",
      primaryClass   = "hep-ph",
      SLACcitation   = "%%CITATION = ARXIV:1210.6832;%%",
}

@article{Sjostrand:2014zea,
      author         = "Sj{\"o}strand, Torbj{\"o}rn and Ask, Stefan and Christiansen,
                        Jesper R. and Corke, Richard and Desai, Nishita and Ilten,
                        Philip and Mrenna, Stephen and Prestel, Stefan and
                        Rasmussen, Christine O. and Skands, Peter Z.",
      title          = "{An introduction to PYTHIA 8.2}",
      journal        = "Comput. Phys. Commun.",
      volume         = "191",
      year           = "2015",
      pages          = "159",
      doi            = "10.1016/j.cpc.2015.01.024",
      eprint         = "1410.3012",
      archivePrefix  = "arXiv",
      primaryClass   = "hep-ph",
      reportNumber   = "LU-TP-14-36, MCNET-14-22, CERN-PH-TH-2014-190,
                        FERMILAB-PUB-14-316-CD, DESY-14-178, SLAC-PUB-16122,
                        --FERMILAB-PUB-14-316-CD",
      SLACcitation   = "%%CITATION = ARXIV:1410.3012;%%"
}

@Article{Lange:2001uf,
      author         = "Lange, D. J.",
      title          = "{The EvtGen particle decay simulation package}",
      booktitle      = "{Proceedings, 7th International Conference on B physics
                        at hadron machines (BEAUTY 2000)}",
      journal        = "Nucl. Instrum. Meth. A",
      volume         = "462",
      year           = "2001",
      pages          = "152",
      doi            = "10.1016/S0168-9002(01)00089-4",
      SLACcitation   = "%%CITATION = NUIMA,A462,152;%%"
}

@Article{Frixione:2008yi,
     author    = "Frixione, Stefano and Laenen, Eric and Motylinski, Patrick and White, Chris and Webber, Bryan R.",
     title     = "{Single-top hadroproduction in association with a \(W\) boson}",
     journal   = "JHEP",
     volume    = "07",
     year      = "2008",
     pages     = "029",
     eprint    = "0805.3067",
     archivePrefix = "arXiv",
     primaryClass  =  "hep-ph",
     doi       = "10.1088/1126-6708/2008/07/029"
}

@Article{Alwall:2014hca,
      author         = "Alwall, J. and Frederix, R. and Frixione, S. and Hirschi,
                        V. and Maltoni, F. and Mattelaer, O. and Shao, H. -S. and
                        Stelzer, T. and Torrielli, P. and Zaro, M.",
      title          = "{The automated computation of tree-level and
                        next-to-leading order differential cross sections, and
                        their matching to parton shower simulations}",
      journal        = "JHEP",
      volume         = "07",
      year           = "2014",
      pages          = "079",
      doi            = "10.1007/JHEP07(2014)079",
      eprint         = "1405.0301",
      archivePrefix  = "arXiv",
      primaryClass   = "hep-ph",
      reportNumber   = "CERN-PH-TH-2014-064, CP3-14-18, LPN14-066, MCNET-14-09,
                        ZU-TH-14-14",
      SLACcitation   = "%%CITATION = ARXIV:1405.0301;%%"
}

@Article{Bahr:2008pv,
      author         = "B{\"a}hr, M. and others",
      title          = "{Herwig++ physics and manual}",
      journal        = "Eur. Phys. J. C",
      volume         = "58",
      year           = "2008",
      pages          = "639",
      doi            = "10.1140/epjc/s10052-008-0798-9",
      eprint         = "0803.0883",
      archivePrefix  = "arXiv",
      primaryClass   = "hep-ph",
      reportNumber   = "CERN-PH-TH-2008-038, CAVENDISH-HEP-08-03, KA-TP-05-2008,
                        DCPT-08-22, IPPP-08-11, CP3-08-05",
      SLACcitation   = "%%CITATION = ARXIV:0803.0883;%%"
}

@Article{Bellm:2015jjp,
      author         = "Bellm, Johannes and others",
      title          = "{Herwig 7.0/Herwig++ 3.0 release note}",
      journal        = "Eur. Phys. J. C",
      volume         = "76",
      year           = "2016",
      number         = "4",
      pages          = "196",
      doi            = "10.1140/epjc/s10052-016-4018-8",
      eprint         = "1512.01178",
      archivePrefix  = "arXiv",
      primaryClass   = "hep-ph",
      reportNumber   = "CERN-PH-TH-2015-289, MAN-HEP-2015-15, IFJPAN-IV-2015-13,
                        HERWIG-2015-01, KA-TP-18-2015, DCPT-15-142, MCNET-15-28,
                        IPPP-15-71, --HERWIG-2015-01",
      SLACcitation   = "%%CITATION = ARXIV:1512.01178;%%"
}

@Article{Nason:2004rx,
      author         = "Nason, Paolo",
      title          = "{A new method for combining NLO QCD with shower Monte Carlo algorithms}",
      journal        = "JHEP",
      volume         = "11",
      pages          = "040",
      doi            = "10.1088/1126-6708/2004/11/040",
      year           = "2004",
      eprint         = "hep-ph/0409146",
      archivePrefix  = "arXiv",
}

@Article{Frixione:2007vw,
      author         = "Frixione, Stefano and Nason, Paolo and Oleari, Carlo",
      title          = "{Matching NLO QCD computations with parton shower
                        simulations: the POWHEG method}",
      journal        = "JHEP",
      volume         = "11",
      pages          = "070",
      doi            = "10.1088/1126-6708/2007/11/070",
      year           = "2007",
      eprint         = "0709.2092",
      archivePrefix  = "arXiv",
      primaryClass   = "hep-ph",
}

@Article{Alioli:2010xd,
      author         = "Alioli, Simone and Nason, Paolo and Oleari, Carlo and Re,
                        Emanuele",
      title          = "{A general framework for implementing NLO calculations in
                        shower Monte Carlo programs: the POWHEG BOX}",
      journal        = "JHEP",
      volume         = "06",
      pages          = "043",
      doi            = "10.1007/JHEP06(2010)043",
      year           = "2010",
      eprint         = "1002.2581",
      archivePrefix  = "arXiv",
      primaryClass   = "hep-ph",
}

@article{Frederix:2012dh,
      author         = "Frederix, Rikkert and Re, Emanuele and Torrielli, Paolo",
      title          = "{Single-top \(t\)-channel hadroproduction in the four-flavour
                        scheme with POWHEG and aMC@NLO}",
      journal        = "JHEP",
      volume         = "09",
      year           = "2012",
      pages          = "130",
      doi            = "10.1007/JHEP09(2012)130",
      eprint         = "1207.5391",
      archivePrefix  = "arXiv",
      primaryClass   = "hep-ph",
      reportNumber   = "ZU-TH-14-12, IPPP-12-54, DCPT-12-108,
                        CERN-PH-TH-2012-206, MCNET-12-10, LPN12-084",
      SLACcitation   = "%%CITATION = ARXIV:1207.5391;%%"
}

@article{Beneke:2011mq,
      author         = "Beneke, M. and Falgari, P. and Klein, S. and Schwinn, C.",
      title          = "{Hadronic top-quark pair production with NNLL threshold
                        resummation}",
      journal        = "Nucl. Phys. B",
      volume         = "855",
      year           = "2012",
      pages          = "695-741",
      doi            = "10.1016/j.nuclphysb.2011.10.021",
      eprint         = "1109.1536",
      archivePrefix  = "arXiv",
      primaryClass   = "hep-ph",
      reportNumber   = "TTK-11-38, ITP-UU-11-26, SPIN-11-19, FR-PHENO-2011-015,
                        SFB-CPP-11-49",
      SLACcitation   = "%%CITATION = ARXIV:1109.1536;%%"
}

@article{Anastasiou:2003ds,
      author         = "Anastasiou, Charalampos and Dixon, Lance and Melnikov,
                        Kirill and Petriello, Frank",
      title          = "{High-precision QCD at hadron colliders: Electroweak
                        gauge boson rapidity distributions at next-to-next-to leading order}",
      journal        = "Phys. Rev. D",
      volume         = "69",
      year           = "2004",
      pages          = "094008",
      doi            = "10.1103/PhysRevD.69.094008",
      eprint         = "hep-ph/0312266",
      archivePrefix  = "arXiv",
      reportNumber   = "SLAC-PUB-10288, UH-511-1042-03",
      SLACcitation   = "%%CITATION = HEP-PH/0312266;%%"
}

@article{Cacciari:2011hy,
      author         = "Cacciari, Matteo and Czakon, Michal and Mangano,
                        Michelangelo and Mitov, Alexander and Nason, Paolo",
      title          = "{Top-pair production at hadron colliders with
                        next-to-next-to-leading logarithmic soft-gluon
                        resummation}",
      journal        = "Phys. Lett. B",
      volume         = "710",
      pages          = "612-622",
      doi            = "10.1016/j.physletb.2012.03.013",
      year           = "2012",
      eprint         = "1111.5869",
      primaryClass   = "hep-ph",
      archivePrefix  = "arXiv",
      reportNumber   = "CERN-PH-TH-2011-277, TTK-11-54",
      SLACcitation   = "%%CITATION = ARXIV:1111.5869;%%",
}

@article{Czakon:2012zr,
      author         = "Czakon, Michal and Mitov, Alexander",
      title          = "{NNLO corrections to top-pair production at hadron
                        colliders: the all-fermionic scattering channels}",
      journal        = "JHEP",
      volume         = "12",
      pages          = "054",
      doi            = "10.1007/JHEP12(2012)054",
      year           = "2012",
      eprint         = "1207.0236",
      archivePrefix  = "arXiv",
      primaryClass   = "hep-ph",
      SLACcitation   = "%%CITATION = ARXIV:1207.0236;%%",
}

@article{Czakon:2013goa,
      author         = "Czakon, Michal and Fiedler, Paul and Mitov, Alexander",
      title          = "{Total Top-Quark Pair-Production Cross Section at
                        Hadron Colliders Through \(O(\alpha_S^4)\)}",
      journal        = "Phys. Rev. Lett.",
      volume         = "110",
      pages          = "252004",
      doi            = "10.1103/PhysRevLett.110.252004",
      year           = "2013",
      eprint         = "1303.6254",
      archivePrefix  = "arXiv",
      primaryClass   = "hep-ph",
      reportNumber   = "CERN-PH-TH-2013-056, TTK-13-08",
      SLACcitation   = "%%CITATION = ARXIV:1303.6254;%%",
}

@article{Baernreuther:2012ws,
      author         = "B{\"a}rnreuther, Peter and Czakon, Michal and Mitov,
                        Alexander",
      title          = "{Percent-Level-Precision Physics at the
                  Tevatron: Next-to-Next-to-Leading Order QCD
                  Corrections to \(q \bar{q} \to t \bar{t} + X\)}",
      journal        = "Phys. Rev. Lett.",
      volume         = "109",
      pages          = "132001",
      doi            = "10.1103/PhysRevLett.109.132001",
      year           = "2012",
      eprint         = "1204.5201",
      archivePrefix  = "arXiv",
      primaryClass   = "hep-ph",
      SLACcitation   = "%%CITATION = ARXIV:1204.5201;%%",
}

@article{Frixione:2007nw,
      author         = "Frixione, Stefano and Ridolfi, Giovanni and Nason, Paolo",
      title          = "{A positive-weight next-to-leading-order Monte Carlo for
                        heavy flavour hadroproduction}",
      journal        = "JHEP",
      volume         = "09",
      pages          = "126",
      doi            = "10.1088/1126-6708/2007/09/126",
      year           = "2007",
      eprint         = "0707.3088",
      archivePrefix  = "arXiv",
      primaryClass   = "hep-ph",
      SLACcitation   = "%%CITATION = ARXIV:0707.3088;%%",
}

@Article{Hoeche:2009rj,
      author         = "H{\"o}che, Stefan and Krauss, Frank and Schumann, Steffen and Siegert, Frank",
      title          = "{QCD matrix elements and truncated showers}",
      journal        = "JHEP",
      volume         = "05",
      pages          = "053",
      doi            = "10.1088/1126-6708/2009/05/053",
      year           = "2009",
      eprint         = "0903.1219",
      archivePrefix  = "arXiv",
      primaryClass   = "hep-ph",
}

@Article{Gleisberg:2008fv,
      author         = "Gleisberg, Tanju and H{\"o}che, Stefan",
      title          = "{Comix, a new matrix element generator}",
      journal        = "JHEP",
      volume         = "12",
      pages          = "039",
      doi            = "10.1088/1126-6708/2008/12/039",
      year           = "2008",
      eprint         = "0808.3674",
      archivePrefix  = "arXiv",
      primaryClass   = "hep-ph",
}

@Article{Schumann:2007mg,
      author         = "Schumann, Steffen and Krauss, Frank",
      title          = "{A parton shower algorithm based on Catani--Seymour dipole factorisation}",
      journal        = "JHEP",
      volume         = "03",
      pages          = "038",
      doi            = "10.1088/1126-6708/2008/03/038",
      year           = "2008",
      eprint         = "0709.1027",
      archivePrefix  = "arXiv",
      primaryClass   = "hep-ph",
}

@article{Hoeche:2012yf,
      author         = "H{\"o}che, Stefan and Krauss, Frank and Sch{\"o}nherr, Marek and
                        Siegert, Frank",
      title          = "{QCD matrix elements + parton showers. The NLO case}",
      journal        = "JHEP",
      volume         = "04",
      year           = "2013",
      pages          = "027",
      doi            = "10.1007/JHEP04(2013)027",
      eprint         = "1207.5030",
      archivePrefix  = "arXiv",
      primaryClass   = "hep-ph",
      reportNumber   = "SLAC-PUB-15191, IPPP-12-52, DCPT-12-104, LPN12-081,
                        FR-PHENO-2012-017, MCNET-12-09, --FR-PHENO-2012-017",
      SLACcitation   = "%%CITATION = ARXIV:1207.5030;%%"
}

@article{Bothmann:2019yzt,
      author = "Bothmann, Enrico and others",
      title = "{Event generation with Sherpa 2.2}",
      journal = "SciPost Phys.",
      volume = "7",
      year = "2019",
      number = "3",
      pages = "034",
      doi = "10.21468/SciPostPhys.7.3.034",
      reportNumber = "FERMILAB-PUB-19-218-T, SLAC-PUB-17433, IPPP/19/42, MCNET-19-11",
      eprint = "1905.09127",
      archivePrefix = "arXiv",
      primaryClass = "hep-ph",
}

@article{Hoeche:2011fd,
      author         = "H{\"o}che, Stefan and Krauss, Frank and Sch{\"o}nherr, Marek and
                        Siegert, Frank",
      title          = "{A critical appraisal of NLO+PS matching methods}",
      journal        = "JHEP",
      volume         = "09",
      year           = "2012",
      pages          = "049",
      doi            = "10.1007/JHEP09(2012)049",
      eprint         = "1111.1220",
      archivePrefix  = "arXiv",
      primaryClass   = "hep-ph",
      reportNumber   = "SLAC-PUB-14661, IPPP-11-67, DCPT-11-134, LPN11-58,
                        FR-PHENO-2011-019, MCNET-11-24",
      SLACcitation   = "%%CITATION = ARXIV:1111.1220;%%"
}

@article{Catani:2001cc,
      author         = "Catani, S. and Krauss, F. and Webber, B. R. and Kuhn, R.",
      title          = "{QCD Matrix Elements + Parton Showers}",
      journal        = "JHEP",
      volume         = "11",
      year           = "2001",
      pages          = "063",
      doi            = "10.1088/1126-6708/2001/11/063",
      eprint         = "hep-ph/0109231",
      archivePrefix  = "arXiv",
      reportNumber   = "CERN-TH-2000-367, CAVENDISH-HEP-00-03",
      SLACcitation   = "%%CITATION = HEP-PH/0109231;%%"
}

@article{Cascioli:2011va,
      author         = "Cascioli, Fabio and Maierh{\"o}fer, Philipp and Pozzorini, Stefano",
      title          = "{Scattering Amplitudes with Open Loops}",
      journal        = "Phys. Rev. Lett.",
      volume         = "108",
      year           = "2012",
      pages          = "111601",
      doi            = "10.1103/PhysRevLett.108.111601",
      eprint         = "1111.5206",
      archivePrefix  = "arXiv",
      primaryClass   = "hep-ph",
      reportNumber   = "ZU-TH-23-11, LPN11-66",
      SLACcitation   = "%%CITATION = ARXIV:1111.5206;%%"
}

@article{Buccioni:2019sur,
    author = "Buccioni, Federico and Lang, Jean-Nicolas and Lindert, Jonas M. and Maierh{\"o}fer, Philipp and Pozzorini, Stefano and Zhang, Hantian and Zoller, Max F.",
    title = "{OpenLoops 2}",
    eprint = "1907.13071",
    archivePrefix = "arXiv",
    primaryClass = "hep-ph",
    reportNumber = "IPPP/19/62, FR-PHENO-2019-12, PSI-PR-19-15, ZU-TH 37/19",
    doi = "10.1140/epjc/s10052-019-7306-2",
    journal = "Eur. Phys. J. C",
    volume = "79",
    number = "10",
    pages = "866",
    year = "2019"
}

@article{Denner:2016kdg,
      author         = "Denner, Ansgar and Dittmaier, Stefan and Hofer, Lars",
      title          = "{\textsc{Collier}: A fortran-based complex one-loop library in
                        extended regularizations}",
      journal        = "Comput. Phys. Commun.",
      volume         = "212",
      year           = "2017",
      pages          = "220-238",
      doi            = "10.1016/j.cpc.2016.10.013",
      eprint         = "1604.06792",
      archivePrefix  = "arXiv",
      primaryClass   = "hep-ph",
      reportNumber   = "FR-PHENO-2016-003, ICCUB-16-016",
      SLACcitation   = "%%CITATION = ARXIV:1604.06792;%%"
}

@article{Kallweit:2014xda,
    author = "Kallweit, Stefan and Lindert, Jonas M. and Maierh{\"o}fer, Philipp and Pozzorini, Stefano and Sch{\"o}nherr, Marek",
    title = "{NLO electroweak automation and precise predictions for W+multijet production at the LHC}",
    eprint = "1412.5157",
    archivePrefix = "arXiv",
    primaryClass = "hep-ph",
    reportNumber = "LPN14-127, IPPP-14-107, DCPT-14-214, MCNET-14-26, ZU-TH-42-14, MITP-14-102",
    doi = "10.1007/JHEP04(2015)012",
    journal = "JHEP",
    volume = "04",
    pages = "012",
    year = "2015"
}

@article{Czakon:2011xx,
  author         = "Czakon, Michal and Mitov, Alexander",
  title          = "{Top++: A program for the calculation of the top-pair cross-section at hadron colliders}",
  journal        = "Comput. Phys. Commun.",
  volume         = "185",
  year           = "2014",
  pages          = "2930",
  doi            = "10.1016/j.cpc.2014.06.021",
  eprint         = "1112.5675",
  archivePrefix  = "arXiv",
  primaryClass   = "hep-ph",
  reportNumber   = "CERN-PH-TH-2011-303, TTK-11-58",
}

@article{Agostinelli:2002hh,
      author         = "Agostinelli, S. and others",
      title          = "{\textsc{Geant4} -- a simulation toolkit}",
      journal        = "Nucl. Instrum. Meth. A",
      volume         = "506",
      year           = "2003",
      pages          = "250",
      doi            = "10.1016/S0168-9002(03)01368-8",
      reportNumber   = "SLAC-PUB-9350, FERMILAB-PUB-03-339",
      SLACcitation   = "%%CITATION = NUIMA,A506,250;%%"
}

@Article{Read:2002hq,
    Author = {Read, Alexander L.},
    Journal = {J. Phys. G},
    Pages = {2693},
    doi = "10.1088/0954-3899/28/10/313",
    Title = "{Presentation of search results: the \(CL_S\) technique}",
    Volume = {28},
    Year = {2002}
}


%% file: ATLAS.bib
@Article{PERF-2007-01,
    author         = "{ATLAS Collaboration}",
    title          = "{The ATLAS Experiment at the CERN Large Hadron Collider}",
    journal        = "JINST",
    volume         = "3",
    year           = "2008",
    pages          = "S08003",
    doi            = "10.1088/1748-0221/3/08/S08003",
    primaryClass   = "hep-ex",
}

@Article{SOFT-2010-01,
    author         = "{ATLAS Collaboration}",
    title          = "{The ATLAS Simulation Infrastructure}",
    journal        = "Eur. Phys. J. C",
    volume         = "70",
    year           = "2010",
    pages          = "823",
    doi            = "10.1140/epjc/s10052-010-1429-9",
    eprint         = "1005.4568",
    archivePrefix  = "arXiv",
    primaryClass   = "physics.ins-det",
}

@Article{TOPQ-2012-19,
    author         = "{ATLAS Collaboration}",
    title          = "{Search for \(W' \rightarrow t\bar{b}\) in the lepton plus jets final state in proton--proton collisions at a centre-of-mass energy of \(\sqrt{s} = 8\,\text{TeV}\) with the ATLAS detector}",
    journal        = "Phys. Lett. B",
    volume         = "743",
    year           = "2015",
    pages          = "235",
    doi            = "10.1016/j.physletb.2015.02.051",
    reportNumber   = "CERN-PH-EP-2014-232",
    eprint         = "1410.4103",
    archivePrefix  = "arXiv",
    primaryClass   = "hep-ex",
}

@Article{EXOT-2013-14,
    author         = "{ATLAS Collaboration}",
    title          = "{Search for \(W' \rightarrow tb \rightarrow qqbb\) decays in \(pp\) collisions at \(\sqrt{s} = 8\,\text{TeV}\) with the ATLAS detector}",
    journal        = "Eur. Phys. J. C",
    volume         = "75",
    year           = "2015",
    pages          = "165",
    doi            = "10.1140/epjc/s10052-015-3372-2",
    reportNumber   = "CERN-PH-EP-2014-152",
    eprint         = "1408.0886",
    archivePrefix  = "arXiv",
    primaryClass   = "hep-ex",
}

@Article{PERF-2014-03,
    author         = "{ATLAS Collaboration}",
    title          = "{Performance of pile-up mitigation techniques for jets in \(pp\) collisions at \(\sqrt{s} = 8\,\text{TeV}\) using the ATLAS detector}",
    journal        = "Eur. Phys. J. C",
    volume         = "76",
    year           = "2016",
    pages          = "581",
    doi            = "10.1140/epjc/s10052-016-4395-z",
    reportNumber   = "CERN-PH-EP-2015-206",
    eprint         = "1510.03823",
    archivePrefix  = "arXiv",
    primaryClass   = "hep-ex",
}

@Article{PERF-2014-07,
    author         = "{ATLAS Collaboration}",
    title          = "{Topological cell clustering in the ATLAS calorimeters and its performance in LHC Run~1}",
    journal        = "Eur. Phys. J. C",
    volume         = "77",
    year           = "2017",
    pages          = "490",
    doi            = "10.1140/epjc/s10052-017-5004-5",
    reportNumber   = "CERN-PH-EP-2015-304",
    eprint         = "1603.02934",
    archivePrefix  = "arXiv",
    primaryClass   = "hep-ex",
}

@Article{PERF-2015-09,
    author         = "{ATLAS Collaboration}",
    title          = "{Jet reconstruction and performance using particle flow with the ATLAS Detector}",
    journal        = "Eur. Phys. J. C",
    volume         = "77",
    year           = "2017",
    pages          = "466",
    doi            = "10.1140/epjc/s10052-017-5031-2",
    reportNumber   = "CERN-EP-2017-024",
    eprint         = "1703.10485",
    archivePrefix  = "arXiv",
    primaryClass   = "hep-ex",
}

@Article{PERF-2015-10,
    author         = "{ATLAS Collaboration}",
    title          = "{Muon reconstruction performance of the ATLAS detector in proton--proton collision data at \(\sqrt{s} = 13\,\text{TeV}\)}",
    journal        = "Eur. Phys. J. C",
    volume         = "76",
    year           = "2016",
    pages          = "292",
    doi            = "10.1140/epjc/s10052-016-4120-y",
    reportNumber   = "CERN-EP-2016-033",
    eprint         = "1603.05598",
    archivePrefix  = "arXiv",
    primaryClass   = "hep-ex",
}

@Article{EXOT-2016-04,
    author         = "{ATLAS Collaboration}",
    title          = "{Search for heavy Higgs bosons \(A/H\) decaying to a top quark pair in \(pp\) collisions at \(\sqrt{s} = 8\,\text{TeV}\) with the ATLAS detector}",
    journal        = "Phys. Rev. Lett.",
    volume         = "119",
    year           = "2017",
    pages          = "191803",
    doi            = "10.1103/PhysRevLett.119.191803",
    reportNumber   = "CERN-EP-2017-134",
    eprint         = "1707.06025",
    archivePrefix  = "arXiv",
    primaryClass   = "hep-ex",
}

@Article{EXOT-2016-18,
    author         = "{ATLAS Collaboration}",
    title          = "{Search for vector-boson resonances decaying to a top quark and bottom quark in the lepton plus jets final state in \(pp\) collisions at \(\sqrt{s} = 13\,\text{TeV}\) with the ATLAS detector}",
    journal        = "Phys. Lett. B",
    volume         = "788",
    year           = "2019",
    pages          = "347",
    doi            = "10.1016/j.physletb.2018.11.032",
    reportNumber   = "CERN-EP-2018-142",
    eprint         = "1807.10473",
    archivePrefix  = "arXiv",
    primaryClass   = "hep-ex",
}

@Article{TRIG-2016-01,
    author         = "{ATLAS Collaboration}",
    title          = "{Performance of the ATLAS trigger system in 2015}",
    journal        = "Eur. Phys. J. C",
    volume         = "77",
    year           = "2017",
    pages          = "317",
    doi            = "10.1140/epjc/s10052-017-4852-3",
    reportNumber   = "CERN-EP-2016-241",
    eprint         = "1611.09661",
    archivePrefix  = "arXiv",
    primaryClass   = "hep-ex",
}

@Article{EXOT-2017-02,
    author         = "{ATLAS Collaboration}",
    title          = "{Search for \(W' \rightarrow tb\) decays in the hadronic final state using \(pp\) collisions at \(\sqrt{s} = 13\,\text{TeV}\) with the ATLAS detector}",
    journal        = "Phys. Lett. B",
    volume         = "781",
    year           = "2018",
    pages          = "327",
    doi            = "10.1016/j.physletb.2018.03.036",
    reportNumber   = "CERN-EP-2017-340",
    eprint         = "1801.07893",
    archivePrefix  = "arXiv",
    primaryClass   = "hep-ex",
}

@Article{HIGG-2017-03,
    author         = "{ATLAS Collaboration}",
    title          = "{Search for the standard model Higgs boson produced in association with top quarks and decaying into a \(b\bar{b}\) pair in \(pp\) collisions at \(\sqrt{s} = 13\,\text{TeV}\) with the ATLAS detector}",
    journal        = "Phys. Rev. D",
    volume         = "97",
    year           = "2018",
    pages          = "072016",
    doi            = "10.1103/PhysRevD.97.072016",
    reportNumber   = "CERN-EP-2017-291",
    eprint         = "1712.08895",
    archivePrefix  = "arXiv",
    primaryClass   = "hep-ex",
}

@Article{PERF-2017-01,
    author         = "{ATLAS Collaboration}",
    title          = "{Electron reconstruction and identification in the ATLAS experiment using the 2015 and 2016 LHC proton--proton collision data at \(\sqrt{s} = 13\,\text{TeV}\)}",
    journal        = "Eur. Phys. J. C",
    volume         = "79",
    year           = "2019",
    pages          = "639",
    doi            = "10.1140/epjc/s10052-019-7140-6",
    reportNumber   = "CERN-EP-2018-273",
    eprint         = "1902.04655",
    archivePrefix  = "arXiv",
    primaryClass   = "physics.ins-det",
}

@Article{DAPR-2018-01,
    author         = "{ATLAS Collaboration}",
    title          = "{ATLAS data quality operations and performance for 2015--2018 data-taking}",
    journal        = "JINST",
    volume         = "15",
    year           = "2020",
    pages          = "P04003",
    doi            = "10.1088/1748-0221/15/04/P04003",
    reportNumber   = "CERN-EP-2019-207",
    eprint         = "1911.04632",
    archivePrefix  = "arXiv",
    primaryClass   = "physics.ins-det",
}

@Article{EGAM-2018-01,
    author         = "{ATLAS Collaboration}",
    title          = "{Electron and photon performance measurements with the ATLAS detector using the 2015--2017 LHC proton--proton collision data}",
    journal        = "JINST",
    volume         = "14",
    year           = "2019",
    pages          = "P12006",
    doi            = "10.1088/1748-0221/14/12/P12006",
    reportNumber   = "CERN-EP-2019-145",
    eprint         = "1908.00005",
    archivePrefix  = "arXiv",
    primaryClass   = "hep-ex",
}

@Article{FTAG-2018-01,
    author         = "{ATLAS Collaboration}",
    title          = "{ATLAS \(b\)-jet identification performance and efficiency measurement with \(t\bar{t}\) events in \(pp\) collisions at \(\sqrt{s} = 13\,\text{TeV}\)}",
    journal        = "Eur. Phys. J. C",
    volume         = "79",
    year           = "2019",
    pages          = "970",
    doi            = "10.1140/epjc/s10052-019-7450-8",
    reportNumber   = "CERN-EP-2019-132",
    eprint         = "1907.05120",
    archivePrefix  = "arXiv",
    primaryClass   = "hep-ex",
}

@Article{JETM-2018-02,
    author         = "{ATLAS Collaboration}",
    title          = "{In situ calibration of large-radius jet energy and mass in \(13\,\text{TeV}\) proton--proton collisions with the ATLAS detector}",
    journal        = "Eur. Phys. J. C",
    volume         = "79",
    year           = "2019",
    pages          = "135",
    doi            = "10.1140/epjc/s10052-019-6632-8",
    reportNumber   = "CERN-EP-2018-191",
    eprint         = "1807.09477",
    archivePrefix  = "arXiv",
    primaryClass   = "hep-ex",
}

@Article{JETM-2018-03,
    author         = "{ATLAS Collaboration}",
    title          = "{Performance of top-quark and \(W\)-boson tagging with ATLAS in Run~2 of the LHC}",
    journal        = "Eur. Phys. J. C",
    volume         = "79",
    year           = "2019",
    pages          = "375",
    doi            = "10.1140/epjc/s10052-019-6847-8",
    reportNumber   = "CERN-EP-2018-192",
    eprint         = "1808.07858",
    archivePrefix  = "arXiv",
    primaryClass   = "hep-ex",
}

@Article{JETM-2018-05,
    author         = "{ATLAS Collaboration}",
    title          = "{Jet energy scale and resolution measured in proton--proton collisions at \(\sqrt{s} = 13\,\text{TeV}\) with the ATLAS detector}",
    journal        = "Eur. Phys. J. C",
    volume         = "81",
    year           = "2020",
    pages          = "689",
    doi            = "10.1140/epjc/s10052-021-09402-3",
    reportNumber   = "CERN-EP-2020-083",
    eprint         = "2007.02645",
    archivePrefix  = "arXiv",
    primaryClass   = "hep-ex",
}

@Article{JETM-2018-06,
    author         = "{ATLAS Collaboration}",
    title          = "{Optimisation of large-radius jet reconstruction for the ATLAS detector in \(13\,\text{TeV}\) proton--proton collisions}",
    journal        = "Eur. Phys. J. C",
    volume         = "81",
    year           = "2020",
    pages          = "334",
    doi            = "10.1140/epjc/s10052-021-09054-3",
    reportNumber   = "CERN-EP-2020-134",
    eprint         = "2009.04986",
    archivePrefix  = "arXiv",
    primaryClass   = "hep-ex",
}

@Article{MUON-2018-03,
    author         = "{ATLAS Collaboration}",
    title          = "{Muon reconstruction and identification efficiency in ATLAS using the full Run~2 \(pp\) collision data set at \(\sqrt{s} = 13\,\text{TeV}\)}",
    journal        = "Eur. Phys. J. C",
    volume         = "81",
    year           = "2021",
    pages          = "578",
    doi            = "10.1140/epjc/s10052-021-09233-2",
    reportNumber   = "CERN-EP-2020-199",
    eprint         = "2012.00578",
    archivePrefix  = "arXiv",
    primaryClass   = "hep-ex",
}

@Article{PIX-2018-001,
    author         = "Abbott, B. and others",
    title          = "{Production and integration of the ATLAS Insertable B-Layer}",
    journal        = "JINST",
    volume         = "13",
    year           = "2018",
    pages          = "T05008",
    doi            = "10.1088/1748-0221/13/05/T05008",
    eprint         = "1803.00844",
    archivePrefix  = "arXiv",
    primaryClass   = "physics.ins-det",
}

@Article{FTAG-2019-02,
    author         = "{ATLAS Collaboration}",
    title          = "{Calibration of the light-flavour jet mistagging efficiency of the \(b\)-tagging algorithms with \(Z\)+jets events using \(139\,\text{fb}^{-1}\) of ATLAS proton--proton collision data at \(\sqrt{s} = 13\,\text{TeV}\)}",
    journal        = "Eur. Phys. J. C",
    volume         = "83",
    year           = "2023",
    pages          = "728",
    doi            = "10.1140/epjc/s10052-023-11736-z",
    reportNumber   = "CERN-EP-2022-211",
    eprint         = "2301.06319",
    archivePrefix  = "arXiv",
    primaryClass   = "hep-ex",
}

@Article{FTAG-2019-07,
    author         = "{ATLAS Collaboration}",
    title          = "{ATLAS flavour-tagging algorithms for the LHC Run~2 \(pp\) collision dataset}",
    journal        = "Eur. Phys. J. C",
    volume         = "83",
    year           = "2022",
    pages          = "681",
    doi            = "10.1140/epjc/s10052-023-11699-1",
    reportNumber   = "CERN-EP-2022-226",
    eprint         = "2211.16345",
    archivePrefix  = "arXiv",
    primaryClass   = "physics.data-an",
}

@Article{FTAG-2020-08,
    author         = "{ATLAS Collaboration}",
    title          = "{Measurement of the \(c\)-jet mistagging efficiency in \(t\bar{t}\) events using \(pp\) collision data at \(\sqrt{s} = 13\,\text{TeV}\) collected with the ATLAS detector}",
    journal        = "Eur. Phys. J. C",
    volume         = "82",
    year           = "2022",
    pages          = "95",
    doi            = "10.1140/epjc/s10052-021-09843-w",
    reportNumber   = "CERN-EP-2021-166",
    eprint         = "2109.10627",
    archivePrefix  = "arXiv",
    primaryClass   = "hep-ex",
}

@Article{EXOT-2021-36,
    author         = "{ATLAS Collaboration}",
    title          = "{Search for vector-boson resonances decaying into a top quark and a bottom quark using \(pp\) collisions at \(\sqrt{s} = 13\) TeV with the ATLAS detector}",
    year           = "2023",
    reportNumber   = "CERN-EP-2023-149",
    eprint         = "2308.08521",
    archivePrefix  = "arXiv",
    primaryClass   = "hep-ex",
}

@Booklet{ATL-SOFT-PUB-2021-001,
    author         = "{ATLAS Collaboration}",
    title          = "{The ATLAS Collaboration Software and Firmware}",
    howpublished   = "{ATL-SOFT-PUB-2021-001}",
    url            = "https://cds.cern.ch/record/2767187",
    year           = "2021",
}

@Booklet{ATL-SOFT-PUB-2023-001,
    author         = "{ATLAS Collaboration}",
    title          = "{ATLAS Computing Acknowledgements}",
    howpublished   = "{ATL-SOFT-PUB-2023-001}",
    url            = "https://cds.cern.ch/record/2869272",
    year           = "2023",
}

@Report{ATLAS-TDR-19,
    author         = "{ATLAS Collaboration}",
    title          = "{ATLAS Insertable B-Layer: Technical Design Report}",
    type           = "ATLAS-TDR-19; CERN-LHCC-2010-013",
    year           = "2010",
    url            = "https://cds.cern.ch/record/1291633",
    related        = "ATLAS-TDR-19-addm",
    relatedstring  = "Addendum:",
}


%% file: CMS.bib
@Article{CMS-B2G-12-010,
    author         = "{CMS Collaboration}",
    title          = "{Search for \(W' \rightarrow tb\) decays in the lepton+jets final state in \(pp\) collisions at \(\sqrt{s} = 8\,\text{TeV}\)}",
    journal        = "JHEP",
    volume         = "05",
    year           = "2014",
    pages          = "108",
    doi            = "10.1007/JHEP05(2014)108",
    reportNumber   = "CERN-PH-EP-2014-011",
    eprint         = "1402.2176",
    archivePrefix  = "arXiv",
    primaryClass   = "hep-ex",
}

@Article{CMS-EXO-12-001,
    author         = "{CMS Collaboration}",
    title          = "{Search for a \(W'\) boson decaying to a bottom quark and a top quark in \(pp\) collisions at \(\sqrt{s} = 7\,\text{TeV}\)}",
    journal        = "Phys. Lett. B",
    volume         = "718",
    year           = "2013",
    pages          = "1229",
    doi            = "10.1016/j.physletb.2012.12.008",
    reportNumber   = "CERN-PH-EP-2012-192",
    eprint         = "1208.0956",
    archivePrefix  = "arXiv",
    primaryClass   = "hep-ex",
}

@Article{CMS-B2G-17-010,
    author         = "{CMS Collaboration}",
    title          = "{Search for heavy resonances decaying to a top quark and a bottom quark in the lepton+jets final state in proton--proton collisions at \(13\,\text{TeV}\)}",
    journal        = "Phys. Lett. B",
    volume         = "777",
    year           = "2018",
    pages          = "39",
    doi            = "10.1016/j.physletb.2017.12.006",
    reportNumber   = "CERN-EP-2017-196",
    eprint         = "1708.08539",
    archivePrefix  = "arXiv",
    primaryClass   = "hep-ex",
}

@Article{CMS-B2G-20-005,
    author         = "{CMS Collaboration}",
    title          = "{Search for \(W'\) bosons decaying to a top and a bottom quark at \(\sqrt{s} = 13\,\text{TeV}\) in the hadronic final state}",
    journal        = "Phys. Lett. B",
    volume         = "820",
    year           = "2021",
    pages          = "136535",
    doi            = "10.1016/j.physletb.2021.136535",
    reportNumber   = "CERN-EP-2021-049",
    eprint         = "2104.04831",
    archivePrefix  = "arXiv",
    primaryClass   = "hep-ex",
}


%% file: ConfNotes.bib
@Booklet{ATLAS-CONF-2018-023,
    author         = "{ATLAS Collaboration}",
    title          = "{\(E_{\text{T}}^{\text{miss}}\) performance in the ATLAS detector using 2015--2016 LHC \(pp\) collisions}",
    howpublished   = "{ATLAS-CONF-2018-023}",
    url            = "https://cds.cern.ch/record/2625233",
    year           = "2018",
}

@Booklet{ATLAS-CONF-2019-021,
    author         = "{ATLAS Collaboration}",
    title          = "{Luminosity determination in \(pp\) collisions at \(\sqrt{s} = 13\,\text{TeV}\) using the ATLAS detector at the LHC}",
    howpublished   = "{ATLAS-CONF-2019-021}",
    url            = "https://cds.cern.ch/record/2677054",
    year           = "2019",
}

@Booklet{ATLAS-CONF-2020-022,
    author         = "{ATLAS Collaboration}",
    title          = "{Measurement of the ATLAS Detector Jet Mass Response using Forward Folding with \(80\,\text{fb}^{-1}\) of \(\sqrt{s} = 13\,\text{TeV}\) \(pp\) data}",
    howpublished   = "{ATLAS-CONF-2020-022}",
    url            = "https://cds.cern.ch/record/2724442",
    year           = "2020",
}


%% file: PubNotes.bib
@Booklet{ATL-PHYS-PUB-2014-021,
    author         = "{ATLAS Collaboration}",
    title          = "{ATLAS Pythia~8 tunes to \(7~\text{TeV}\) data}",
    howpublished   = "{ATL-PHYS-PUB-2014-021}",
    url            = "https://cds.cern.ch/record/1966419",
    year           = "2014",
}

@Booklet{ATL-PHYS-PUB-2016-020,
    author         = "{ATLAS Collaboration}",
    title          = "{Studies on top-quark Monte Carlo modelling for Top2016}",
    howpublished   = "{ATL-PHYS-PUB-2016-020}",
    url            = "https://cds.cern.ch/record/2216168",
    year           = "2016",
}

@Booklet{ATL-PHYS-PUB-2020-017,
    author         = "{ATLAS Collaboration}",
    title          = "{Boosted hadronic vector boson and top quark tagging with ATLAS using Run 2 data}",
    howpublished   = "{ATL-PHYS-PUB-2020-017}",
    url            = "https://cds.cern.ch/record/2724149",
    year           = "2020",
}
